\documentclass[sigconf]{acmart}

\usepackage{tabu}                      
\usepackage{booktabs}                  
\usepackage{balance}
\usepackage{todonotes}
\usepackage{booktabs}
\usepackage{multirow}
\usepackage{rotating,tabularx}
\usepackage{xcolor,colortbl}

\usepackage{arydshln}
\usepackage{multirow}
\usepackage[sf]{subfigure}
\usepackage{verbatim}

\usepackage{multirow}
\usepackage{nicematrix}

\usepackage{graphicx}
\usepackage{caption}

\author{Hanseob Kim}
\affiliation{%
 \institution{Korea University}
\city{Seoul}
 \country{Republic of Korea}}
\email{khseob0715@korea.ac.kr}

\author{Bin Han}
\affiliation{%
 \institution{University of Southern California}
\city{Los Angeles, CA}
 \country{USA}}
\email{binhan@usc.edu}

\author{Jieun Kim}
\affiliation{%
 \institution{\hspace*{-1em}\mbox{Korea Institute of Science and Technology}}
\city{Seoul}
 \country{Republic of Korea}}
\email{092296@kist.re.kr}

\author{\mbox{Muhammad Firdaus Syawaludin}}
\affiliation{%
 \institution{University of Indonesia}
\city{Depok}
 \country{Indonesia}}
\email{m.firdaus04@ui.ac.id}

\author{Gerard Jounghyun Kim}
\affiliation{%
 \institution{Korea University}
 \city{Seoul}
 \country{Republic of Korea}
}
\email{gjkim@korea.ac.kr}

\author{Jae-In Hwang}
\affiliation{%
 \institution{\hspace*{-1em}\mbox{Korea Institute of Science and Technology}}
 \city{Seoul}
 \country{Republic of Korea}
}
\email{hji@kist.re.kr}

\AtBeginDocument{%
  }

\setcopyright{acmlicensed}
\copyrightyear{2024}
\acmYear{2024}

\acmConference[CHI '24]{Proceedings of the CHI Conference on Human Factors in Computing Systems}{May 11--16, 2024}{Honolulu, HI, USA}
\acmBooktitle{Proceedings of the CHI Conference on Human Factors in Computing Systems (CHI '24), May 11--16, 2024, Honolulu, HI, USA}
\acmDOI{10.1145/3613904.3642917}
\acmISBN{979-8-4007-0330-0/24/05}

\begin{document}

\title[Engaged and Affective Virtual Agents]{Engaged and Affective Virtual Agents: Their Impact on Social Presence, Trustworthiness, and Decision-Making in the Group Discussion}



\renewcommand{\shortauthors}{Kim et al.}

\begin{abstract} 
This study investigates how different virtual agent (VA) behaviors influence subjects' perceptions and group decision-making.
Participants carried out 
experimental group discussions with a VA exhibiting varying levels of engagement and affective behavior.
Engagement refers to the VA's focus on the group task, whereas affective behavior reflects the VA's emotional state.
The findings revealed that VA's engagements effectively captured participants' attention even in the group setting and enhanced group synergy, thereby facilitating more in-depth discussion and producing better consensus.
On the other hand, VA's affective behavior negatively affected the
perceived social presence and trustworthiness. Consequently, 
in the context of group discussion, participants preferred the engaged and non-affective VA to the non-engaged and affective VA.
The study provides valuable insights for improving the VA's behavioral design as a team member for collaborative tasks.
\end{abstract}




\maketitle

\setlength\doublerulesep{0.8pt}

\section{Introduction}

Ways to design and effectively utilize interactive virtual agents (VAs) for various social and professional activities are being extensively investigated, for instance, online gatherings~\cite{gatto2022met},
e-learning~\cite{hoque2013mach,sajjadi2020effect, daher2020embodied}, and even medical counseling~\cite{devault2014simsensei, kim2019effects}.
These studies explored the design space of the VAs' verbal and non-verbal social cues as an effective avenue to enhance their social presence~\cite{devault2014simsensei, nam2022emotionally}.
Through their tone of voice and facial expressions, the VA can evoke a range of emotions or affectiveness~\cite{celsoDecisionMaking, GerdIeeeVR2023, de2011impact}.
The VA's level of engagement can also be sensed through its gazing behavior and body gestures~\cite{ali2020automatic, kum2022can}. 
Such engagement and affectiveness have been proven to be key components of VAs in the modulation of how humans perceive and interact with them, thereby improving the overall quality of communication~\cite{PaivaEtAl2017_EmpathyInVirtualAgents, huang2011virtual, devault2014simsensei}.

In this context, our study delves into the design of VAs' behaviors for collaborative 
``group'' tasks (vs. dyadic), anticipating such use cases to become more common in the near future.
The success of group activities may be influenced by 
several factors, including the aforementioned engagement and affect, but possibly in a way different from the one-to-one setting. 
Fostering the emotional involvement~\cite{fischer2016social, Unterweger2022_EmotionalInvolvement} can be beneficial to group discussions, leading to a quicker arrival at a consensus and an enhanced sense of unity and connection among participants~\cite{knierim2017emotion, paakkanen2021responding}.
This, in turn, would strengthen the overall group productivity and encourage the members to reassess their viewpoints~\cite{clarke2010emotional, Kerkhof2006_MakingADifference}.

However, emotional involvement has potential drawbacks as well. 
Heightened emotions can impede learning and disrupt group dynamics, sometimes
leading to misinterpretation or neglect of messages~\cite{celsoDecisionMaking, trevors2016identity, GerdIeeeVR2023}. 
Furthermore, it can bring about subjective and personal arguments, 
hampering the group's ability to reach an agreement amicably~\cite{abiodun2014organizational}. 
It may become imperative for group members to consciously restrain and suppress their emotions to maintain social relationships, a process known as emotional suppression~\cite{gross1993emotional}.
Therefore, understanding this dichotomy is crucial in designing VAs' behaviors that effectively contribute to group discussions while mitigating emotional disruptions. 

While previous studies have shown that VAs with rich emotions and compelling behavior are well received~\cite{PaivaEtAl2017_EmpathyInVirtualAgents} 
and how interactive features are needed to provide rapport to users~\cite{devault2014simsensei}, 
these studies were mostly confined to dyadic interactions rather than group settings. 
When considering the design of engaging VAs for group interactions, additional factors need to be taken into account (e.g., shifting conversational focus, relatively distributing and directing attention~\cite{sebo2020robots, visualattentionInGroup, divekar2019you}).
Moreover, in the group discussion, humans typically adhere to ``social'' norms~\cite{zeman1996display, kramer2002communication} and restrain their emotions to enhance group performance~\cite{gross1993emotional}.
Thus, developing specific provisions, requirements, or guidelines would be required for the VA to facilitate group discussions in the right and natural way.

To examine the impact of engaging and affective behaviors in groups, we conducted a human-subject study in which a VA and two human participants carried out group discussions and made group decisions.
This study aimed to explore how various degrees of the VA's engagement and affective behaviors influence users' behavior and perceptions of VAs.
Specifically, we sought to understand the dynamics of human-VA interactions within a group. 
The experiment was designed to address the following research questions (RQs):

\begin{itemize}
    \item \textbf{RQ1:} 
    What is the effect of the level of the VAs' engagement in group discussions? 

    \item \textbf{RQ2:}
    What is the effect of the level of the VAs' emotional expression in group discussions? 
   
    \item \textbf{RQ3:} 
    What type of a VA is appropriate for facilitating the group's decision-making in terms of its engaging and/or affective behaviors?
\end{itemize}

\section{Related Works}
\label{relatedworks}

\subsection{Virtual Agents with Engaging Behavior}
\label{sec:2.1}
Embodied conversational agents (referred to as VAs~\cite{von2010doesn, bailenson2006transformed}) have distinct features compared to AI systems without physical or graphical form~\cite{kangsooReducing, kim2019effects, wang2019exploring}, such as Siri, Bixby, and ChatGPT.
One such key feature of VAs, particularly those with human-like appearances, is the ability to engage through visual cues displaying 
enthusiastic participation during user-agent interaction~\cite{kim2019effects, divekar2019you, 8613756}.
This engaging behavior has been shown to attract users' attention and enhance users' experience with VAs~\cite{gratch2007can}. 
For example, DeVault et al.~\cite{devault2014simsensei} demonstrated the effectiveness of the VA's listening gestures, such as nodding and gazing, in building rapport during clinical interviews. 
Another line of research~\cite{ali2020automatic, kum2022can} focused on improving the naturalness of conversational interactions with hand gestures and synchronizing them with speech. Research has generally shown that accurately portraying the VA's reactions with body movements in response to current situations can enhance the social presence and the overall interaction experience~\cite{walker2019influence, wang2019exploring, hanseob2021visualeffect, myunho2016wobblytable}.

It is clear that engaging behavior is crucial for successful interactions between VAs and users. 
However, most existing research predominantly has focused on the 
dyadic (one-to-one) interactions, with only a limited look at the
one-to-many~\cite{ter2011design, rehm2005they} and many-to-many~\cite{yumak2014tracking} scenarios.
This indicates a notable gap in understanding VAs in group activities with human users. 

Research in Human-Robot Interaction has 
found that in groups, the presence of other humans detracted attention from robots, potentially reducing their effectiveness~\cite{sebo2020robots}.  
Such a finding raises a similar possibility - i.e., the VA's engaging behavior might have a diminished impact on user perception and group task performance (related to RQ1).
Given the potential of VAs to become pervasive and interacting with them 
more commonplace in the future, exploring ways to supplement the VA's abilities in groups is necessary.

\subsection{Virtual Agents with Affective Behaviors}
Aside from the engaging quality, the emotional/affective behavior of the VA not only serves as an alternative to captivate users' interest within group settings but also acts as another potent element for improving users' interactive and communicative experience~\cite{GerdIeeeVR2023, fox2015avatars, jones2014expressing}.
Affective expressions encompass both direct verbal cues~\cite{nam2022emotionally, torre2019effect} and nonverbal ones, including facial expressions~\cite{choi2012affective, fox2015avatars, kimmel2023let, GerdIeeeVR2023}, bodily gestures~\cite{randhavane2019eva, nayak2005emotional, jones2014expressing}, and voice tones~\cite{moridis2012affective, torre2020if, elkins2013sound}. 

Numerous studies have shown that VAs exhibiting a broad range of emotions and possessing unique personalities can enhance the user's perception of VAs~\cite{randhavane2019eva, moridis2012affective}.
Nam et al.~\cite{nam2022emotionally} found that VA's dynamic emotions positively affected realism, learning effectiveness, and usefulness in the nursing training experiences.
Moridis et al.~\cite{moridis2012affective} observed that when a VA mirrors 
the user's emotions (e.g., happiness, sadness, and fear) through facial and vocal expression, it can effectively transform and reinforce one's emotions.
The VA's rich facial expressions increased the user's belief, cooperation~\cite{scharlemann2001value, choi2012affective}, and their co-presence~\cite{kimmel2023let}.
Furthermore, body gestures (e.g., gaze, gait, and posture) for VAs corresponding to happy, sad, or angry emotions can enhance the social presence~\cite{randhavane2019eva}, friendliness~\cite{jones2014expressing}, and trustworthiness~\cite{devault2014simsensei}.

However, it is important to consider the context in which VAs will be utilized~\cite{choi2012affective, torre2019effect, celsoDecisionMaking, GerdIeeeVR2023}. 
Research has shown that in collaborative tasks, negative emotions, particularly anger, can reduce the perceived trustworthiness of the VA, leading to decreased cooperation~\cite{torre2019effect, celsoDecisionMaking, choi2012affective}.
In social contexts, humans often inhibit or suppress their emotions, adopting a neutral tone – known as emotional suppression~\cite{gross1993emotional} – 
to foster collaboration~\cite{xavier2005you}, social harmony~\cite{zeman1996display}, and positive relationships, especially within organizational settings~\cite{grandey2000emotional}. 
Thus, a more judicious approach is necessary when designing affective VAs for social contexts (related to RQ2), as the discrepancy between VA-to-human and human-to-human interactions may potentially lead to adverse effects for users~\cite{melo2010influence}.

\subsection{Virtual Agent's Roles in Group Decision-Making}
Research on interpersonal relationships provides the basis for 
hypotheses about the roles and important features of interactive VAs 
in group settings for ``enhancing group synergy (the ability to function more effectively as a group than divided)'' or 
``influencing decision-making'' (related to RQ3).
Zhou et al.~\cite{zhou2020emotional} emphasized the need for emotional intelligence (EI), defined as the ability to understand and manage the emotions of groups to promote group performance. Melsic et al.~\cite{meslec2013too} stated that if one person is more rational than others, they can lead to an increase in group synergy. 
This, in turn, can enhance the group's ability to understand problems through collaborative elements and improve decision-making. 
Thus, designing VAs with EI and engaging with rational behaviors could be a way to improve both group synergy and performance.

However, Chang et al.~\cite{competitivewithrobot} found that humans acted more competitively towards robots in groups (despite having both positive and negative emotions), resulting in a limited impact on decision-making.
This result stems from the in-group bias, whereby people tend to favor in-group (i.e., humans) to out-group (i.e., robots) members~\cite{rudman2004gender}. 
Therefore, there is a possibility that individuals perceive VAs as out-group and their interactions as a form of competition, which could limit the VA's ability to contribute to group synergy or influence decision-making. 
Designing VA to mimic in-group members with engaging/affective behaviors might be an alternative; however, this could have a limited impact on group dynamics.

On the contrary, Kleef et al.~\cite{van2004interpersonal} noted that individuals tended to concede more to negative counterparts than to positive ones. 
This implies that if VAs display negative emotions, they may influence decision-making more. 
However, such behavior is more about just avoiding conflict with the VA without much trust and may not necessarily result in group synergy.
We posit that the design of VAs should be contingent on the 
specific roles and objectives in the group, e.g., whether it is to enhance group synergy or just to influence decision-making (e.g., VA as the leader or persuader). 
In summary, this study aims to investigate and establish design guidelines for tailoring VAs based on their roles in group activities.

\begin{figure*}[t]{
  \centering
  \includegraphics[width=1\linewidth]{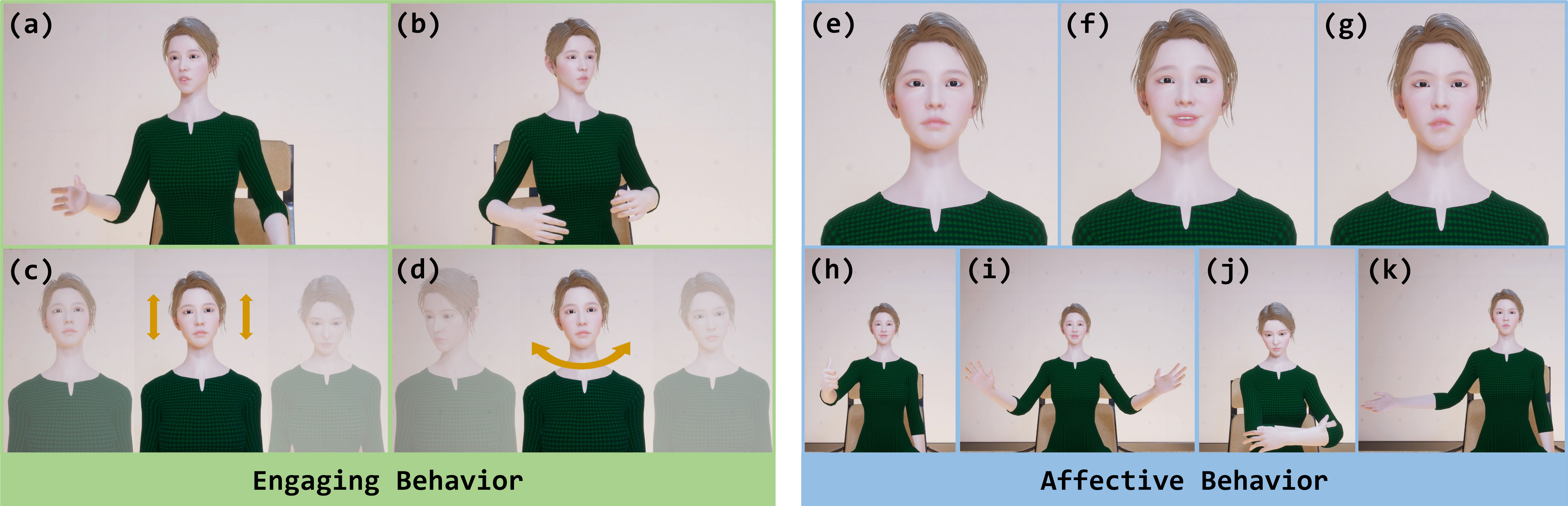}
  \vspace{-1.8em}
  \Description{
Figure 1: The two main factors/traits in this ``virtual agent for group discussion" design space study: (1) Engaging behavior and (2) Affective behavior.
"Engaging" behavior is the VA's level of engagement in the group, expressed non-verbally.
Examples include directing gaze and posture toward interlocutors while using co-speech gestures (a, b); and exhibiting listening gestures, such as a head nod and shake, in response to the interlocutor's statements (c, d).
"Affective" behavior is the expression of VA's emotions through verbal and non-verbal behavior.
Examples of facial expressions include neutral (e), happy (f), and angry (g); positive gestures, such as thumbs up (h) and cheering arms (i); and negative gestures, such as crossed arms (j) and a hand shrug (k).
Affective behavior can also be shown by the tone of voice and emotional utterance, which may be positive or negative.}
  \caption{
 The two main factors/traits in this ``virtual agent for group discussion'' design space study: (1) Engaging behavior and (2) Affective behavior.  
 ``Engaging'' behavior is the VA's level of engagement in the group, expressed non-verbally.
  Examples include directing gaze and posture toward interlocutors while using co-speech gestures (a, b); 
  and exhibiting listening gestures, such as a head nod and shake, in response to the interlocutor's statements (c, d). 
  ``Affective'' behavior is the expression of VA's emotions through verbal and non-verbal behavior.
Examples of facial expressions include neutral (e), happy (f), and angry (g); positive gestures, such as thumbs up (h) and cheering arms (i); and negative gestures, such as crossed arms (j) and a hand shrug (k).
Affective behavior can also be shown by the tone of voice and emotional utterance, which may be positive or negative.
}
\vspace{-0.5em}
\label{fig:figure.1}
}
\end{figure*}

\section{Experiment}
\label{experiment}

\subsection{Experiment Design}
\label{independentvar}

This section presents a human-subject study that investigates how engaging and affective behaviors of VAs
affect group decision-making. 
The experiment followed a 2 $\times$ 2 within-subject design, focusing on two independent variables: 
(1) whether the VA exhibited engaging behavior or not (E/NE), 
and (2) whether the VA expressed affective behavior or not (A/NA), as illustrated in Figure~\ref{fig:figure.1}. 
Further details are explained in subsequent sections.

\subsubsection{Engaging and Non-Engaging behavior (E/NE)}

Engaging behavior (E) aims to make the user aware of the VA's active participation in the group discussion. 
Our study incorporated four non-verbal behaviors, which also have been commonly utilized in previous research~\cite{ali2020automatic, huang2011virtual, devault2014simsensei} - namely, \textit{mutual gaze, directed posture, listening gestures, and speaking gestures}.
In the setting of a group discussion, the VA primarily takes on two roles: listening to the participants and speaking with them.

While listening, the VA exhibits active engagement by altering its gazes and posture towards the current speaker in a natural manner, as shown in Figure~\ref{fig:figure.1} (a) and (b). 
This behavior is triggered by a speaker detection algorithm, detailed in Section~\ref{sec:speaker identification}.
Furthermore, the VA utilizes listening gestures, like nodding or shaking their head, to indicate attentiveness and interest during group discussions, as depicted in Figure~\ref{fig:figure.1} (c) and (d).

While speaking, the VA provides co-speech gestures to express 
active participation (see Figure~\ref{fig:figure.1} (a) and (b)). 
To generate semantically relevant gestures, 
we employed a large-scale text-to-gesture rule map~\cite{ali2022} with cleaned motion capture data (mostly by arms and hand movements) and sentence-level embeddings. 
The sentence level embeddings were extracted using the Sentence-BERT~\cite{reimers-2019-sentence-bert} from transcripts compiled from over 250 hours of (motion-captured) videos by different speakers on diverse topics (e.g., current affairs, topical lectures, religion, and politics).
The rule map was applied following the scripts of the experimental scenario to generate the appropriate accompanying gestures and display them toward the counterpart participant when the VA started speaking.

In contrast, the VA in Non-Engaging behavior (NE) interacts with participants while staring at the front during group discussions, without any of the behaviors mentioned earlier being applied.

\subsubsection{Affective and Non-Affective behavior (A/NA)}

Affective behavior (A) is designed to convey the VA's emotions during the group discussion.
Affective behavior involves both verbal and non-verbal cues, such as facial expressions, emotional gestures, tone of voice, and emotional utterances.  
These behaviors have been identified as representative behaviors based on previous studies~\cite{choi2012affective, randhavane2019eva,devault2014simsensei, nam2022emotionally}.
During the group discussion, the affective VA expresses ``positive'' emotions when participants' statements align with the VA's opinions, which are considered correct answers. 
On the other hand, the VA exhibits ``negative'' emotions when participants provide wrong answers.
It should be noted that the VA always provides the correct answer; however, participants are unaware of the accuracy or validity of the VA's statement.

    \textbf{Facial expressions: }
    Three types of facial animations were employed (see Figure~\ref{fig:figure.1} (e), (f), and (g)) - happy/positive, angry/negative, and neutral.
    The animation was generated by modifying the blend shapes of the character's 3D model, using the facial expression dataset~\cite{lucey2010extended} as a guide. Additionally, commercial animation assets were utilized to ensure a smooth and seamless transition~\cite{SALSA}.

    \textbf{Emotional gestures:}
    We prepared 16 emotional gesture animations
    from Mixamo~\cite{Mixamo}, comprising 8 positive and 8 negative emotions. 
    The positive gestures included a thumbs-up and cheering arms (see Figure~\ref{fig:figure.1} (h) and (i)), 
    while the negative gestures included crossing arms and a hand shrug (see Figure~\ref{fig:figure.1} (j) and (k)).
    These gestures were based on the Ultimate Guide to Body Language~\cite{BodyLanguage}.
    To provide variation in intensity, each emotional aspect was divided into two levels: weak and strong - according to how ``big'' the actions were. 
    Thus, the final animation set included four animations for each of the following categories: 
    4 positive-weak, 4 positive-strong, 4 negative-weak, and 4 negative-strong. 
    During the experiment, our system selected the emotional state based on the participant's statements and chose the intensity in a 
    counter-balanced manner.
    
    \textbf{Voice tone: } 
    The text-to-speech (TTS) API was utilized to generate speech files featuring a female voice~\cite{NaverCloudAPI}. 
    This API allowed for the modulation of the voice's tone to reflect different emotional states, which could be categorized as positive, negative, or neutral.
    
    \textbf{Emotional utterance: }
    A set of 20 emotional utterances were used to articulate the statements spoken by the VA, either in agreement or disagreement.
    These utterances were categorized into two groups: 10 positive (agreement) and 10 negative (disagreement) reactions. 
    Examples of positive and negative utterances were, 
    ``Oh, it's nice that we have the same opinion!'' 
    and ``Ugh, it's frustrating that your opinion is different from mine,'' respectively. 
    To assess the effectiveness of these utterances in conveying the associated emotion, an off-the-shelf sentiment analysis API~\cite{NaverCloudAPI} was employed, in line with the methods used in related previous work~\cite{nam2022emotionally}.
    The analysis showed that all positive utterances indeed had positive sentiment scores of over 99\%, and all negative utterances likewise had negative sentiment scores of over 99\%.
    In the experiment, emotional utterances were made upon either a correct or incorrect statement by participants, then followed by a statement relevant to the discussion task     (see Section \ref{sec:task} for details).

In comparison, the VA displaying Non-Affective (NA) behavior did not exhibit any of the previously mentioned affective or emotional behaviors during the group discussion. This led to the participants perceiving and interacting with the NA's VA in a neutral, detached, and emotionless way.

\subsection{Experimental Task: Group Decision-Making}
\label{sec:task}

We employed a group decision-making scenario as the experimental task, specifically prioritizing 15 important items for survival. 
This task is commonly used in team-building activities and similar VA studies~\cite{walker2019influence, kangsooReducing, torre2019effect}.
The study included four distressing scenarios: \textit{Desert}~\cite{desertSurvivalTask}, \textit{Winter}~\cite{johnson1991joining}, \textit{Sea}~\cite{nemiroff2001lost}, and \textit{Outer space}~\cite{nasasurvivalTask}. 
This task can be characterized as collaborative as a final decision needs to be made through collective effort, but it is also mildly competitive since participants tend to vie against each other 
and hope that one's opinion would be taken into account.
Also, note that the participants have never seen \textit{The Other Participant} (opposite human subject) and the VA before. 
As such, we assume that participants would be reserved and try to adhere to customary social norms and etiquette.

\begin{figure}[!b]
\centering
\includegraphics[trim={0 0 0 0}, clip, width=1\linewidth]{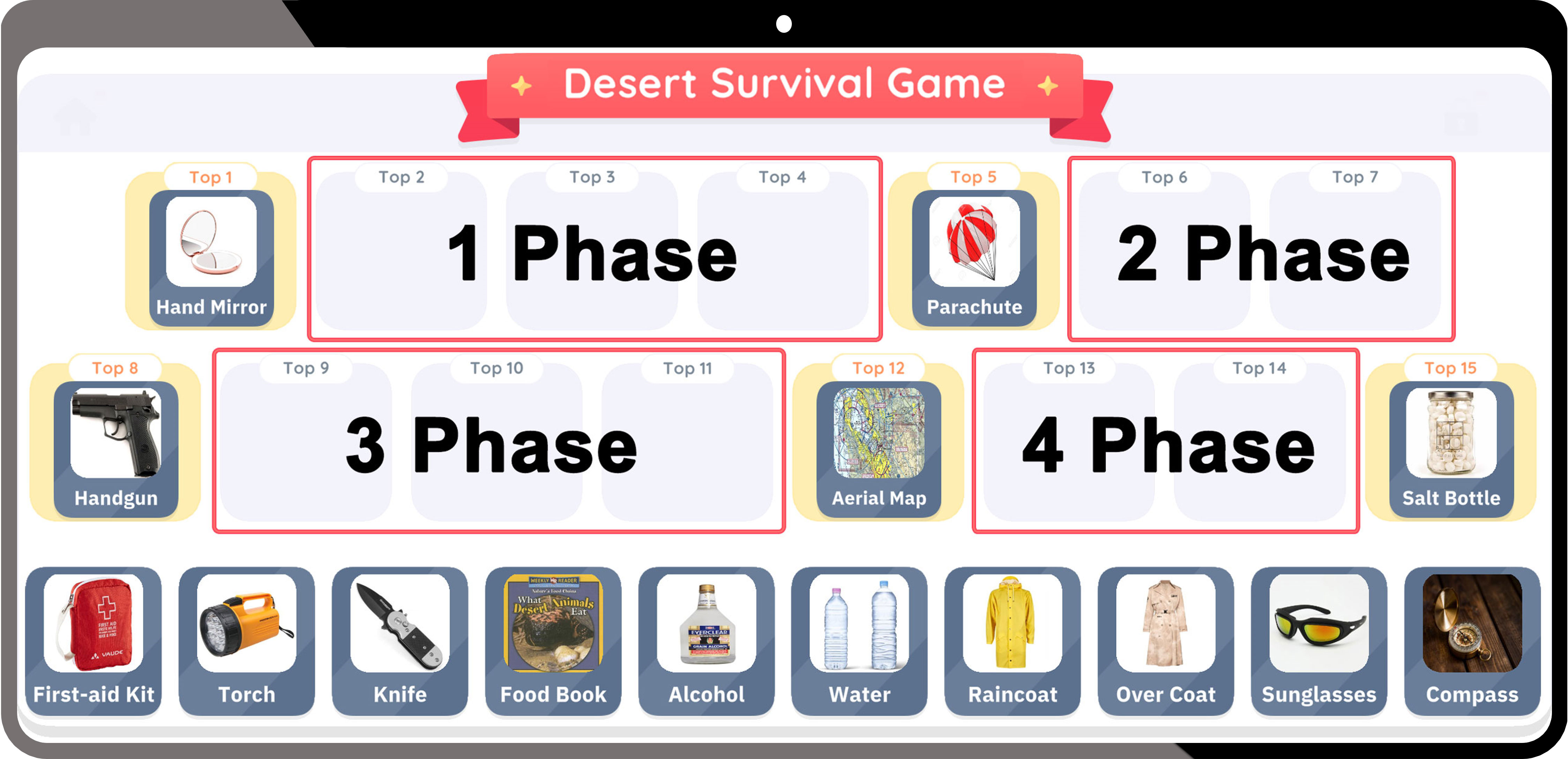}
\caption{
The mobile app for easily arranging and setting the priority orders of survival items as an experimental task. \label{fig:gameboard}
}
\Description{
The mobile application for easily arranging and setting the priority orders of survival items as an experimental task (or discussion topics).
The target items were laid out at the bottom, and they were to be dragged and filled into the proper positions in four different phases of the treatment session.}
\end{figure}

During the experiment, a group of three-person, consisting of two human subjects and one VA, discussed prioritizing the survival items. 
To aid participant comprehension of the situation and enhance participation in discussions, a mobile application was developed (see Figure~\ref{fig:gameboard}). 
The prioritized objects could be easily arranged using a drag-and-drop interface on the mobile application.
Also, to reduce the total time of experiment and alleviate the difficulty of discussion, the priority orders of five objects (among the fifteen) were given (1\textsuperscript{st}, 5\textsuperscript{th}, 8\textsuperscript{th}, 12\textsuperscript{th}, and 15\textsuperscript{th}) - the participants only had to decide on the priority order for the remaining 10 items.
This application was installed on an 11-inch tablet and provided to each participant in - Step 1 [Individual Selection] (see Section~\ref{sec:procedure}).
Afterward, to encourage concentration toward interlocutors (including a VA), we retrieved the tablet before the start of the group discussion. 
Instead, we placed a laptop (see Figure~\ref{fig:physical setup}) to allow participants to refer to it (as \textit{Task object}) and check the currently arranged items.

\textbf{VA's statements for discussions:}
One important provision for carrying out the task in a fair manner was to ensure that participants experienced the same level of expertise from the VA by the statements (about the priority-setting task itself) they made in following up with the discussion.  
The VA was designed to provide arguments based on the ``correct'' priorities (i.e., the VA knew and always provided certain feedback toward the correct answer).

We pre-generated a total of 840 statements (105 agreement and 105 disagreement statements for each of the four survival tasks) to ensure appropriate responses to participants' statements.
Each statement’s length (130-150 characters) was adjusted so that the VA would not speak for more than 20 seconds (also considering combining the preceding emotional utterances described in Section~\ref{independentvar}).
It should be noted that the affective VA follows a specific speaking pattern.
It initiates by expressing emotional utterances and subsequently provides statements about the task itself.

In addition, as the experimental design incorporated the VA's affective behavior as a condition, we ensured that the VA's statements were neutral and devoid of emotion. 
To achieve this, the same sentiment analyzer API~\cite{NaverCloudAPI} was used to proofread the statements for sufficient neutrality.
As a result, all statements were judged by the API to be 99\% neutral (see Table~\ref{tableAppendixSentiment} of the Appendix).

\subsection{Experimental Set-Up and Control}
\label{section313}

\subsubsection{Virtual Agent Specification} 

The realistic 3D model of VA was meticulously crafted, featuring detailed facial characteristics achieved through facial blend shapes ~\cite{blendshapereference}.
To enhance the VA's realism, various body animations—such as idling, listening, and speaking—were created using Mixamo~\cite{Mixamo}, along with the incorporation of motion capture data~\cite{ali2022}.
The VA's voice files were pre-generated from the CLOVA API~\cite{NaverCloudAPI}.
A lip-sync asset named SALSA~\cite{SALSA} was employed to ensure the VA's natural mouth movements and facial expressions. 
All these elements were rendered using the Unity 3D engine with a high-definition render pipeline.

\subsubsection{Interlocutor detection algorithm} 
\label{sec:speaker identification} 
Both mutual gaze and directed posture are essential when the VA is speaking and listening~\cite{senju2009eye,muller2018robust}. 
To incorporate the VA's non-verbal engaging cues, such as naturally looking at or turning towards the current speaker and listener, an \textit{Interlocutor detection (ItD)} algorithm was implemented.
This algorithm operates differently based on the VA's role.

When the role of the VA is the listener, the ItD algorithm processes the input video streams at 960 $\times$ 540 resolution and applies the dlib library~\cite{king2009dlib}, known for its accurate face detection capabilities~\cite{kazemi2014one}, to focus on the landmarks around the mouth (49\textsuperscript{th} to the 68\textsuperscript{th} points) and identify the speaking entity. 
In addition, the Pyaudio~\cite{pham2006pyaudio} detects sound invocation and its directionality, aiding the speaker identification.
Upon identifying the speaker,
the VA redirects its body and gaze toward the identified speaker, simulating attentive listening as part of the active engagement behavior.

While the VA serves as a speaker, 
the ItD algorithm identifies which of the two human participants is more attentively engaging with the VA.
Using the dlib library~\cite{king2009dlib}, the ItD algorithm analyzes head pose data to evaluate gaze patterns, discerning whether a participant is looking towards the VA and measuring the duration of their gaze.
We categorized three different gaze scenarios, applying the collected data to each scenario as follows:

\begin{itemize}
    \item \textbf{Both are looking at the VA:} 
    The ItD identifies which participant is the preferred interlocutor and adjusts the VA's direction to focus on one of them. 
    The preferred participant is determined as the one who has spent ``less'' time gazing at the VA thus far.
    This ensures an equitable distribution of eye contact with each participant.
    
    \item \textbf{One participant is looking at the VA:} 
    If the ItD detects that only one participant is looking at the VA, the VA will adjust its direction to face that participant.
    
    \item \textbf{Both are not looking at the VA:} 
    In cases where neither participant is looking at the VA, 
    the ItD adjusts the VA's posture to face the participant who has shown a higher level of engagement with it thus far.
    This decision is based on the expectation that this participant is more likely to soon reestablish eye contact with the VA.
\end{itemize}

It should be noted that the ItD algorithm was utilized only with the treatment discussed with Engagement VA (i.e., E-A, E-NA).
In contrast, NE's VA maintains a fixed gaze straight ahead without any responsive behaviors.

\begin{figure}[t]
\centering
\includegraphics[trim={0em 0em 0em 0em}, clip, width=1\columnwidth]{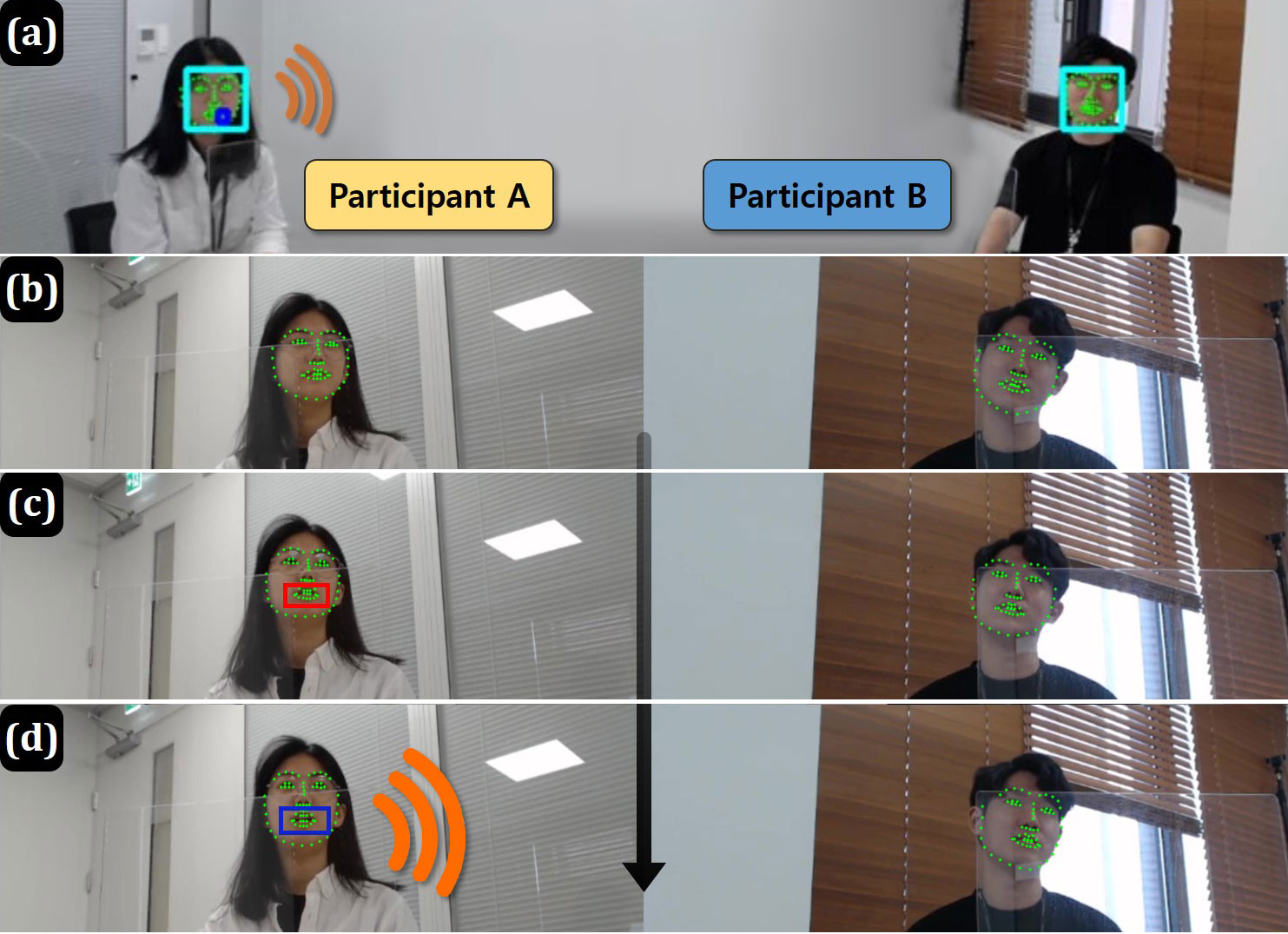}
\caption{Interlocutor Detection (ItD) algorithm:
(a) Participant A is identified as the speaker because (c) Participant A has an open mouth and (d) sound is detected from Participant A, (b) despite both participants  looking at the VA.~\label{fig:interloctor}}
\Description{The Interlocutor Detection (ItD) algorithm works as follows:
(a) The algorithm determines that Participant A is the current speaker. This decision is based on two key observations:
(c) Participant A is observed to have an open mouth.
(d) Sound is detected emanating from Participant A.
(b) Importantly, this identification is made for Participant A as the current speaker,
despite the fact that both participants are directing their attention towards the Virtual Agent.
}
\vspace{-2em}
\end{figure}

\subsubsection{Semi-Wizard of Oz control}

To administer the experiment so far with minimal external bias, the VA or system needs to have the capabilities of nearly human-like natural language understanding, conversational responses, and associated bodily gestures and animations, which in itself is a more than challenging task.  

Thus, we employed the semi-Wizard of Oz method, in which the controller, or ``Wizard'' (the person behind the curtain), interpreted the participant's statements (i.e., unstructured natural language) to determine the items they wanted to be ranked.
It should be noted that the VA's verbal and non-verbal response behaviors were automatically generated and selected by the system once a human controller interpreted the participants' statements.
To facilitate this process, a client-server application was developed to enable the controller to easily and quickly designate the item to be promoted (for survival discussion) and to provide the VA's statements.

\begin{figure*}[t] 
\centering
\includegraphics[width=1\linewidth]{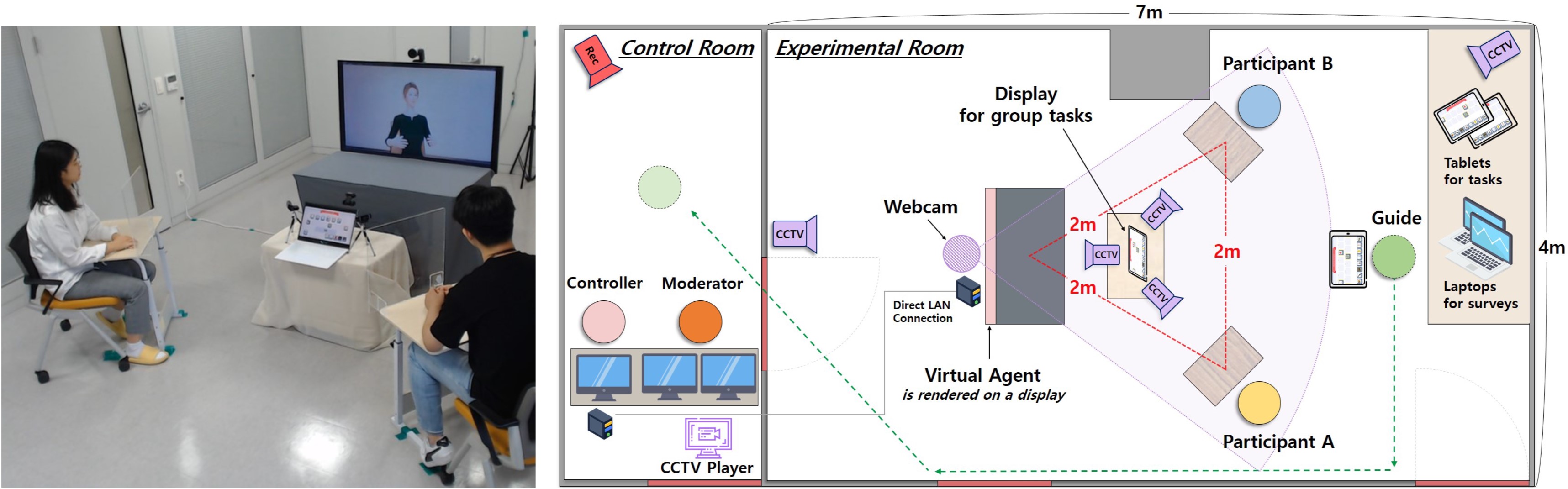}
\caption{The left photo shows the group discussion; The right diagram depicts the control and experimental rooms.~\label{fig:physical setup}}
\Description{Displayed on the left side is a photo of the configuration of seats for a three-person debate.
The figure on the right depicts the spatial arrangement of both the control and experimental rooms.}
\end{figure*}

\subsubsection{Physical Set-up}

Figure~\ref{fig:physical setup} shows the setup in the experimental and the control rooms.
The experimental room had one table and two small desks/chairs.
On the table, we placed a large 55-inch display to render the VA's upper body at a size comparable to that of an average female's physical stature.
The VA had a nearly realistic appearance and was rendered on a desktop with GPUs to ensure smooth performance without frame drops.

As a way to prevent the participants from forming a team (rather than working individually and semi-competitively) against the VA, the two participants were seated at the two far ends of the triangular-shaped space (see Figure~\ref{fig:physical setup} (a)).
The distance between the table and each desk was set to 2 meters, as the social zone (within 3.6 meters) for office gatherings was considered to be larger than the personal space (within 1.2 meters) suggested by the widely known proxemics~\cite{hall1966hidden}. 
We also placed another small table in the center to display the ongoing group task results (the help app) on the laptop and three cameras to observe and record how participants behave to VA.
During the experiment, experimenters were in the control room.
Multiple PCs were placed in the control room to remotely control the VA and moderate the interactive discussion via a moderator (experimenter).  
Details follow in the next section.

\subsection{Experimental Procedure}
\label{sec:procedure}
\begin{figure}[t]
\centering
\includegraphics[width=1\linewidth]{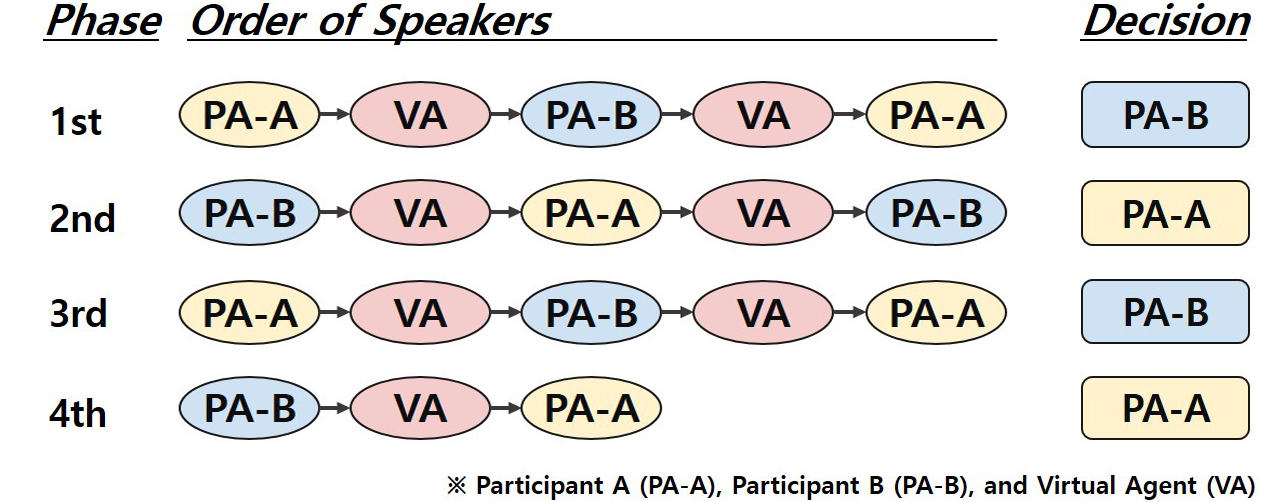}
\vspace{-1em}
\caption{The speaking order among two participants and the VA for each of the four phases within the single treatment.
The order of PA-A and PA-B was counter-balanced.
\label{fig:interactionorder}}
\Description{The speaking order among the two participants and the virtual agent (VA) for each phase of the four is as follows:
(Phase 1) Participant A speaks first. The VA follows. Participant B follows. The VA follows. The Participant A responds last. Participant B decided on item priority.
(Phase 2) PA B -> VA -> PA A -> VA -> PA B / Decideded by PA A
(Phase 3) PA A -> VA -> PA B -> VA -> PA A / Decideded by PA B
(Phase 4) PA B -> VA -> PA A / Decideded by PA A}
\end{figure}

The experimental guide provided a brief explanation of the experimental procedure and instructed to fill out the pre-questionnaire with a laptop.
After that, the guide gave each participant a tablet with the survival tasks application installed and asked them to wait until the moderator started the experiment. 
As described, the moderator was located in the control room and delivered instructions via microphone. 
Participants received instructions via an audio device and proceeded with the assigned tasks.
Each treatment consisted of the following three steps:

\begin{itemize}
    \item \textbf{Step 1 [Individual selection]: }
     The moderator instructed participants to look at their individually assigned tablet screen (which displayed a description of the distressing situation) and then described the distress situation.
     After the instruction, participants individually selected item priorities they deemed important within 3 minutes.
     Once the participants completed their individual selections, the experimental guide retrieved the participants' tablets and exited the experimental room for a three-person group discussion.
     This step lasted about 5 minutes.
     
    \item \textbf{Step 2 [Group discussion]: }
    The second step involved conducting a group discussion to prioritize items.
    To make the duration of the experiment manageable, the participants had already been given the rankings of the 1\textsuperscript{st}, 5\textsuperscript{th}, 8\textsuperscript{th}, 12\textsuperscript{th}, and 15\textsuperscript{th} items. Participants were also instructed to make the discussion to rank items in pre-designated groups in four phases: (Phase 1) 2\textsuperscript{nd}, 3\textsuperscript{rd}, and 4\textsuperscript{th}; (Phase 2) 6\textsuperscript{th} and 7\textsuperscript{th} items; (Phase 3) 9\textsuperscript{th}, 10\textsuperscript{th}, and 11\textsuperscript{th}; and (Phase 4)
    13\textsuperscript{th} and 14\textsuperscript{th} items.
    For each phase, both the participants and VA followed a pre-determined speaking order (see Figure~\ref{fig:interactionorder}). 
    They were given 30 seconds to provide their reasoning for the item rankings 
    (The VA was designed to deliver a statement for about 20 seconds, as described in Section~\ref{sec:task}).
    The final speaker (either a human participant) for the given phase made the final decision about the item rankings.
    Depending on the phase, each participant had one or two opportunities to speak, balanced by changing the order of Participant A (PA-A) and Participant B (PA-B) in subsequent phases (see Figure~\ref{fig:interactionorder}).
    Note that the order of PA-A and PA-B was alternated for each treatment to ensure counter-balanced speaking turns, 
    i.e., PA-A was the first to speak in the first and third treatments, whereas PA-B began speaking in the second and fourth treatments.
    In Step 2, the duration of the group discussion was about 12 minutes.
    \item \textbf{Step 3 [Post-treatment questionnaire]: }
    At the end of the discussion, the experiment guide entered the experimental room and provided each participant with a laptop to complete the post-treatment questionnaire.
\end{itemize}

After all treatments, participants were asked to fill out a post-experiment questionnaire to provide subjective rankings and any feedback.
The order of four treatments was counter-balanced using the balanced Latin square method.
Each treatment took about 20 minutes, followed by a 10-minute break, resulting in a total duration of about 2 hours.

\subsection{Dependent Variables}

There are several dependent variables we looked at, broadly categorized into three types:
(1) validity assessments of the study design and implemented factors;
(2) objective measures of the impact of the VA's behavior on participants' decision and attentive behavior; 
and lastly, (3) subjective measures of the influence of the VA's behavior on participants' subjective perception (which has a potentially indirect relevance to the objective measures)
in group discussions.

\subsubsection{Experimental Validity Measures}

The experiment's validity and any derived conclusions hinge on whether users properly perceived the manipulated independent variables, namely, the VA's engaging and affective behaviors as described and implemented so far. 
Likewise, the task difficulty and the VA's level of expertise need to be maintained equally, more or less, across four different task scenarios. 
For this purpose, we administered a validity questionnaire, designed based on previous work~\cite{Harms2006InternalCA} and modified for our purpose, to assess the following perceptual qualities of the test system (after each treatment): 
(1) the level of VA's engagement, and (2) the level of VA's affectiveness on a 7-point Likert scale.
After the whole experiment, participants were requested to assign rankings for the following perceptual aspects: 
(1) VA's engagement,
(2) VA's affectiveness,
(3) Task's difficulty, 
and (4) VA's expertise.

\subsubsection{Factor Effects - Objective}

To explore the subconscious changes in which participant's behavior is affected by the VA during the discussion, two objective and quantitative measures were used:

\textbf{Task score:} 
This measures the correctness of the final decision made after a group discussion.
A standardized scoring mechanism~\cite{Togetheralone} was used to tabulate the score (+d for a given answer where d is the difference to the correct, e.g., if an item was ranked 6th rather than 2nd (correct answer), $6 - 2 = 4$ is added to the total score - thus, 0 represents the best).

\textbf{Attention allocation:}
The participant's behavior captured on video was examined for one's level of gaze toward the VA. 
This was based on the premise that eye gaze can serve as a proxy for measuring attention~\cite{sidner2004look}, and consequently could correlate to the exhibited VA's behavior during the group discussion.

\vspace{-0.5em}
\subsubsection{Factor Effects - Subjective}
The study measured the impact of independent variables on participants' perceptions of VAs using post-treatment questionnaires.
Refer to Table~\ref{TableAppendixQuestionnaire} in the Appendix for the questionnaire details. 
It was based on previous research~\cite{bailenson2003interpersonal,yumak2014modelling,jian2000foundations,hart2006nasa}, and modified for this experiment's needs. 
The questionnaire used a 7-point Likert scale and focused on examining the following perceptual/subjective qualities:

\textbf{Social presence:}
This aims to measure the extent to which participants perceive the VA 
as an active and socially engaged member in group discussions.
These aspects are considered important in fluid groups and coordinated decision-making.  
In particular, we focused on its continuity,
responsiveness, and membership.

\textbf{Social influence:} 
This refers to the extent to which the VA's opinion was reflected in the group decision-making process. 
Also, the study looked into how the emotional state of the VA might have affected the participant's behavior, e.g., positively, if decisions were made without any concern for the VA's emotions or vice versa.

\textbf{Trustworthiness:}
This can affect the participant's willingness to follow the VA's guidance.
The positive aspect focuses on evaluating the credibility and confidence of the VA, while the negative aspect measures the level of doubt regarding the VA's behavior.

\textbf{Task load \small{(Mental demand, Performance, and Frustration)}:} 
Another important factor to consider is the mental (or psychological) load induced by the various behaviors of the VA.
After all, social activities can be inherently stressful, especially considering the task was at least mildly competitive.

Lastly, in the post-experiment questionnaire, 
we included a question asking participants to rank their preferred VAs as team members during group discussions.
This subjective dependent variable is referred to as \textbf{``Preference''}.

\vspace{-0.5em}
\subsection{Participants}
\label{sec:participant}
We recruited 32 participants (14 females, 18 males, aged 22--45, $M$ = 27.4, $SD$ = 4.9) from the local community. 
The experiment was conducted in pairs with 16 teams consisting of 12 mixed teams, 3 men's teams, and 1 women's team.
Each pair had never met before.
All participants had normal/corrected vision and hearing.
Prior to the experiment, we collected participants' prior knowledge of survival tasks ($M$ = 3.3, $SD$ = 1.2) and their experience with VA technology ($M$ = 3.6, $SD$ = 1.4) through a pre-questionnaire using a 7-point Likert scale. 
Individuals who already possessed proficient experience in the field of survival (e.g., personnel in the military or medicine) were excluded from the recruiting process.
Given the four conditions and
and statistical parameter settings
from the 2-way repeated measures (RM) ANOVA with the effect size of 0.25 (medium), the significance level of 0.05,
and the power of 0.8,
the G*power, a popular statistical power analysis tool~\cite{faul2009statistical}, 
determined the minimum sample size to be 32 
(critical $F =$ 3.917).
This study received approval from the Institutional Review Board (IRB No. KIST-202209-HR-015). 
Participants received a gift of \$20.

\section{Results}
\label{result}

\subsection{Experimental Validity Measures}
\label{sec:validityMeasureResult}

\subsubsection{Level of Engagement and Affectiveness}
\label{sec:vali_art}

The collected data were checked for the normal distribution 
using the Shapiro-Wilk test~\cite{shapiro}, and for the sphericity assumption using the Mauchly's test~\cite{mauchly1940significance}.
Both tests showed that data deviated from the normal distribution and also violated the sphericity assumption.
Consequently, we performed the aligned rank transform (ART)~\cite{ARTmethodCitation}, which is a non-parametric approach 
for a 2-way repeated measure (RM) ANOVA.
For the post-hoc, we used the ART-C~\cite{ART-C}, with p-values adjusted using the Bonferroni correction. 
The statistical significance level was set at 5\% for all analyses.

\textbf{Level of Engagement:}
The Mauchly’s test showed sphericity violation ($p$ = .024) and all conditions showed normal distributions except for the E-A (Shapiro-Wilk test $p$ = .008).
Only the Engage factor exhibited a significant effect 
($F$ = 12.57; $p <$ .001***).
This finding suggests that participants perceived distinct levels of engagement based on whether the VAs exhibited engaging or non-engaging behavior. 
Conversely, there were neither significant effects by the Affective factor ($F$ = 0.56; $p$ = .45) nor any interaction ($F$ = 3.32; $p$ = .07).
The post-hoc revealed two statistical different pairs (E-NA $>$ NE-NA; $p =$ .002**) and (E-A $>$ NE-NA; $p =$ .04*).

Given that the interaction effect was close to significance ($p$ = .07), 
it is postulated that combining the Engage
factor with the Affective factor could enhance the VA's engagement.
Distinguishing between engagement and affectiveness can be subjective and not always clear-cut. 
The affective behavior might have been perceived as another form of engagement~\cite{choi2012affective}.
As a result, the following measure of VA's affectiveness is relatively more noticeable.

\textbf{Level of Affectiveness:}
The response data did not show sphericity ($p$ < .001) and also 
diverted from the normal distribution in all conditions ($p <$ .05).
The Affective factor solely influenced VA's affectiveness ($F$ = 234.21; $p <$ .001***), whereas the Engage factor ($F$ = 0.09; $p$ = .76) and interaction effect ($F$ = 0.01; $p$ = .88) showed no statistical significance.
The post-hoc test indicated that significant differences existed in four treatment pairs: (E-A $>$ E-NA; $p <$ .001***), (E-A $>$ NE-NA; $p <$ .001***), (E-NA $<$ NE-A; $p <$ .001***), and (NE-A $>$ NE-NA; $p <$ .001***).
This indicates that the participants strongly perceived different levels of VA’s affectiveness, but were not influenced by Engage factor.

\subsubsection{The rank of perceptual aspects}
\label{sec:4.1.2}

The following ordinal ranks were assessed through the post-experiment questionnaire after all the treatments and analyzed using the 
Friedman tests~\cite{friedman1940comparison}.
For the post-hoc, the Conover's test~\cite{conover1979multiple} with Bonferroni correction was used.
All tests were performed at a 5\% significance level.

\begin{figure}[t]
\centering
\subfigure[Engagement Rank]{\includegraphics[width=0.49\linewidth]{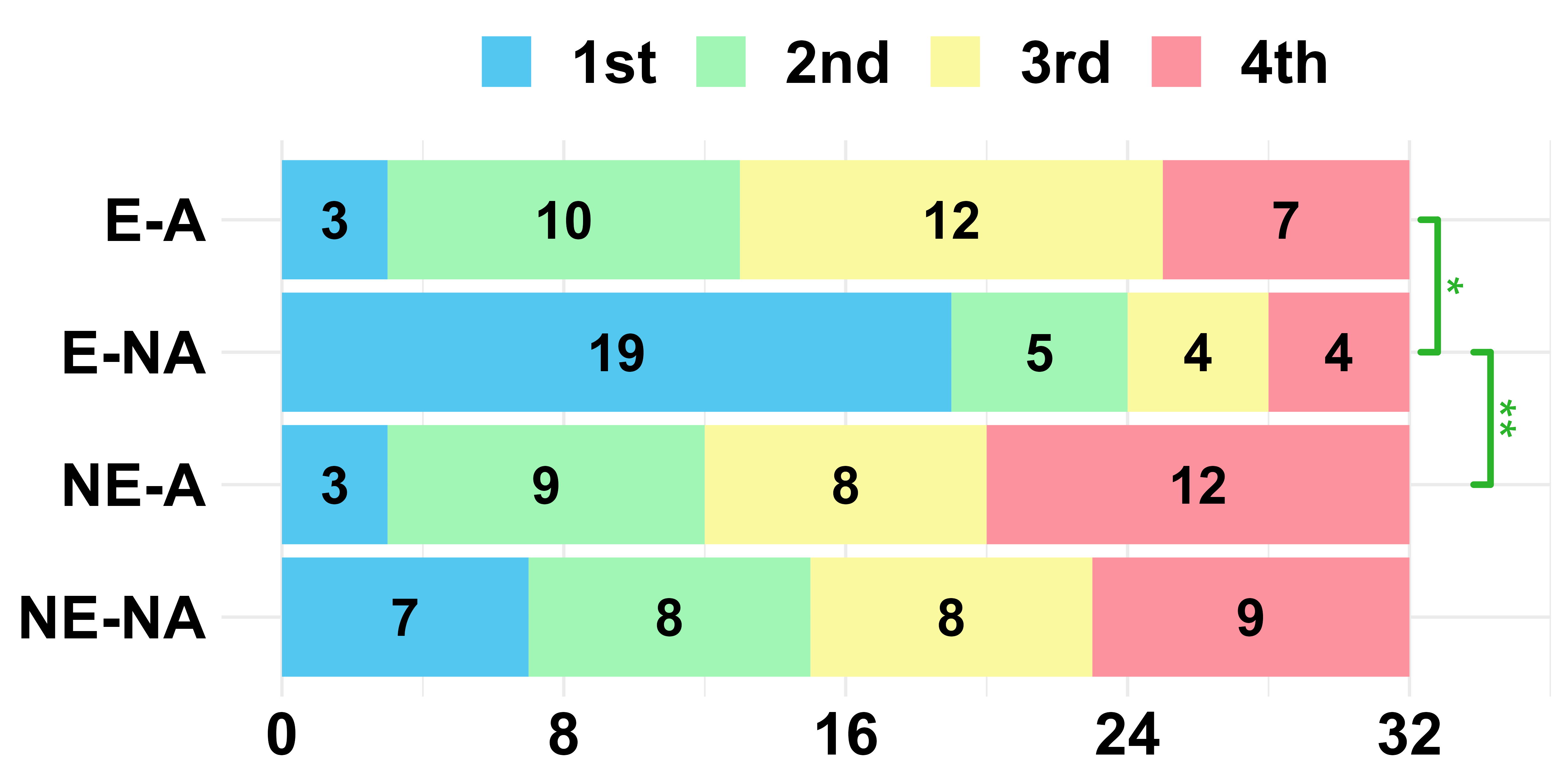}} 
\hfill
\subfigure[Affectiveness Rank]{\includegraphics[width=0.49\linewidth]{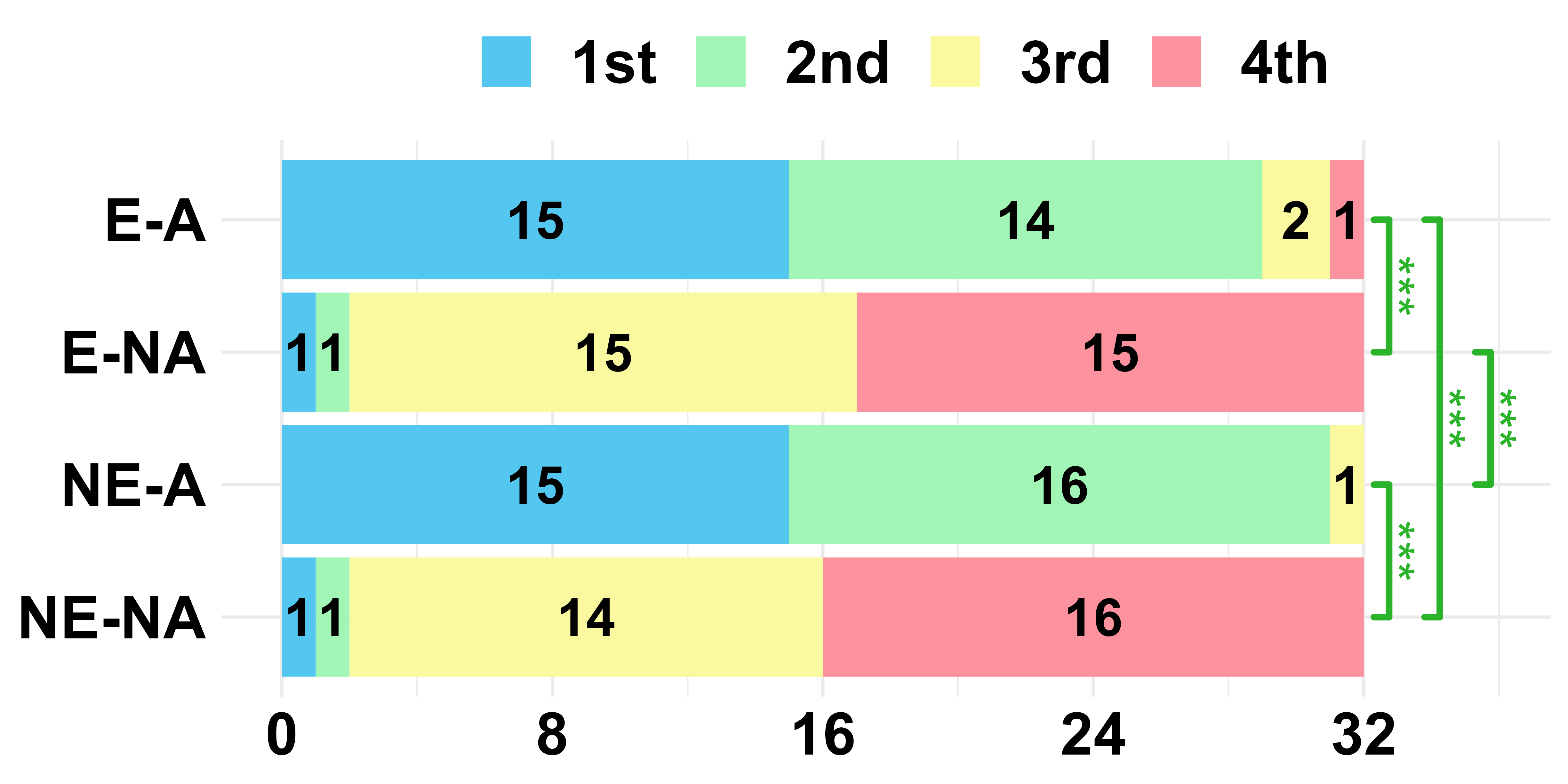}} 
\\ \vspace{-0.25em}
\subfigure[Preference Rank]{\includegraphics[width=0.49\linewidth]{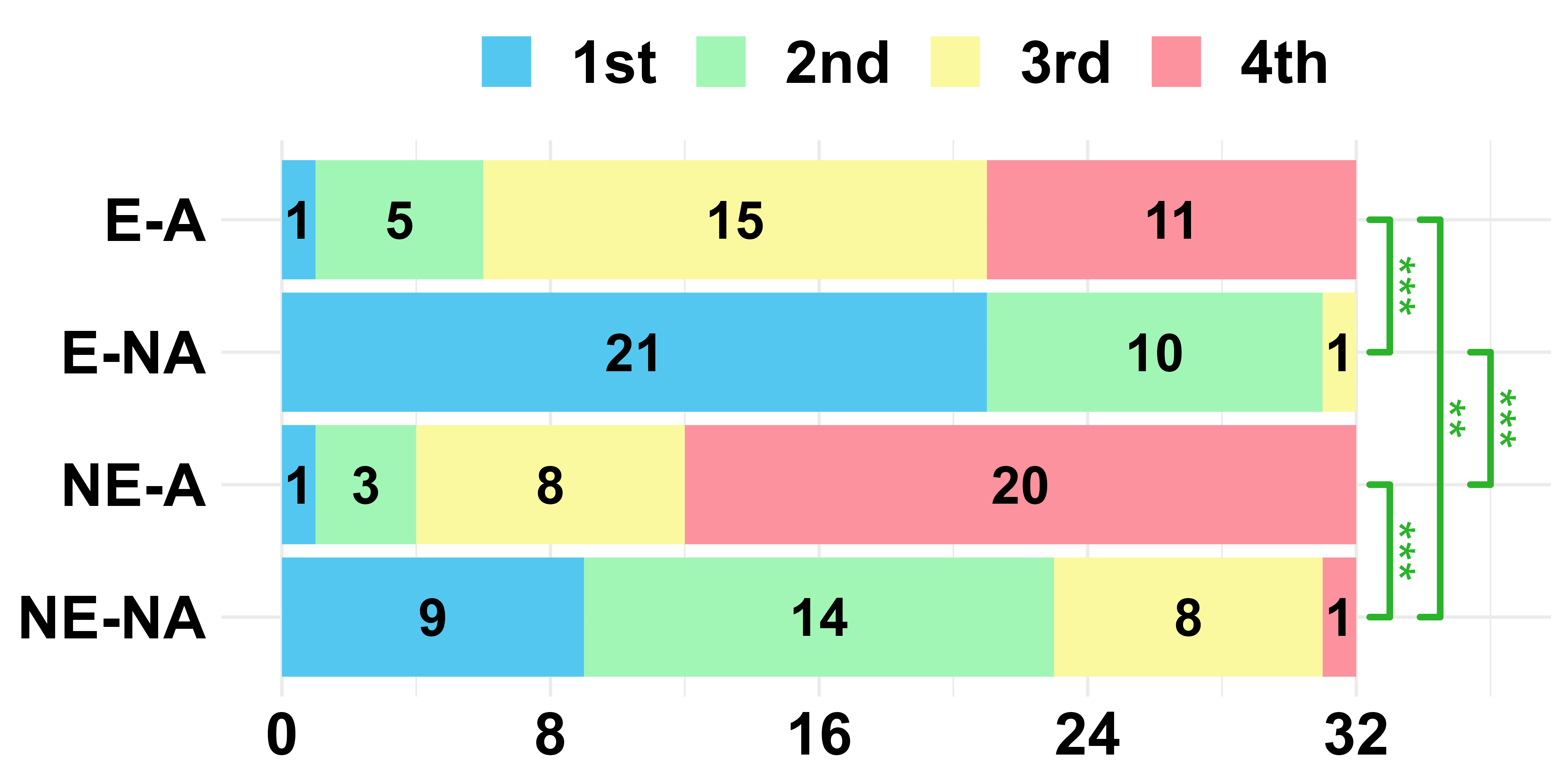}} 
\vspace{-1em}
\caption{
Results for ranking data. The X-axis indicates the number of votes (* $p$\,\textless\,.05; ** $p$\,\textless\,.01; and *** $p$\,\textless\,.001).\label{fig:rank order}}
\Description{Results for ranking data under the four conditions. The X-axis indicates the number of votes.
}
\vspace{-1em}
\end{figure}

\textbf{VA's Engagement Rank:}
The results of the ranking votes (from the 1st to 4th) are shown in 
Figure~\ref{fig:rank order} (a).
We found a statistically significant effect ($\chi^2$\,=\,14.17; $p =$ .003**).
The post-hoc revealed that only two pairs had statistical differences: 
(E-A $<$ E-NA; $p$ = .029*) and (E-NA $>$ NE-A; $p$ = .005**).
One pair nearly reached significance (E-NA $>$ NE-NA; $p$ = .083).
Thus, subjects likely perceived E-NA’s VA as the
most engaging VA in group discussions.

\textbf{VA's Affectiveness Rank:}
The results are shown in Figure~\ref{fig:rank order} (b).
The Friedman test yielded a statistically significant effect ($\chi^2$\,=\,61.01; $p <$ .001***).
The post-hoc indicated that there were four significantly different pairs:
(E-A $>$ E-NA; $p <$ .001***), 
(E-A $>$ NE-NA; $p <$ .001***), 
(E-NA $<$ NE-A; $p <$ .001***), and 
(NE-A $>$ NE-NA; $p <$ .001***).
Moreover, the number of votes for the first place (15)
was the same for the affective VAs (i.e., E-A, NE-A), 
and the affective VAs had higher votes compared to those (1) of the 
non-affective VAs (i.e., E-NA, NE-NA).
Thus, subjects likely perceived that the affective VAs (i.e., E-A, NE-A) had sufficient affectiveness compared to the non-affective VAs (i.e., E-NA, NE-NA).

\textbf{The rank of Task's difficulty/VA's expertise:}
We did not find any difference in task difficulties ($\chi^2$\,=\,1.8; $p =$ .61) 
and VA's expertise ($\chi^2$\,=\,3.48; $p =$ .32).
The findings indicated that the task's difficulty and VA's expertise did not impact any of the following objective/subjective measures, thus confirming the study's validity.

Based on the validity measures, it is important to highlight that two main factors/traits (i.e., engaging/affective behavior) were designed well to enable subjects to perceive varying levels of engagement and affectiveness in different treatments.
This finding confirms their suitability for exploring the objectives of the study.

\subsection{Factor Effects - Objective Measures}
\label{sec:4.2}

\subsubsection{Task Score}
\label{Sec:TaskScore}

\begin{figure}[t]
\centering
\includegraphics[trim={0 0 0 0.5em}, clip, width=0.9\linewidth]{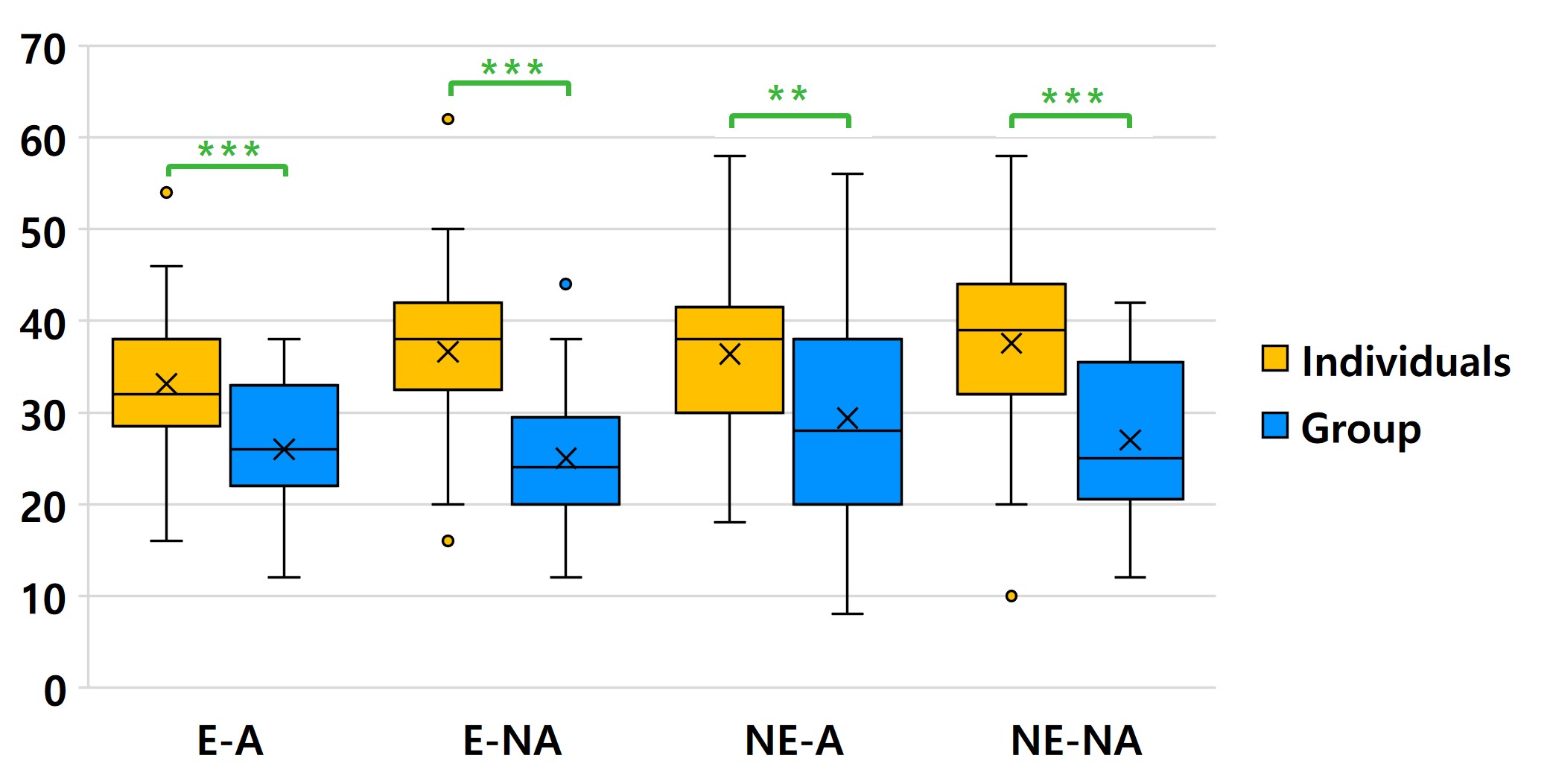}
\vspace{-1em}
\caption{
Boxplots of the individual and group scores.
Lower scores indicate better results (${}^{**}$\  $p<.01$; ${}^{***}$\ $p<.001$).}
\label{fig:groupscore}
\Description{Boxplots for the individual and group scores for survival tasks for each condition.
Lower scores indicate better performance.}
\vspace{-1em}
\end{figure}

We computed the task score based on the items' priorities as determined individually 
and by groups, respectively.
Figure~\ref{fig:groupscore} displays the results, where the score signifies the deviation from the correct answer, i.e., with lower scores indicating better performance.
Note that the distribution of individual scores excludes the VA's scores, as the VA is programmed to always provide the correct answer.
The task scores were analyzed in four different ways: 
(1) Comparison of group scores among treatments; 
(2) Individual vs. group scores within each treatment; 
(3) Effect of group discussion toward individual performance; 
and (4) Reflection of opinion/view of a particular member (VA or human).

\textbf{(1) Group Scores:}
The Shapiro-Wilk test verified that the data followed a normal distribution, and Mauchly's test confirmed the sphericity assumption ($p$ = .68).
However, it is important to acknowledge that despite the adequate number of individual 
subjects (32, as suggested by the G*Power, also see Section~\ref{sec:participant}), there are only 
16 paired group samples, which might have impacted the reliability of the analysis results
with regards to group behavior.
A post-hoc power analysis indicated that the statistical power of this measurement stood at 0.5, indicating potential limitations to detect true effects.
Due to the small sample size, the ART method with a 5\% significance level was used instead of the 2-way RM ANOVA.
The results showed no significant effect for the Engage factor ($F$ = 0.89; $p$ = .35), the Affective factor ($F$ = 0.46; $p$ = .49), nor their interaction ($F$ = 0.01; $p$ = .90).

\textbf{(2) Individual vs. Group Scores:}
To assess the statistical differences between individual and group scores within each condition, 
the paired samples t-test was considered, 
i.e., comparing individual (pre-test) and group (post-test) scores. 
However, in each condition, there were 32 samples for individual scores, 
whereas group scores were derived from the pairs, resulting in 16.
So, for the comparison, we applied two imputation methods for the missing values.

One was to calculate the mean of individual scores and compare it with group scores (i.e., comparing 16 samples).
The other was to project each subject's group scores 
using one's own score but weighted relatively to that of the other members in the group
(i.e., allowing comparison among 32 samples). The following equation was used 
to attain the imputed score: 
$\text{PA-A's group score} = \text{Group score} \times \left( \frac{\text{PA-A's individual}}{\text{PA-A's individual} + \text{PA-B's individual}} \right) \times 2$.
Either way, there would be effectively two types of scores for each individual: the score before the group discussion and after reaching a consensus (which is set to that of the group one belongs to).
Data normality was checked by the Shapiro-Wilk test, and the Student's t-test or the Wilcoxon signed-rank test was applied for further analysis with the significance level set at 5\%.

With the 16 sample comparison for each treatment (as mean imputation), the Student's t-test (due to data normality) showed: E-A ($t$ = 4.9; $p <$ .001***); E-NA ($t$ = 9.9; $p <$ .001***); NE-A ($t$ = 2.5; $p =$ .020**); and NE-NA ($t$ = 5.7; $p <$ .001***).
However, with the limited number of samples, the statistical power is only 0.5.

With the 32 sample comparison (with weighted imputation), the Wilcoxon signed-rank test for E-A (Shapiro-Wilk $p$ = .009), and the Student's t-test for the remaining treatments showed: E-A ($z$ = 4.7; $p <$ .001***); E-NA ($t$ = 12.3; $p <$ .001***); NE-A ($t$ = 3.3; $p =$ .002**); and NE-NA ($t$ = 7.8; $p <$ .001***).

Despite slight differences in the statistical figures, both analyses generally showed similar levels of statistical significance. 
That is, the group scores significantly outperformed the individual across all treatments (see Figure~\ref{fig:groupscore}).
It is notable that the NE-A exhibited a slightly lesser effect than the others.

\textbf{(3) Effect of Group Discussion:}
The possible effect of the group discussion was indirectly assessed by
counting the number of occurrences in which the individual score 
(or the score of the group) was higher or lower after the 64 group discussions.  
Note that the score of the individual, after the group discussion with the consensus, was set to that of the group one belongs to.
The data showed that in 57 instances, the outcome of the discussion went closer to the 
ground truth answer (improved performance).  
The seven cases for which the direction went the other way (lowered performance)
came from the treatments of NE-A (6) and NE-NA (1).
These findings suggest that VA with the non-engaged and affective behavior (i.e., NE-A's VA) 
may have adversely affected participants' decision-making.
This may also explain the negative effect of the NE-A's VA on group scores reported in the 
previous section.

\textbf{(4) Reflection of Particular Member's View:}
Whether a particular group member was able to exert one's view more strongly was assessed by
counting the cases for which the given participant's (PA-A, PA-B, and VA) 
opinion was the most (or least) reflected in the final decision, following the method adopted 
in the similar work by ~\cite{meslec2013too}. 
The results are summarized in Table~\ref{fig:two_tables}.

The VA's opinion was perceived as the most reflected in the NE-A condition
(but also as the least reflected at the same time), while in the NE-NA condition, the VA's opinion was clearly the least reflected.
Although not definitive, these figures corroborate that NE-A's VA may have negatively impacted the user's experience.

\definecolor{lightgreen}{rgb}{0.72, 0.89, 0.80}

\begin{table*}[h!]
\renewcommand{\arraystretch}{1.2}
\caption{The number of cases in which the participant (PA-A, PA-B, or VA) had the most or least opinions reflected in the final decision across four treatments.  
Note that each condition should total 16 teams' data; however, as the PA-A, PA-B, and VA could have equal contributions to the final decision, thus the total can be more than 16.
The group score is represented as G.Score.\label{fig:two_tables}}
\vspace{-0.25em}
\Description{The number of cases in which the participant (PA-A, PA-B, or VA) had the most or least opinions reflected in the final decision across four conditions.
Note that each test condition should total 16 teams' data; however, as it was possible for the PA, PB, and VA to have equal contributions to the final decision,
thus the total can be more than 16.}
\centering
\resizebox{1\linewidth}{!}{
\begin{tabular}{lwc{0.7cm}wc{0.7cm}wc{0.7cm}>{\columncolor{lightgray!60}}c
wc{0.7cm}wc{0.7cm}wc{0.7cm}>{\columncolor{lightgray!60}}c
wc{0.7cm}wc{0.7cm}wc{0.7cm}>{\columncolor{lightgray!60}}c
wc{0.7cm}wc{0.7cm}wc{0.7cm}>{\columncolor{lightgray!60}}c}
\toprule[1pt]\midrule[0.3pt]
 & \multicolumn{4}{c}{\textbf{E-A}} & \multicolumn{4}{c}{\textbf{E-NA}} & \multicolumn{4}{c}{\textbf{NE-A}} & \multicolumn{4}{c}{\textbf{NE-NA}} \\ 

\cmidrule(lr{-0.02em}){2-5}
\cmidrule(lr{-0.02em}){6-9}
\cmidrule(lr{-0.02em}){10-13}
\cmidrule(lr{-0.02em}){14-17}
\textbf{Groups} & \textbf{PA-A} & \textbf{PA-B} & \textbf{VA} & \cellcolor{lightgray!60}{\textbf{G. Score}} & \textbf{PA-A} & \textbf{PA-B} & \textbf{VA} & \cellcolor{lightgray!60}{\textbf{G. Score}} & \textbf{PA-A} & \textbf{PA-B} & \textbf{VA} & \cellcolor{lightgray!60}{\textbf{G. Score}} & \textbf{PA-A} & \textbf{PA-B} & \textbf{VA} & \cellcolor{lightgray!60}{\textbf{G. Score}} \\
\midrule[1pt]
Group 1 & 3rd & \cellcolor{lightgreen!70}{1st} & 2nd & \cellcolor{lightgray!60}{22} & \cellcolor{lightgreen!70}{1st} & 3rd & \cellcolor{lightgreen!70}{1st} & \cellcolor{lightgray!60}{22} & 2nd & \cellcolor{lightgreen!70}{1st} & 3rd & \cellcolor{lightgray!60}{56} & 2nd & 3rd & \cellcolor{lightgreen!70}{1st} & 20 \\
Group 2 & 2nd & \cellcolor{lightgreen!70}{1st} & 3rd & 34 & 2nd & \cellcolor{lightgreen!70}{1st} & 3rd & 38 & 2nd & \cellcolor{lightgreen!70}{1st} & 3rd & 38 & 3rd & 2nd & \cellcolor{lightgreen!70}{1st} & 26 \\
Group 3 & 2nd & \cellcolor{lightgreen!70}{1st} & 3rd & 34 & \cellcolor{lightgreen!70}{1st} & 2nd & 3rd & 32 & 3rd & 2nd & \cellcolor{lightgreen!70}{1st} & 20 & 3rd & \cellcolor{lightgreen!70}{1st} & 2nd & 30 \\
Group 4 & 2nd & 3rd & \cellcolor{lightgreen!70}{1st} & 28 & \cellcolor{lightgreen!70}{1st} & 3rd & 2nd & 28 & \cellcolor{lightgreen!70}{1st} & 3rd & 2nd & 34 & \cellcolor{lightgreen!70}{1st} & 2nd & 3rd & 38 \\
Group 5 & \cellcolor{lightgreen!70}{1st} & 3rd & 2nd & 22 & \cellcolor{lightgreen!70}{1st} & 3rd & 2nd & 28 & \cellcolor{lightgreen!70}{1st} & 3rd & 2nd & 26 & \cellcolor{lightgreen!70}{1st} & 2nd & 3rd & 36 \\
Group 6 & 3rd & \cellcolor{lightgreen!70}{1st} & 2nd & 22 & 3rd & \cellcolor{lightgreen!70}{1st} & 2nd & 18 & 2nd & 3rd & \cellcolor{lightgreen!70}{1st} & 20 & \cellcolor{lightgreen!70}{1st} & \cellcolor{lightgreen!70}{1st} & 3rd & 34 \\
Group 7 & 3rd & 2nd & \cellcolor{lightgreen!70}{1st} & 20 & 3rd & \cellcolor{lightgreen!70}{1st} & 2nd & 20 & 3rd & 2nd & \cellcolor{lightgreen!70}{1st} & 12 & \cellcolor{lightgreen!70}{1st} & 2nd & 3rd & 12 \\
Group 8 & 2nd & \cellcolor{lightgreen!70}{1st} & 3rd & 34 & 2nd & 3rd & \cellcolor{lightgreen!70}{1st} & 12 & 3rd & 3rd & \cellcolor{lightgreen!70}{1st} & 30 & \cellcolor{lightgreen!70}{1st} & 2nd & 3rd & 24 \\
Group 9 & \cellcolor{lightgreen!70}{1st} & 3rd & \cellcolor{lightgreen!70}{1st} & 22 & \cellcolor{lightgreen!70}{1st} & 3rd & 2nd & 24 & \cellcolor{lightgreen!70}{1st} & \cellcolor{lightgreen!70}{1st} & 3rd & 26 & 2nd & \cellcolor{lightgreen!70}{1st} & 3rd & 42 \\
Group 10 & \cellcolor{lightgreen!70}{1st} & 2nd & 3rd & 30 & \cellcolor{lightgreen!70}{1st} & 3rd & 2nd & 12 & 2nd & \cellcolor{lightgreen!70}{1st} & 3rd & 40 & 2nd & \cellcolor{lightgreen!70}{1st} & 3rd & 32 \\
Group 11 & 3rd & \cellcolor{lightgreen!70}{1st} & 2nd & 18 & 3rd & \cellcolor{lightgreen!70}{1st} & 2nd & 26 & 3rd & 2nd & \cellcolor{lightgreen!70}{1st} & 8 & \cellcolor{lightgreen!70}{1st} & 3rd & \cellcolor{lightgreen!70}{1st} & 16 \\
Group 12 & 3rd & \cellcolor{lightgreen!70}{1st} & 2nd & 12 & \cellcolor{lightgreen!70}{1st} & 2nd & 3rd & 44 & 2nd & \cellcolor{lightgreen!70}{1st} & 3rd & 38 & 2nd & \cellcolor{lightgreen!70}{1st} & 3rd & 38 \\
Group 13 & 3rd & \cellcolor{lightgreen!70}{1st} & 2nd & 26 & 3rd & \cellcolor{lightgreen!70}{1st} & 2nd & 20 & 2nd & \cellcolor{lightgreen!70}{1st} & 3rd & 52 & 2nd & 3rd & \cellcolor{lightgreen!70}{1st} & 22 \\
Group 14 & 3rd & 2nd & \cellcolor{lightgreen!70}{1st} & 26 & 3rd & \cellcolor{lightgreen!70}{1st} & 2nd & 24 & 3rd & 2nd & \cellcolor{lightgreen!70}{1st} & 14 & 2nd & 3rd & \cellcolor{lightgreen!70}{1st} & 22 \\
Group 15 & 2nd & \cellcolor{lightgreen!70}{1st} & 3rd & 38 & 3rd & \cellcolor{lightgreen!70}{1st} & 2nd & 30 & \cellcolor{lightgreen!70}{1st} & 2nd & 3rd & 34 & \cellcolor{lightgreen!70}{1st} & 3rd & 2nd & 18 \\
Group 16 & 3rd & \cellcolor{lightgreen!70}{1st} & 3rd & 28 & 2nd & 3rd & \cellcolor{lightgreen!70}{1st} & 22 & \cellcolor{lightgreen!70}{1st} & 3rd & \cellcolor{lightgreen!70}{1st} & 22 & \cellcolor{lightgreen!70}{1st} & 3rd & 2nd & 22 \\
\midrule[1pt]
\cellcolor{lightgreen!70}{\textbf{1st (Most)}}  & \cellcolor{lightgreen!70}{3} & \cellcolor{lightgreen!70}{\textbf{[10]}} & \cellcolor{lightgreen!70}{4} & - & \cellcolor{lightgreen!70}{\textbf{[7]}} & \cellcolor{lightgreen!70}{\textbf{[7]}} & \cellcolor{lightgreen!70}{3} & - & \cellcolor{lightgreen!70}{5} & \cellcolor{lightgreen!70}{6} & \cellcolor{lightgreen!70}{\textbf{[7]}} & - & \cellcolor{lightgreen!70}{\textbf{[8]}} & \cellcolor{lightgreen!70}{5} & \cellcolor{lightgreen!70}{5} & - \\
2nd         & 5 & 3 & 6 & - & 3 & 2 & 10 & - & 6 & 5 & 2 & - & 6 & 5 & 3 & - \\
\cellcolor{red!20}{\textbf{3st (Least)}} & \cellcolor{red!20}{\textbf{[8]}} & \cellcolor{red!20}{3} & \cellcolor{red!20}{6} & - & \cellcolor{red!20}{6} & \cellcolor{red!20}{\textbf{[7]}} & \cellcolor{red!20}{3} & - & \cellcolor{red!20}{5} & \cellcolor{red!20}{5} & \cellcolor{red!20}{\textbf{[7]}} & - & \cellcolor{red!20}{2} & \cellcolor{red!20}{6} & \cellcolor{red!20}{\textbf{[8]}} & - \\ \midrule[0.3pt]\bottomrule[1pt]
\end{tabular}
}
\label{tab:my_label}
\end{table*}

\begin{table*}[t]
\renewcommand{\arraystretch}{1.1}
\caption{The results of statistical analysis on the ratio of gaze distribution for each target (${}^{*}$\   $p<.05$;  ${}^{**}$\  $p<.01$; and ${}^{***}$\ $p<.001$).~\label{table:statisticalresultaboutgaze}}
\Description{The results of statistical analysis on the ratio of gaze distribution for each target
}
\resizebox{1\linewidth}{!}{
\begin{tabular}{lllll}
\toprule[1pt]\midrule[0.3pt]
\textbf{Target of Gaze}        & \textbf{Engage}                 & \textbf{Affective}              & \textbf{Interaction}          & \textbf{post-hoc} \\ \midrule[1pt]
\textbf{Virtual agent}         & \textit{F = 59.32; p $<$ .001}  & \textit{F = 22.39; p < .001} & \textit{F = 2.04; p = .015} & E-A > NE-A***; E-A > NE-NA***; \\
&&&&E-NA > NE-NA; NE-A > NE-NA*** \\
\textbf{Task object}           & \textit{F = 26.25; p $<$ .001}  & \textit{F = 5.57; p = .020}   & \textit{F=  4.79; p = .031}  & E-A < NE-NA***; E-NA < NE-NA***; NE-A < NE-NA*    \\ 
\textbf{The other participant} & \textit{F = 0.02; p = .86}      & \textit{F = 8.51; p = .004}     & \textit{F = 4.18; p = .043}    & E-A < E-NA**                      \\ 
\textbf{Elsewhere}             & \textit{F = 0.73; p = .39}      & \textit{F = 0.03; p = .86}      & \textit{F = 1.04; p = .30}    &                                                                                 \\\midrule[0.3pt]\bottomrule [1pt]
\end{tabular}
}
\end{table*}


\subsubsection{Attention Allocation} 
\label{gazeResult}

The  distribution of the participants' gazes during group discussions (Step 2 in Section~\ref{sec:procedure})
were captured (using an OpenCV-based program) 
focused on four targets: \textit{Virtual Agent}, 
\textit{Task Object} (laptop that shows the ongoing discussion situation), 
\textit{The Other Participant}, and \textit{Elsewhere} (e.g., floor, ceiling). 

\begin{figure}[!b]
\centering
\includegraphics[width=0.83\linewidth]{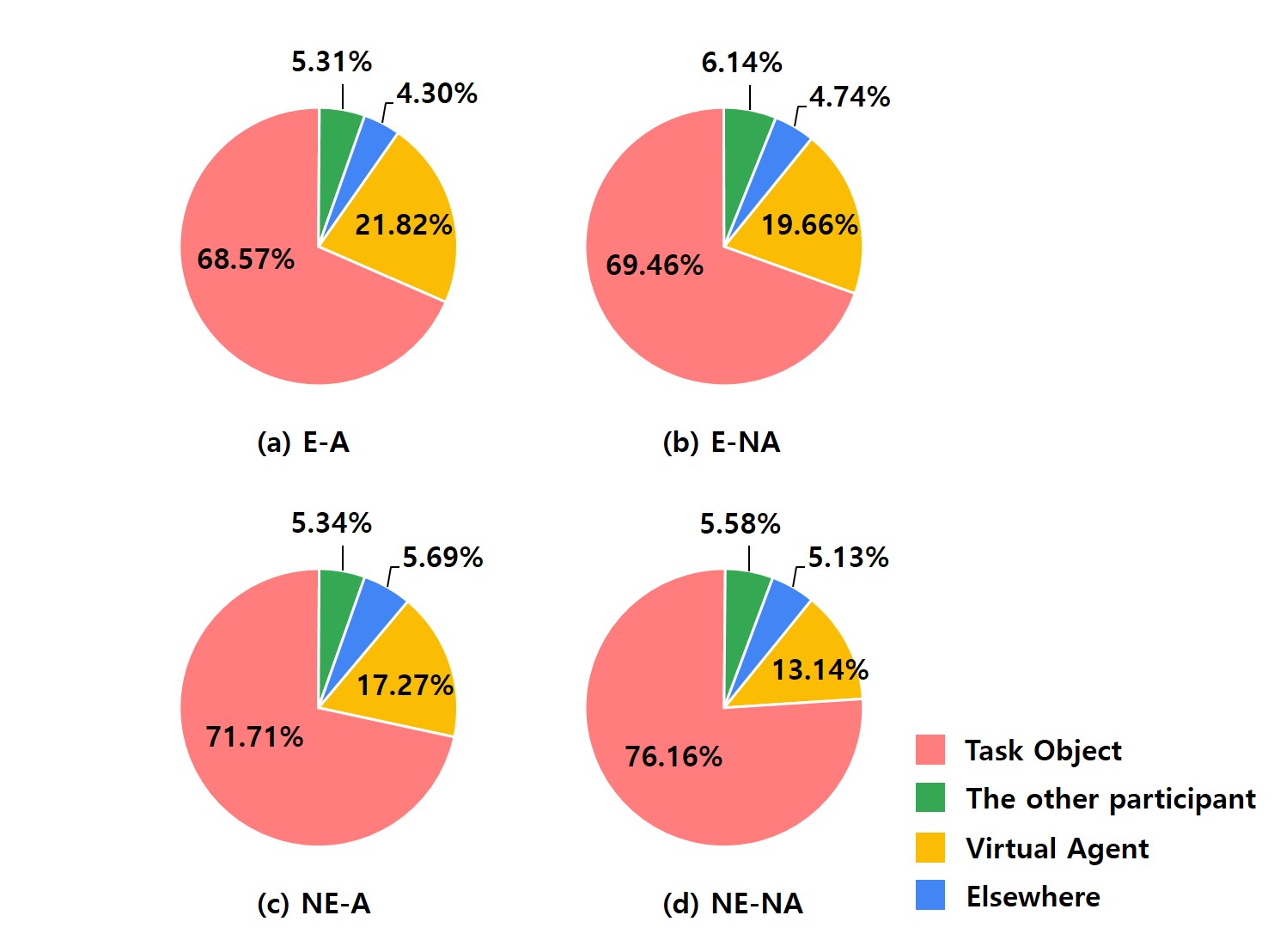}
\vspace{-0.5em}
\caption{
Average gaze distribution for each target.\label{fig:gazebehavior}
}
\Description{The average ratio of gaze distribution for each target.}
\end{figure}

Figure~\ref{fig:gazebehavior} shows the average gaze distribution results for each target.
The Mauchly's test revealed significant sphericity violation only for the \textit{Virtual Agent} ($p$ < .001). The Shapiro-Wilk test identified non-normal distributions for several variables: \textit{Task Object} (E-A: $p$ = .001 and NE-A: $p$ = .048), \textit{The Other Participant} (E-A: $p$ = .003), 
and \textit{Elsewhere} (E-A: $p$ = .034, NE-A: $p$ = .003, and NE-NA: $p$ = .036).
Thus, all categories were analyzed using the ART method followed by the ART-C post-hoc tests with Bonferroni correction with a 5\% significance level.
Significant effects were found for all gaze target categories except \textit{Elsewhere}.
More details are described in Table~\ref{table:statisticalresultaboutgaze}.

In the case of the \textit{Virtual Agent}, 
it was observed that VA's engagement played a definite and positive role in capturing users' attention, even in group settings ($F$ = 59.32; $p <$ .001). 
Moreover, the VA's affective behavior also had a significant effect on attracting users' attention ($F$ = 22.39; $p <$ .001).
However, their interaction did not yield any significant effect  ($F$ = 2.04; $p$ = .155).
The post-hoc analysis showed that attention toward the \textit{Virtual Agent} significantly increased when it displayed engaging behavior compared to non-engaging agents
(E-A $>$ NE-A; $p <$ .001***), 
(E-A $>$ NE-NA; $p <$ .001***), 
and (E-NA $>$ NE-NA; $p <$ .001***).
Furthermore, even without the engaging behavior, VA's affective behavior also significantly increased gaze allocation (NE-A $>$ NE-NA; $p <$ .001***).

Attention to the \textit{Task Object} was also significantly affected by the Engage ($F$ = 26.25; $p <$ .001) and Affective factors ($F$ =  5.57; $p$ = .02), and  
and their interaction ($F = $ 4.72; $p$ = .031).
The post-hoc indicated that the gaze towards \textit{Task Object} significantly decreased when the VA exhibited engaging/affective behavior in three pairs: (E-A < NE-NA; $p <$ .001***), (E-NA < NE-NA; $p <$ .001***), and (NE-A $<$ NE-NA; $p$ = .03*).
This further supports the results with regards to the \textit{Virtual Agent}, which indicated that participants' attention was directed more toward it based on the VA's behavior.

We observed a significant effect on the gaze target for \textit{The Other Participant}, by the Affective factor ($F$ = 8.51; $p$ = .004) and its interaction ($F$ = 4.18; $p$ = .04), 
but not by the Engage factor ($F$ = 0.02; $p$ = .86).
The post-hoc revealed that the subjects were looking significantly more at \textit{The Other Participant} in the E-NA  than in the E-A (E-A $<$ E-NA; $p$ = .003**).
These results suggest that the combined engagement and affective behavior (i.e., E-A) did not always draw attention toward the VA. 
Instead, it often led to avoidance of eye contact with the VA or directed participants' attention toward something else.
Participants may have perceived that they were more comfortable focusing on \textit{The Other Participant} who were less emotionally involved or conforming to the social norms~\cite{torre2019effect}.

\subsection{Factor Effects - Subjective Measures} 
\label{sec:4.3}
\begin{figure*}[h]
\centering
\subcapcentertrue
\subfigure[Social Presence (Continuity)]{\includegraphics[width=0.19\linewidth]{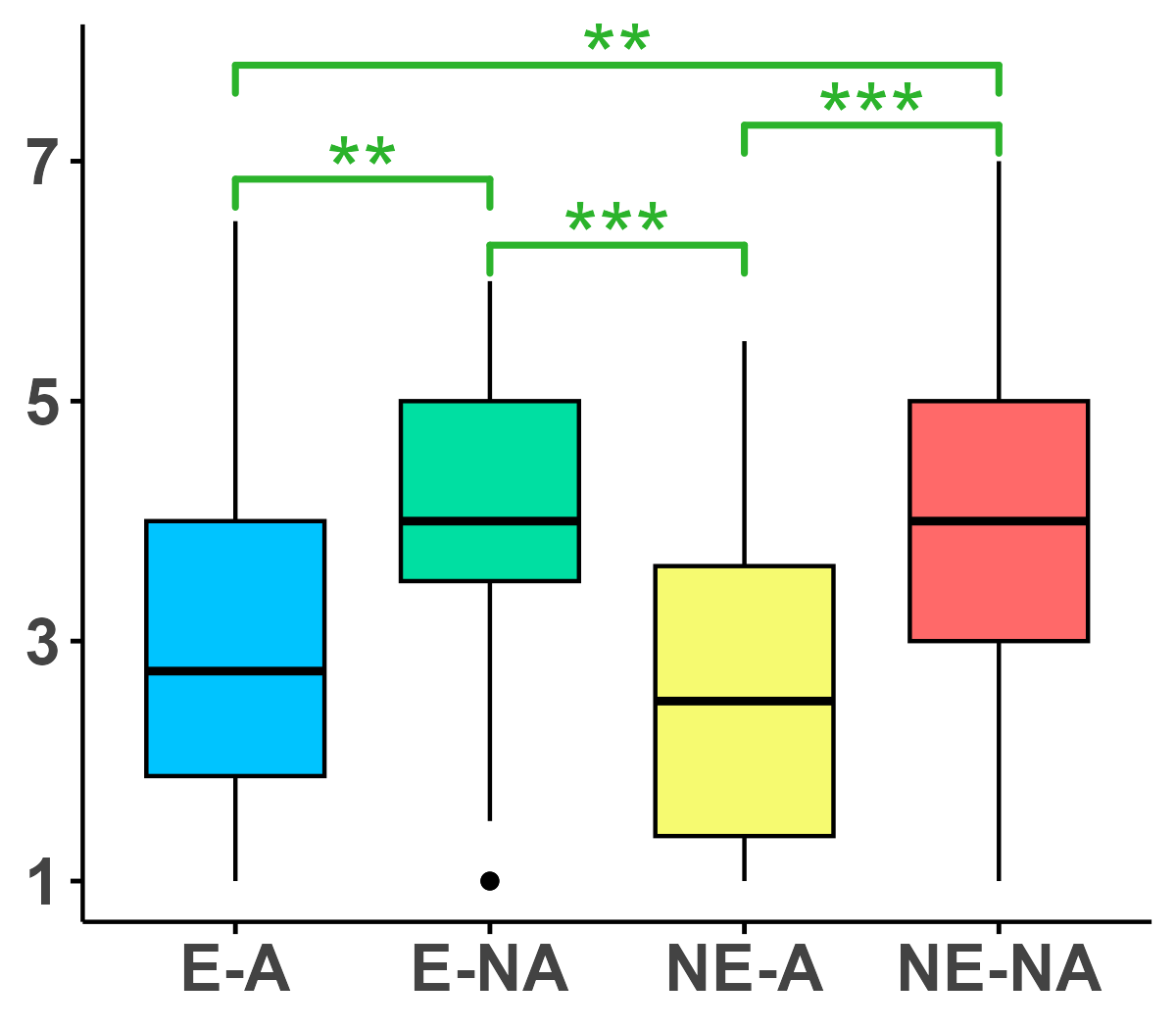}} 
\hfill
\subfigure[Social Presence (Responsiveness)]{\includegraphics[width=0.19\linewidth]{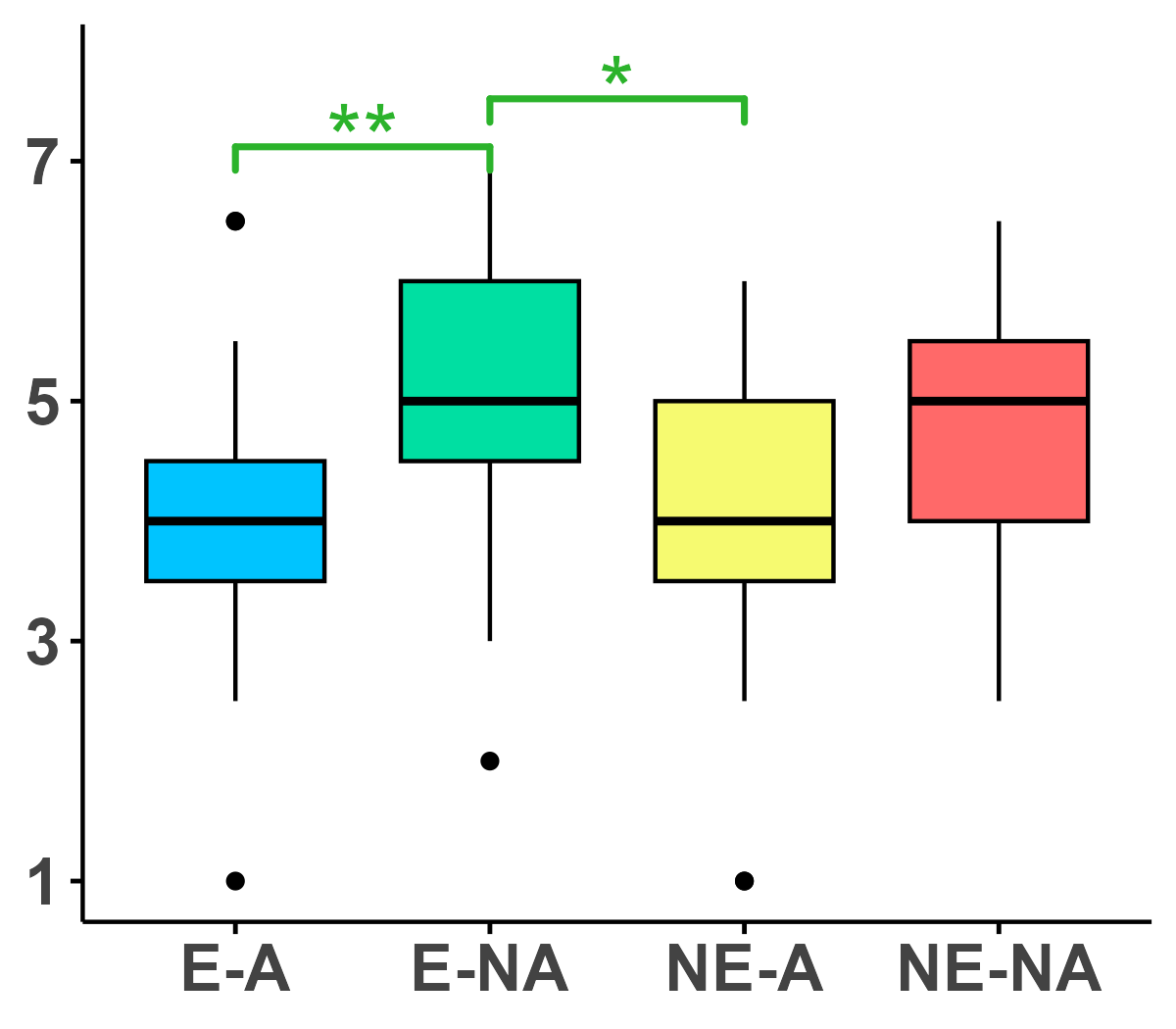}} 
\hfill
\subfigure[Social Presence (Membership)]{\includegraphics[width=0.19\linewidth]{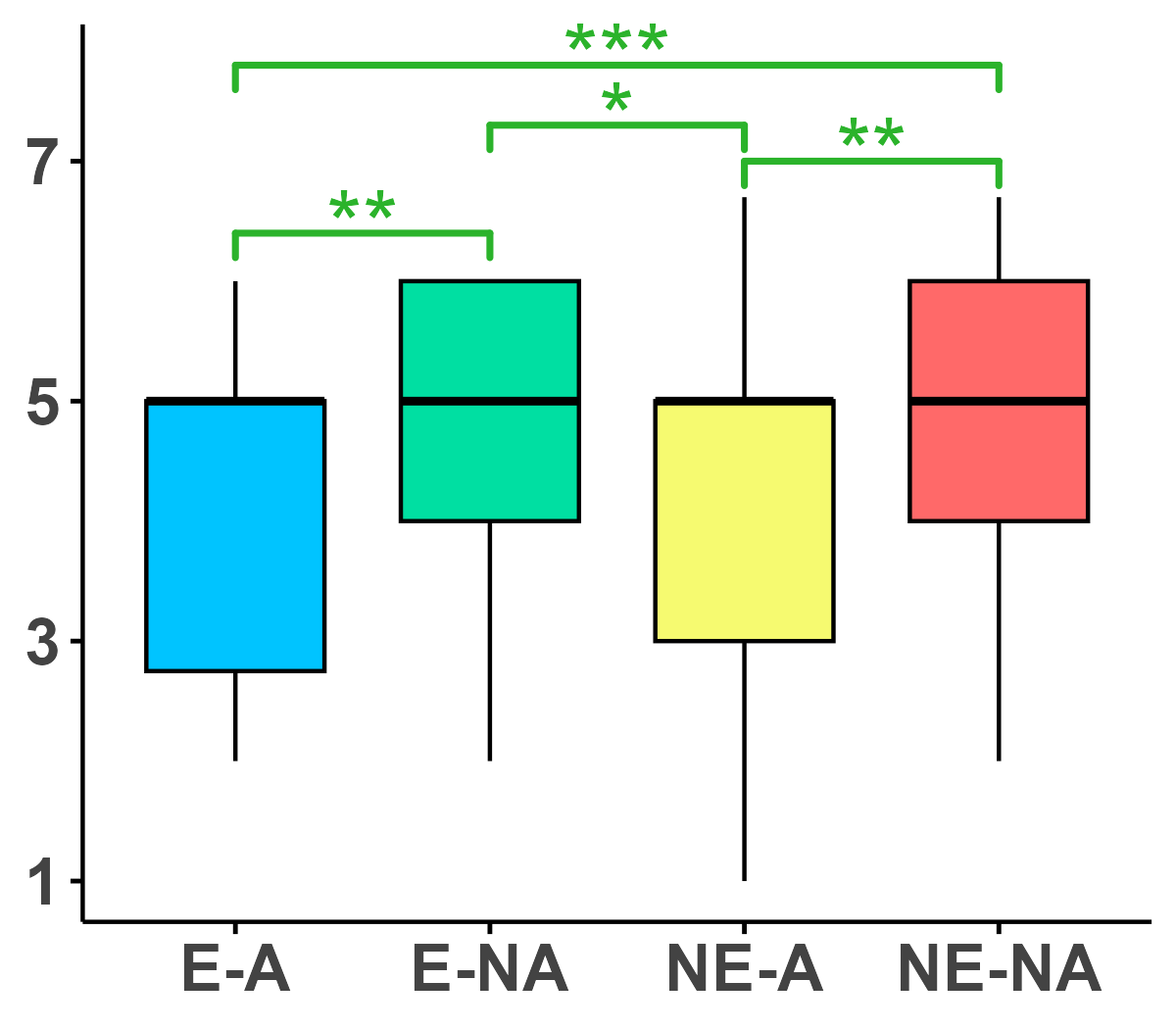}} 
\hfill
\subfigure[Social Influence\, (Positive)]{\includegraphics[width=0.19\linewidth]{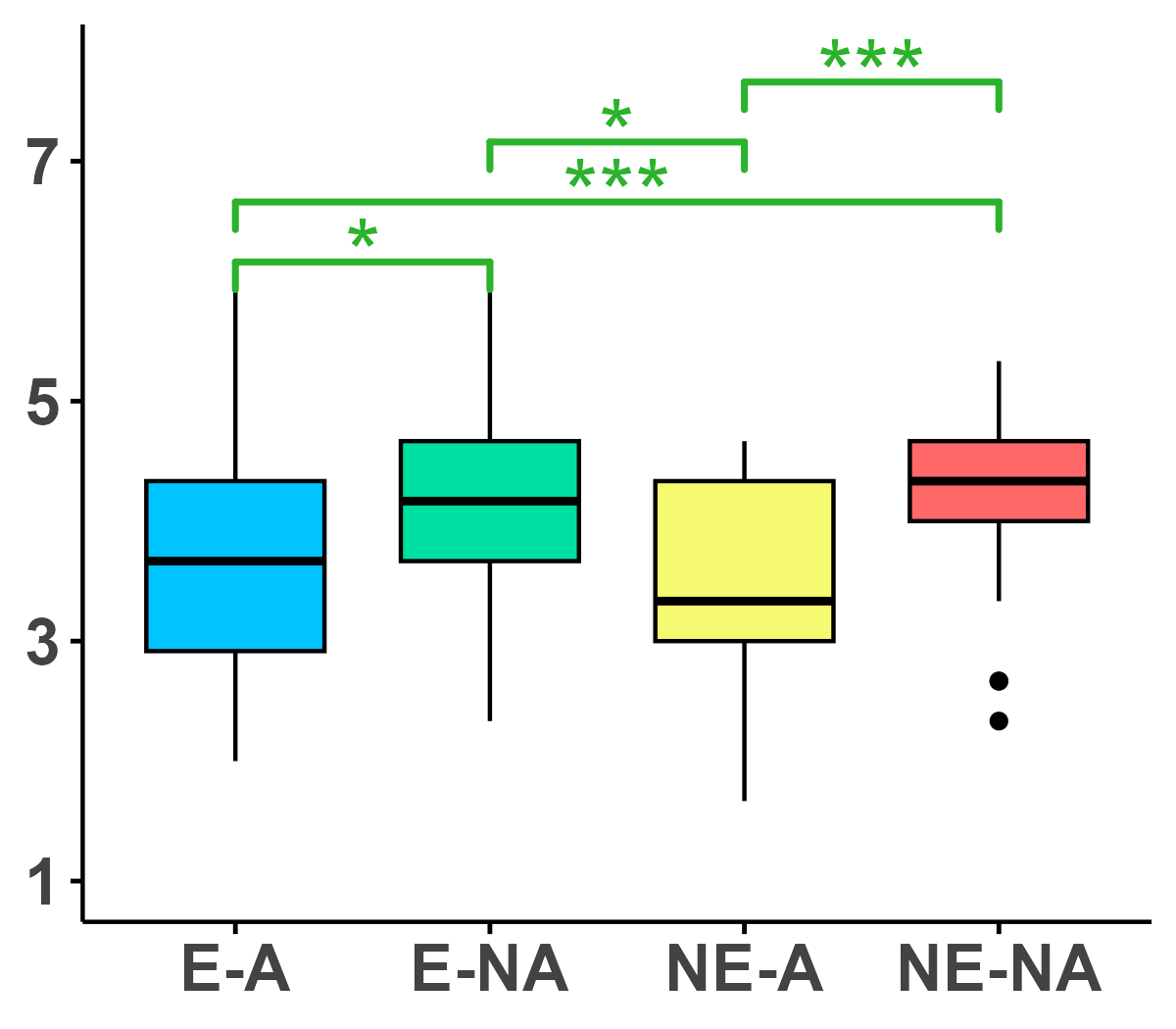}} 
\hfill
\subfigure[Social Influence (Negative)]{\includegraphics[width=0.19\linewidth]{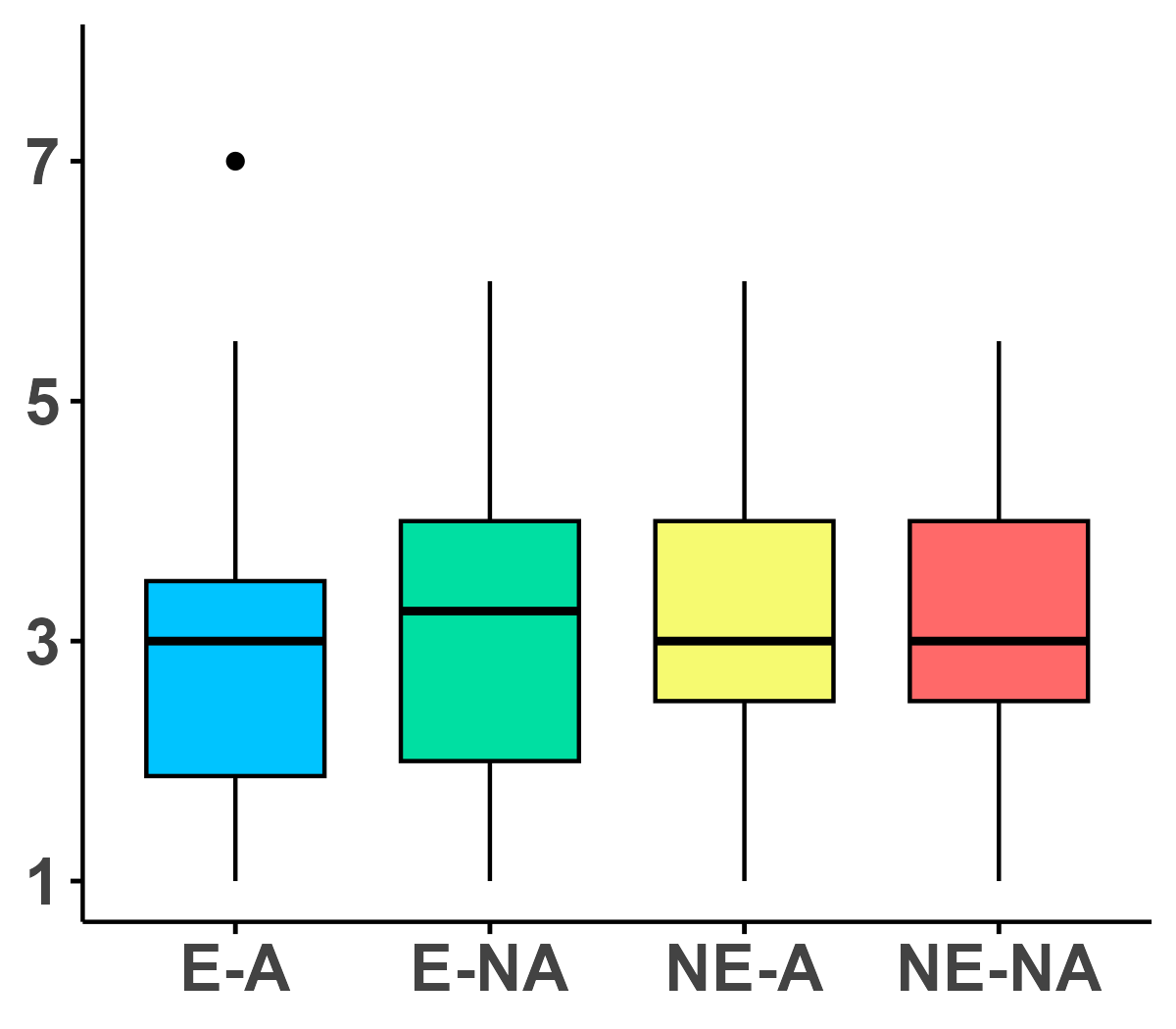}} 
\\ [-0.5em]
\subfigure[Trustworthiness (Positive)]{\includegraphics[width=0.19\linewidth]{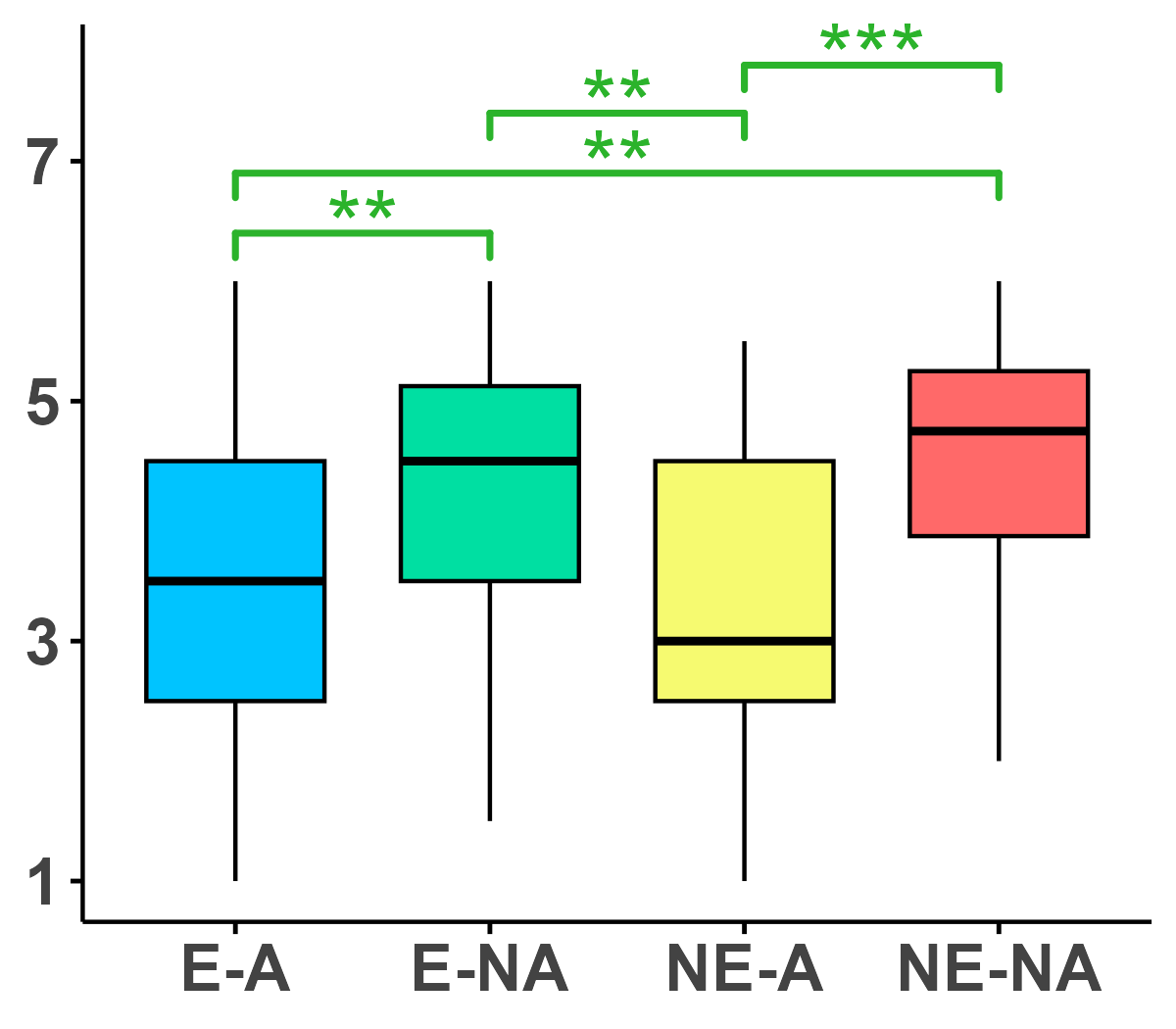}} 
\hfill
\subfigure[Trustworthiness (Negative)]{\includegraphics[width=0.19\linewidth]{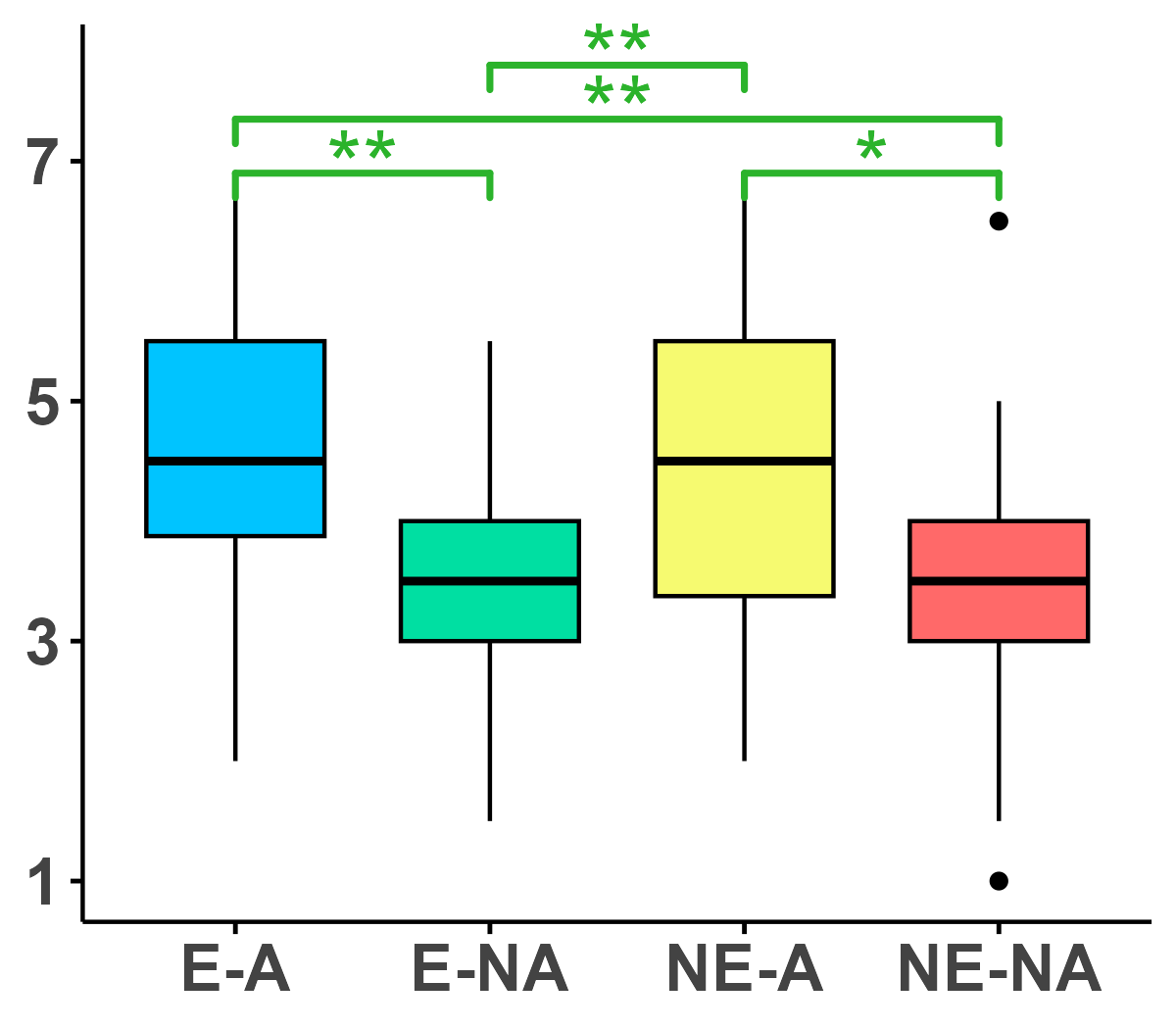}} 
\hfill
\subfigure[Task Load (Mental Demand)]{\includegraphics[width=0.19\linewidth]{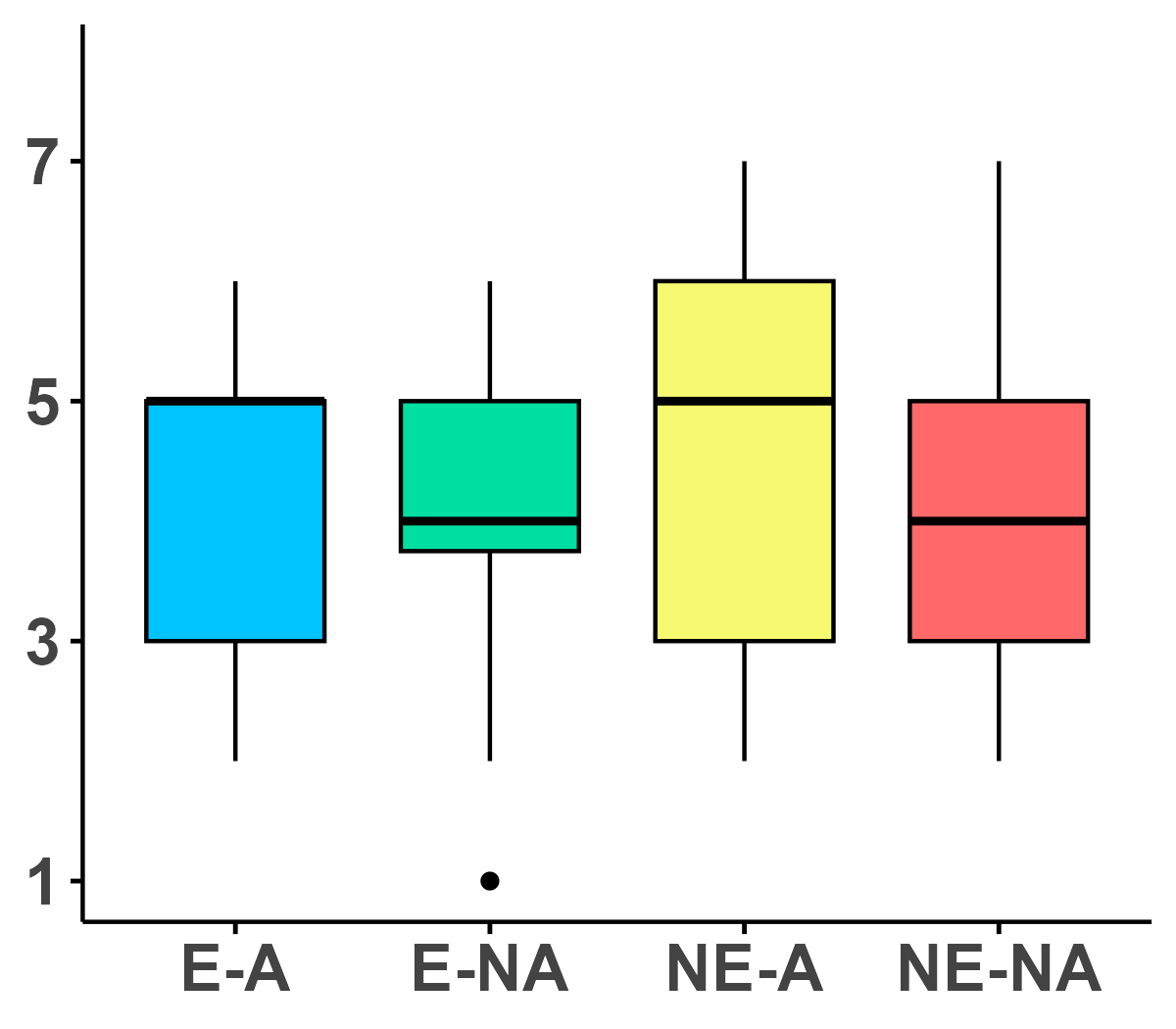}}  
\hfill
\subfigure[Task Load (Performance)]{\includegraphics[width=0.19\linewidth]{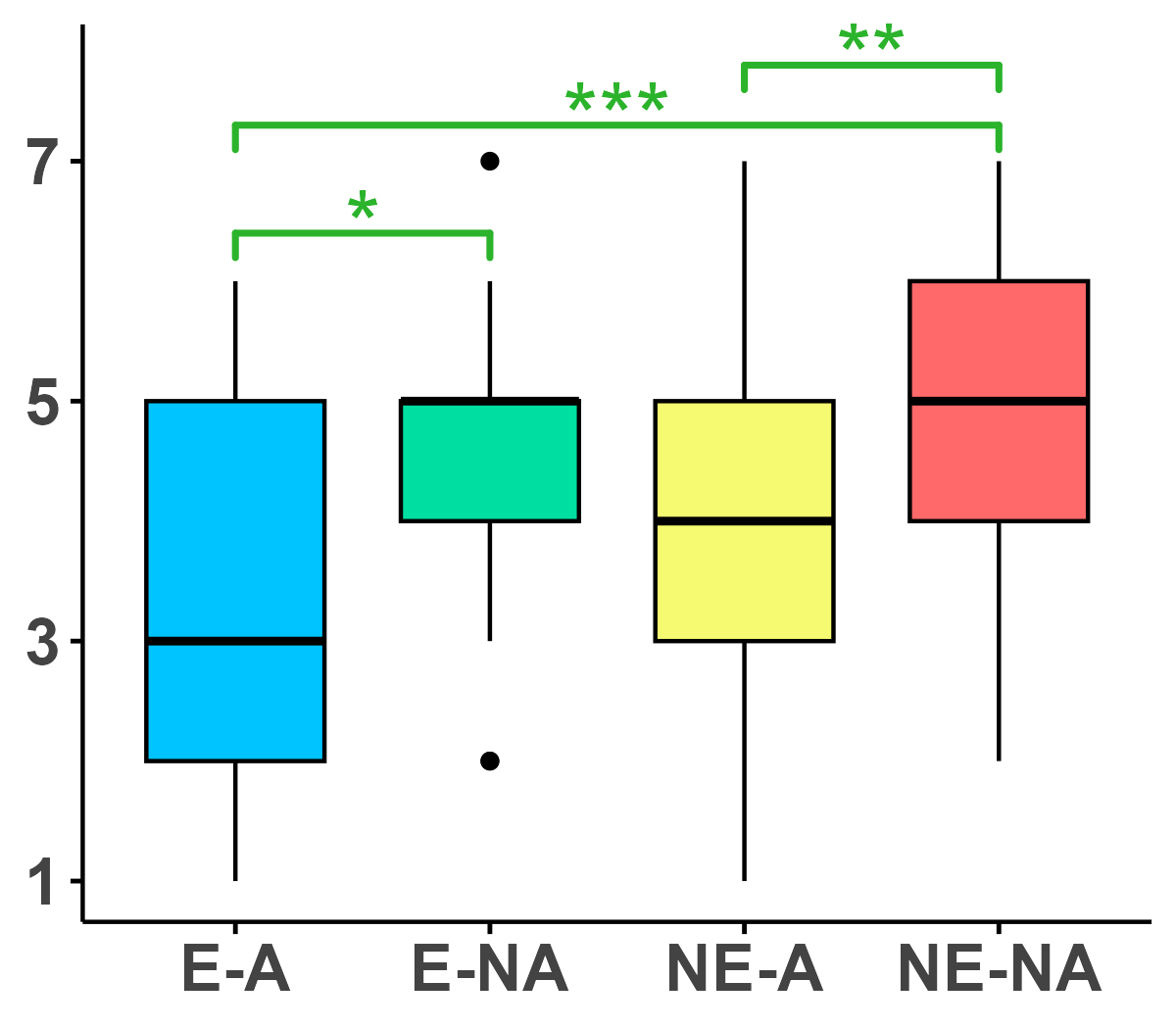}}
\hfill
\subfigure[Task Load (Frustration)]{\includegraphics[width=0.19\linewidth]{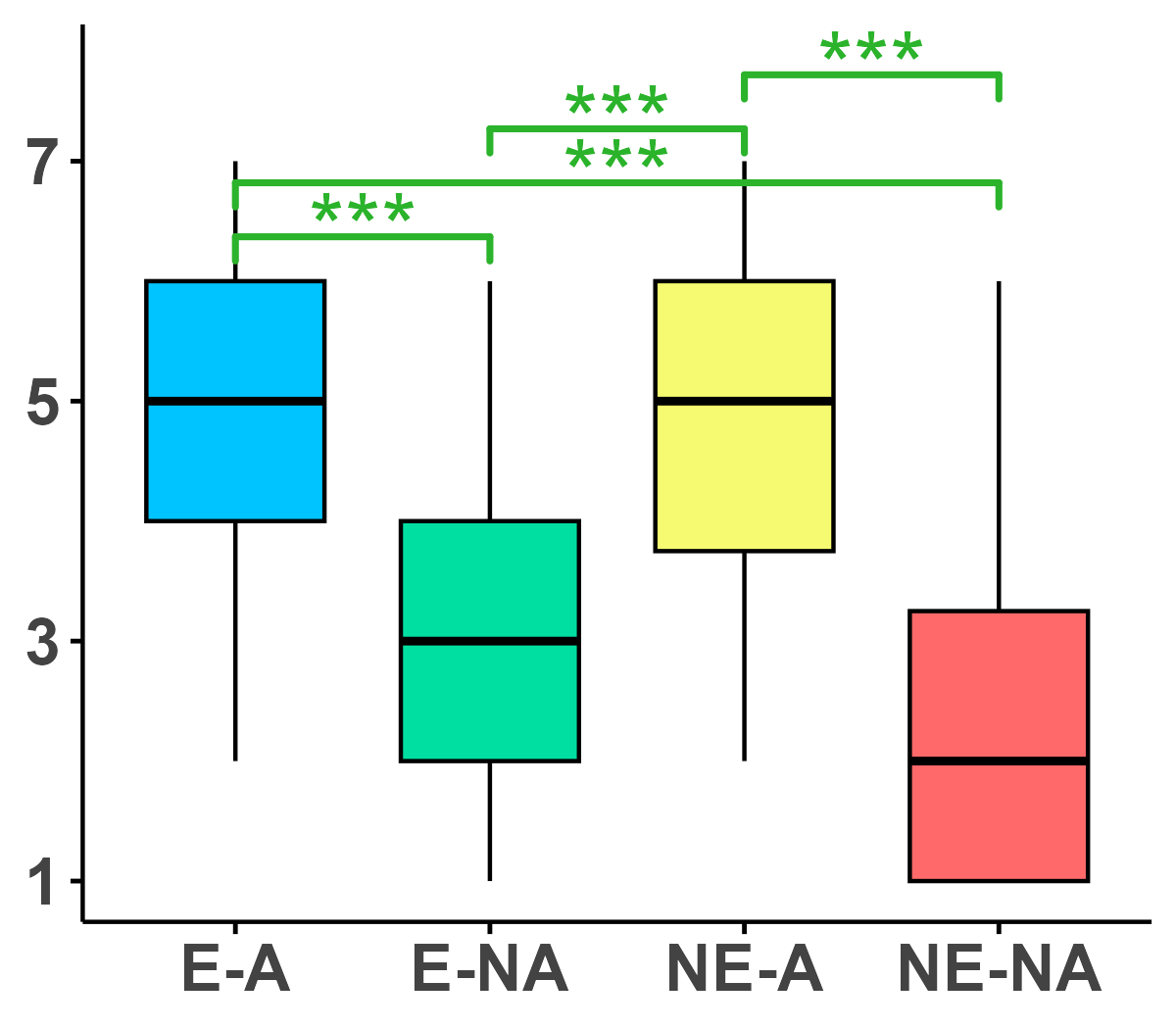}} 
\vspace{-0.5em}
\caption{Boxplots of the subjective measures on a 7-point Likert scale 
(* $p$\,\textless\,.05; ** $p$\,\textless\,.01; and *** $p$\,\textless\,.001).\label{fig:subjectiveART}}
\Description{Boxplots of the subjective measures on a 7-point Likert scale, including social presence, social influence, trustworthiness, and task load.}
\end{figure*}

The same statistical analysis procedure (as described in Section~\ref{sec:vali_art}) was applied.
The Mauchly's test revealed sphericity violations in the following measures: responsiveness ($p$ = .016), membership ($p$ = .002), the negative aspect of social influence ($p$ = .005), the negative aspect of trustworthiness ($p$ = .002), and performance ($p$ = .023).
Furthermore, the Shapiro-Wilk test indicated that
there was at least one treatment data that deviated from the normal distribution.
As such, the ART method was employed for the statistical analysis.
For the post-hoc, the ART-C method was used with Bonferroni correction. 
All analyses were set at the 5\% significance level.
Overall, the VA's affective behavior showed a significantly negative impact on the subjective measures, while the VA's engagement had a positive impact.
The analysis results are summarized in Table~\ref{Table:ARTResults} and Figure~\ref{fig:subjectiveART}.
An in-depth discussion of the subjective variables is presented in Section~\ref{discussion}, but two variables with no statistical differences are described below with several potential reasons.

\begin{table*}[h]
\renewcommand{\arraystretch}{1.2}
\centering
\caption{
Summary of the results for subjective measures: social presence (SP), social influence (SI), trustworthiness (TN), and task load (TL).
These results were analyzed using a two-way non-parametric analysis (${}^{*}$\   $p<.05$;  ${}^{**}$\  $p<.01$; and ${}^{***}$\ $p<.001$).
}
\Description{Summary of the results for subjective measures, including social presence, social influence, trustworthiness, and task load).
These results were analyzed using a two-way non-parametric analysis.
}
\label{Table:ARTResults} 
\resizebox{1\linewidth}{!}{
\begin{tabular}{llllll}
\toprule[1pt]\midrule[0.3pt]
\multicolumn{2}{l}{\textbf{Subjective Measures}}    & \textbf{Engage}           & \textbf{Affective}       & \textbf{Interaction}     & \textbf{post-hoc}                                   \\ \midrule[1pt]
\textbf{SP} & \textbf{Continuity} & $F$ = 0.3, $p$ = .56 & $F$ = 32.6, $p <$ .001 & $F$ = 0.0, $p$ = .86 & 
E-A $<$ E-NA**; E-A $<$ NE-NA**; E-NA $>$ NE-A***; NE-A $<$ NE-NA*** \\
\textbf{} & \textbf{Responsiveness} & $F$ = 0.0, $p$ = .91 & $F$ = 12.8, $p <$ .001 & $F$ = 0.6, $p$ = .41 & 
E-A $<$ E-NA**; E-NA $>$ NE-A* \\ \vspace{0.5em}
\textbf{} & \textbf{Membership} & $F$ = 1.1, $p$ = .29 & $F$ = 21.1, $p <$ .001 & $F$ = 0.0, $p$ = .83 & 
E-A $<$ E-NA**; E-A $<$ NE-NA***; E-NA $>$ NE-A*; NE-A $<$ NE-NA** \\ 
\textbf{SI} & \textbf{Positive} & $F$ = 0.4, $p$ = .48 & $F$ = 25.2, $p <$ .001 & $F$ = 0.1, $p$ = .69 & 
E-A $<$ E-NA*; E-A $<$ NE-NA***; E-NA $>$ NE-A*; NE-A $<$ NE-NA*** \\ \vspace{0.5em}
\textbf{} & \textbf{Negative} & $F$ = 1.2, $p$ = .26 & $F$ = 0.63, $p$ = .42 & $F$ = 0.7, $p$ = .39 &  \\ 
\textbf{TN} & \textbf{Positive} & $F$ = 0.0, $p$ = .94 & $F$ = 30.4, $p <$ .001 & $F$ = 0.4, $p$ = .48 & 
E-A $<$ E-NA**; E-A $<$ NE-NA**; E-NA $>$ NE-A**; NE-A $<$ NE-NA***  \\ \vspace{0.5em}
\textbf{} & \textbf{Negative} & $F$ = 0.0, $p$ = .89 & $F$ = 20.5, $p <$ .001 & $F$ = 0.0, $p$ = .82 & 
E-A $>$ E-NA**; E-A $>$ NE-NA**; E-NA $<$ NE-A**; NE-A $>$ NE-NA* \\ 
\textbf{TL} & \textbf{Mental Demand} & $F$ = 0.0, $p$ = .89 & $F$ = 0.74, $p$ = .39 & $F$ = 1.4, $p$ = .23 & 
\\ 
\textbf{} & \textbf{Performance} & $F$ = 1.7, $p$ = .18 & $F$ = 20.1, $p <$ .001 & $F$ = 0.0, $p$ = .76 & 
E-A $<$ E-NA*; E-A $<$ NE-NA***; NE-A $<$ NE-NA**  \\ 
\textbf{} & \textbf{Frustration} & $F$ = 1.2, $p$ = .25 & $F$ = 76.5, $p <$ .001 & $F$ = 0.9, $p$ = .33 &
E-A $>$ E-NA***; E-A $>$ NE-NA***; E-NA $<$ NE-A***; NE-A $>$ NE-NA*** \\ 
\midrule[0.3pt]\bottomrule[1pt]
\end{tabular}
\vspace{-0.5em}
}
\end{table*}

\textbf{Social influence (Negative):}
The median scores for all conditions were close to 3 (out of 7), as shown in  Figure~\ref{fig:subjectiveART} (e). 
Such scores may be attributed to cognitive dissonance and reactance, which refer to the psychological factors that make individuals reluctant to acknowledge adopting opinions that contradict their beliefs~\cite{brehm2013psychological}.

\textbf{Mental demand:}
The medians of all conditions were larger than 4 (out of 7), as shown in Figure~\ref{fig:subjectiveART} (h). 
The result implies that, regardless of the VA's behaviors, participants experienced psychological distress while conversing with the VA.

\vspace{0.25em}
Finally, the \textbf{\textit{Preference Rankings}}, which were collected from the post-experiment questionnaire, are shown in Figure~\ref{fig:rank order} (c).
Considering the ordinal rank measures, we followed the same analysis process (as described in Section~\ref{sec:4.1.2}).
Friedman test indicated a significant effect ($\chi^2$\,=\,54.0; $p <$ .001***), and the post-hoc revealed significant differences between: 
(E-A $<$ E-NA; $p <$ .001***), 
(E-A $<$ NE-NA; $p $ = .006**), 
(E-NA $>$ NE-A; $p <$ .001***), and 
(NE-A $<$ NE-NA; $p <$ .001***).
The findings suggest that participants had distinct preferences for the VA behavior during group discussions. 
Specifically, in scenarios where social norms were encouraged, the most preferred VA was the one who exhibited engaged behavior while maintaining a rational or emotionally neutral demeanor. 
This implies that a VA characterized by engagement and neutral affect (i.e., E-NA's VA) was perceived as the ideal team member.

\section{Discussion}
\label{discussion}

\subsection{The Level of VA's Engagement in Groups}

Regarding RQ1 (What is the effect of VA's engagement in group discussions?), 
there is a general acknowledgment that
VA's engaging behaviors contribute to fostering successful interactions with individual users~\cite{devault2014simsensei, kangsooReducing, kenny2007building}. 
However, this study prompts an inquiry into how these effects are transferable to group settings. 
We hypothesize that the presence of other humans may instigate different 
social dynamics or at least attenuate the aforementioned effect.

The results of the gaze distribution revealed that participants allocated significantly more attention towards the engaged VA (E-A, E-NA) than the non-engaged ones.
These findings underscore the substantial impact of VA's engagement on enhancing user perception, aligning with observed results from the usual dyadic (1-to-1) interactions~\cite{kangsooReducing, myunho2016wobblytable}.
However, the subjective measures indicate only a modest improvement in the users' 
perceptions of VAs.
The statistical analysis showed that the engaging factor had a minor impact (or non-significant), especially when contrasted with the affective factor.
Although the VA's engagement significantly influenced participants' attention allocation, the group settings may have limited the time available for these engagements to alter user perceptions, as we had hypothesized. 

These limitations likely originate from two main reasons.
Firstly, the user's expectations of the VA may clash with those of other ``human'' members in group settings—an aspect not observed in dyadic interactions.
In prior studies, the VAs were often designed to focus on a single specific behavior~\cite{hanseob2021visualeffect, choi2012affective, ali2020automatic, kum2022can}, a common approach to exploiting the impact of VA engagement. 
In such cases, users might perceive a specific VA as awkward compared to others, and yet conclude that the less awkward VA demonstrates higher engagement. 
However, in our experiment, human counterparts actively participated in interactions with the VA. 
While users may not overtly compare the VA and other participants, it is plausible that they subconsciously made such comparisons regarding the engagement levels of the VAs with other human participants,
and this could be perceived as less acceptable compared to human counterparts.

Another contributing factor is that, in group or discussion settings, user attention directed towards the VA may be divided among other humans~\cite{sebo2020robots, ter2011design} and discussion tools, such as laptops and presentation screens~\cite{rehm2005they}. In fact, participants allocated at least 5\% of their attention to \textit{The Other Participant} and dedicated a significant portion, approximately 70\%, to the \textit{Task Object}. It is natural for participants to often direct their attention to the \textit{Task Object} to facilitate seamless discussions, given that it presents an overview of the ongoing situation—a pattern consistently noted in previous studies~\cite{rehm2005they, wong2023comparing, wang2019exploring}.
An additional explanation could be that the visibility of the \textit{Task Object} naturally encouraged participants to avoid direct eye contact, potentially relieving psychological burdens~\cite{singh2021beholden, phelps2006helping, takano2012tablets}.
This consideration is particularly pertinent given that our study was conducted with unfamiliar participants.

While the presence of other humans and task objects may pose challenges
for humans to engage with VAs, 
their participation in interactions with VAs is unavoidable in group and organizational settings.
It is evident that creating innovative strategies for VAs to capture users' attention even though other humans and task objects exist -- by making it much more informative as the ``go-to'' entity for critical pieces of information in decision-making (e.g., taking up the role of situation monitor of the laptop). 

\subsection{Negative Aspects of VA's Affective Behavior}
Considering RQ2 (What is the effect of the level of VAs’ affectiveness in group discussion?), 
subjective measures showed that VAs with rich emotions were consistently not received well.
In scenarios where a group makes decisions with unfamiliar individuals, participants might perceive the VA's affective behavior as acting odd, ambivalent, or inconsistent because the VA does not conform to social contexts/norms (particularly when compared to real humans)~\cite{qu2014conversations, torre2019effect, GerdIeeeVR2023, gatto2022met}.
The affective VA incorporated emotional utterances which participants may not easily disregard~\cite{torre2019effect}. 
Considering that the participants' primary objective was to complete the task successfully (The great attention to \textit{Task Object} may support this assumption, as shown in Figure~\ref{fig:gazebehavior}).
In this context, emotional utterances, even if correct, can be seen as relatively unrelated to the task and disregardable or even distracting~\cite{veletsianos2012learners}.
Such a VA's behavior, which disregards the social context and manifests their responses positively or negatively (based on the correctness of the statement), may ultimately lead to the negative perception (even if it is correct) and explains why the affective condition yielded relatively lower values in the various subjective measures.

Another potential factor is the formation of competitive relationships between a VA and participants rather than discussions being collaborative.
This aligns with the theory that humans tend to view real humans as part of their in-groups and robots (or virtual agents) as part of out-groups~\cite{rudman2004gender, baron2015representing, competitivewithrobot}.
Support for this possibility can be found in Figure~\ref{fig:subjectiveART} (c), which shows the membership aspect of the affective VAs being statistically lower than that of the non-affective VAs.
Furthermore, the NE-A's score variations were relatively lower compared to other conditions (see Figure~\ref{fig:groupscore}).
Notably, in most instances involving NE-A's VA (6 out of 7 cases), group scores declined compared to individual scores.
These results support the notion that discussions involving affective agents tend to be more competitive than cooperative.
Therefore, it is conceivable that the participants perceived the affective/emotional VA (which does not adhere to social norms) as a competitive out-group member.
Consequently, these factors are believed to be responsible for the negative effects of VAs' affective/emotional behavior.

Based on such findings, we recommend that 
the VAs first establish cooperative and amicable relationships with human interlocutors 
(e.g. through ``casual'' emotional expression),
and then safely continue to employ such affective qualities (as the means of making the VAs look natural and consistent) for the critical task discussions thereafter, particularly in groups or organizational settings.
In line with the work by Veletsianos~\cite{veletsianos2012learners}, 
after building the base trust, 
it would be recommended not to 
act casually and ensure the VA behaves professionally and focuses on the task.

\subsection{Design of VA Behaviors for Group Dynamics}
Considering RQ3 ({What type of a VA 
is appropriate for facilitating the group's decision-making in terms of its engaging and/or affective behaviors?), 
we questioned whether the VA's role in the group was ``to enhance group synergy'' or ``influence decision-making.''

\subsubsection{VA's engagement enhances group synergy}

While no significant difference was found in \textit{Group Scores}, 
significant improvements in group scores from individual scores were observed in all treatments (see Figure~\ref{fig:groupscore}).
This general improvement would be owing to collective intelligence~\cite{hiltz1991group,mchugh2016collective}, as observed in prior studies~\cite{walker2019influence, kangsooReducing}.
Nonetheless, the engaging VA (i.e., E-A, E-NA) had relatively better scores and 
more focused gaze compared to the non-engaging VA (i.e., NE-A, NE-NA), as shown in Figure~\ref{fig:groupscore}.
In summary, the VA's engaging behavior had a positive effect on the group discussion.
In addition, 
E-NA's VA was significantly preferred compared to E-A and NE-A (see Figure~\ref{fig:rank order} (c)).
However, as shown in Table~\ref{fig:two_tables}, 
under E-A and E-NA treatments, VA had relatively little influence on the group's final decisions compared to the NE-A treatment.
Thus, we conclude that the primary role of the VA's engagement would be to facilitate group synergy rather than influence decision-making.

This can be explained by the ``social facilitation'' effect, which states that an individual's performance or behavior can be influenced by the attention/expectation of others~\cite{rosenberg2020robot, harkins1987social, cottrell1968social}.
The VA's engagement behavior, driven by the \textit{ItD} algorithm (see Section~\ref{sec:speaker identification}), was exhibited by naturally leaning towards the direction of the current speaker, making eye contact, and showing head nodding as if the VA had interest and acknowledged with one's opinions~\cite{devault2014simsensei}.
This behavior might have motivated participants to present their views more enthusiastically, seeking to fulfill the VA's expectations.
The behavior of the VA likely had a positive impact on the participants' attitudes during discussions, underscoring the potential of VAs as ``facilitators'' in stimulating and maintaining engaging discussions, potentially fostering group synergy~\cite{rosenberg2020robot, matsuyama2015four}.

\subsubsection{VA's affectiveness influences decision-making}

On the other hand, the VA's affective behavior may play a crucial role in deciding whether participants accept or reject the VA's opinion.
According to the Reflection results described in Section~\ref{Sec:TaskScore} (also refer to Table~\ref{fig:two_tables}),
NE-A's VA's opinion was reflected both the most and least in the final decision.
Although these decisions are likely influenced by the participants' disposition~\cite{putten2010our}, 
participants might have shown resistance to NE-A's VA that lacked interactivity and/or failed to comprehend participants' emotions during the group discussion~\cite{choi2012affective, torre2019effect}. 
This perception could lead to participants rejecting the VA's opinions, even if the VA's statements were correct, 
as evidenced by the 6 out of 7 discussions with deteriorated group performance involving NE-A's VA (see result (3) in Section~\ref{Sec:TaskScore}).
From another perspective, 
participants may accept the VA's opinion more readily due to a desire to avoid impasse and conflict situations, regardless of the VA's confidence/trustworthiness~\cite{bond2003role, van2004interpersonal}.
Thus, the role of the VA's affectiveness is to influence ``passive'' decision-making, regardless of the main objective to derive the right group decision.

We propose another possible role for the affective VAs, namely ``convictive and emotionally charged mediator''. 
Such a trait might be fitting for, e.g., middle managers who, among many things, may be responsible to convey and push through the upper management directions and agendas, and persuade/lead the subordinates as amicably and passionately as possible~\cite{cooper2011managing}.
As it is adamant to maintain a healthy and agreeable relationship across the organizational hierarchy~\cite{hinkin1994examination}, such influencing mediation becomes critical~\cite{martin2016leader, yukl1989managerial}.
Affective VAs can assume (or help fulfill) the role of fostering acceptance sustained for extended duration (as indicated by our findings and prior research~\cite{veletsianos2012learners, van2004interpersonal}).
Other studies~\cite{pauw2022avatar, lucas2014s, lucas2017reporting} have shown that emotional VAs can build rapport and have a sense of anonymity (in which they do not spread one's stories to others due to they are artificial). 
This encourages more self-disclosure than human interlocutors because individuals feel comfortable and not threatened in their social evaluations~\cite{lucas2014s, pauw2022avatar}. 
Thus, VAs would act as a bridge in communication between superiors and subordinates. 
They can clearly convey directives and strategic objectives from superiors to subordinates and, conversely, relay feedback, concerns, and ideas from subordinates to superiors as well.


To summarize, this study presents the following contributions:

\begin{itemize}
    \item We found that the VA's engagements could capture user interest in group settings, but its impact on user's subjective perception was minimal.
    \item We found that the VA's affective behavior adversely affected users' perceptions during group discussions, yet it could influence the decision-making process.
    \item We proposed that as potential roles, VA's engagement could facilitate active group participation, which might positively affect group synergy and performance, and their affective behavior could mediate the group's decision-making process, leading to smoother and more directed decision outcomes.
\end{itemize}

\subsection{Limitations and Future Works}

Our study has several limitations in two design aspects: (1) group settings and (2) VA's behavior.
First of all, the study only examined interactions between three individuals - two humans and one VA, which may not reflect the complexity and dynamics of multi-user settings in actual organizations (with many more users or crowds).
Moreover, the group's performance was evaluated using only one specific task, the ``survival game,'' which may not encompass all group activities, such as staffing, lecturing, and reporting in actual organizations. 
Another limitation is that the participants were strangers to each other. This may not be consistently replicated in groups where social relationships have already been formed.

Concerning the design of VA's behavior, 
the study used pre-set emotional responses for the VA, which were decided before the experiment. 
This approach, by not including the context of ongoing discussions, may increase artificiality, leading to unnatural interactions and a diminished emotional connection between the VA and users.
For example, PA2 noted that \textit{``some responses felt programmed and did not flow naturally.''}
Furthermore, the study considered only a limited range of behavioral factors/types. 
There are many other engaging/affective behaviors that VAs can exhibit, such as gender~\cite{gendervirtualagent, 9258960}, age~\cite{ageprefer2020}, and visual style~\cite{horstmann2021just, wang2019exploring}.

Further research is needed to explore more diverse and complex group settings, including larger groups and those with pre-existing social relationships, to better reflect the dynamics of real-world organizations.
In addition, it is crucial to investigate the impact of dynamically generated emotional responses that are contextually relevant to ongoing discussions, aiming to enhance naturalness and emotional engagement.
Lastly, exploring a wider spectrum of behavioral factors will provide a more comprehensive understanding of how VAs can support organizational activities in other contexts.

\section{Conclusion}
\label{conclusion}
This paper investigated the influence of different VAs' engaging and affective behaviors on individuals' subjective perceptions and behavioral changes during group decision-making scenarios.
The study revealed that participants highly preferred the Engaged and Non-Affective VA (i.e., E-NA’s VA) as a team member, in contrast to the Non-Engaged and Affective VA (i.e., NE-A's VA), which was less favored.
Specifically, we observed that the engaging behaviors of VAs attracted participants’ attention in group settings, positively fostering group synergy to enhance group performance.
On the other hand, subjective measures indicated that VAs exhibiting affective behaviors resulted in less social presence, social influence, and trustworthiness compared to those of non-affective VAs. 
This outcome was attributed to participants' distrust in the words and actions of affective VAs, leading to a more mentally demanding decision-making process.
This study offers valuable insights for enhancing the role of VAs as team members, serving as a guide for designing VAs in group settings and organizational settings.

%
\begin{acks}
This work was supported 
by the Korea Institute of Science and Technology (KIST) Institutional Program (Project No. 2E32991)
and the Basic Research Laboratory Program funded by the National Research Foundation of Korea (NRF) (Grant No. 2022R1A4A1018869).
\end{acks}

\balance

\begin{thebibliography}{119}


\ifx \showCODEN    \undefined \def \showCODEN     #1{\unskip}     \fi
\ifx \showDOI      \undefined \def \showDOI       #1{#1}\fi
\ifx \showISBNx    \undefined \def \showISBNx     #1{\unskip}     \fi
\ifx \showISBNxiii \undefined \def \showISBNxiii  #1{\unskip}     \fi
\ifx \showISSN     \undefined \def \showISSN      #1{\unskip}     \fi
\ifx \showLCCN     \undefined \def \showLCCN      #1{\unskip}     \fi
\ifx \shownote     \undefined \def \shownote      #1{#1}          \fi
\ifx \showarticletitle \undefined \def \showarticletitle #1{#1}   \fi
\ifx \showURL      \undefined \def \showURL       {\relax}        \fi
\providecommand\bibfield[2]{#2}
\providecommand\bibinfo[2]{#2}
\providecommand\natexlab[1]{#1}
\providecommand\showeprint[2][]{arXiv:#2}

\bibitem[Abiodun(2014)]%
        {abiodun2014organizational}
\bibfield{author}{\bibinfo{person}{Ashimi~Rashidat Abiodun}.} \bibinfo{year}{2014}\natexlab{}.
\newblock \showarticletitle{Organizational conflicts: Causes, effects and remedies}.
\newblock \bibinfo{journal}{\emph{International Journal of Academic Research in Economics and Management Sciences}} \bibinfo{volume}{3}, \bibinfo{number}{6} (\bibinfo{year}{2014}), \bibinfo{pages}{118}.
\newblock


\bibitem[Ahmed et~al\mbox{.}(2023)]%
        {9258960}
\bibfield{author}{\bibinfo{person}{Imtiaj Ahmed}, \bibinfo{person}{Ville~J. Harjunen}, \bibinfo{person}{Giulio Jacucci}, \bibinfo{person}{Niklas Ravaja}, \bibinfo{person}{Tuukka Ruotsalo}, {and} \bibinfo{person}{Michiel~M. Spapé}.} \bibinfo{year}{2023}\natexlab{}.
\newblock \showarticletitle{Touching Virtual Humans: Haptic Responses Reveal the Emotional Impact of Affective Agents}.
\newblock \bibinfo{journal}{\emph{IEEE Transactions on Affective Computing}} \bibinfo{volume}{14}, \bibinfo{number}{1} (\bibinfo{year}{2023}), \bibinfo{pages}{331--342}.
\newblock
\urldef\tempurl%
\url{https://doi.org/10.1109/TAFFC.2020.3038137}
\showDOI{\tempurl}


\bibitem[Ali and Hwang(2022)]%
        {ali2022}
\bibfield{author}{\bibinfo{person}{Ghazanfar Ali} {and} \bibinfo{person}{Jae-In Hwang}.} \bibinfo{year}{2022}\natexlab{}.
\newblock \showarticletitle{Improving Co-Speech Gesture Rule-Map Generation via Wild Pose Matching with Gesture Units.}. In \bibinfo{booktitle}{\emph{SIGGRAPH Asia 2022 Posters}} \emph{(\bibinfo{series}{SA '22})}. \bibinfo{publisher}{ACM}, Article \bibinfo{articleno}{4}, \bibinfo{numpages}{2}~pages.
\newblock
\showISBNx{9781450394628}
\urldef\tempurl%
\url{https://doi.org/10.1145/3550082.3564185}
\showDOI{\tempurl}


\bibitem[Ali et~al\mbox{.}(2020)]%
        {ali2020automatic}
\bibfield{author}{\bibinfo{person}{Ghazanfar Ali}, \bibinfo{person}{Myungho Lee}, {and} \bibinfo{person}{Jae-In Hwang}.} \bibinfo{year}{2020}\natexlab{}.
\newblock \showarticletitle{Automatic text-to-gesture rule generation for embodied conversational agents}.
\newblock \bibinfo{journal}{\emph{Computer Animation and Virtual Worlds}} \bibinfo{volume}{31}, \bibinfo{number}{4-5} (\bibinfo{year}{2020}), \bibinfo{pages}{e1944}.
\newblock
\urldef\tempurl%
\url{https://doi.org/10.1002/cav.1944}
\showDOI{\tempurl}


\bibitem[Bailenson and Beall(2006)]%
        {bailenson2006transformed}
\bibfield{author}{\bibinfo{person}{Jeremy~N Bailenson} {and} \bibinfo{person}{Andrew~C Beall}.} \bibinfo{year}{2006}\natexlab{}.
\newblock \showarticletitle{Transformed social interaction: Exploring the digital plasticity of avatars}.
\newblock In \bibinfo{booktitle}{\emph{Avatars at work and play: Collaboration and interaction in shared virtual environments}}. \bibinfo{publisher}{Springer}, \bibinfo{pages}{1--16}.
\newblock


\bibitem[Bailenson et~al\mbox{.}(2003)]%
        {bailenson2003interpersonal}
\bibfield{author}{\bibinfo{person}{Jeremy~N Bailenson}, \bibinfo{person}{Jim Blascovich}, \bibinfo{person}{Andrew~C Beall}, {and} \bibinfo{person}{Jack~M Loomis}.} \bibinfo{year}{2003}\natexlab{}.
\newblock \showarticletitle{Interpersonal distance in immersive virtual environments}.
\newblock \bibinfo{journal}{\emph{Personality and social psychology bulletin}} \bibinfo{volume}{29}, \bibinfo{number}{7} (\bibinfo{year}{2003}), \bibinfo{pages}{819--833}.
\newblock


\bibitem[Baron and Dunham(2015)]%
        {baron2015representing}
\bibfield{author}{\bibinfo{person}{Andrew~Scott Baron} {and} \bibinfo{person}{Yarrow Dunham}.} \bibinfo{year}{2015}\natexlab{}.
\newblock \showarticletitle{Representing ‘us’ and ‘them’: Building blocks of intergroup cognition}.
\newblock \bibinfo{journal}{\emph{Journal of Cognition and Development}} \bibinfo{volume}{16}, \bibinfo{number}{5} (\bibinfo{year}{2015}), \bibinfo{pages}{780--801}.
\newblock


\bibitem[Bond and Bunce(2003)]%
        {bond2003role}
\bibfield{author}{\bibinfo{person}{Frank~W Bond} {and} \bibinfo{person}{David Bunce}.} \bibinfo{year}{2003}\natexlab{}.
\newblock \showarticletitle{The role of acceptance and job control in mental health, job satisfaction, and work performance.}
\newblock \bibinfo{journal}{\emph{Journal of applied psychology}} \bibinfo{volume}{88}, \bibinfo{number}{6} (\bibinfo{year}{2003}), \bibinfo{pages}{1057}.
\newblock


\bibitem[Brehm and Brehm(2013)]%
        {brehm2013psychological}
\bibfield{author}{\bibinfo{person}{Sharon~S Brehm} {and} \bibinfo{person}{Jack~W Brehm}.} \bibinfo{year}{2013}\natexlab{}.
\newblock \bibinfo{booktitle}{\emph{Psychological reactance: A theory of freedom and control}}.
\newblock \bibinfo{publisher}{Academic Press}.
\newblock


\bibitem[Chang et~al\mbox{.}(2012)]%
        {competitivewithrobot}
\bibfield{author}{\bibinfo{person}{Wan-Ling Chang}, \bibinfo{person}{Jeremy~P. White}, \bibinfo{person}{Joohyun Park}, \bibinfo{person}{Anna Holm}, {and} \bibinfo{person}{Selma Šabanović}.} \bibinfo{year}{2012}\natexlab{}.
\newblock \showarticletitle{The effect of group size on people's attitudes and cooperative behaviors toward robots in interactive gameplay}. In \bibinfo{booktitle}{\emph{2012 IEEE RO-MAN: The 21st IEEE International Symposium on Robot and Human Interactive Communication}}. \bibinfo{pages}{845--850}.
\newblock
\urldef\tempurl%
\url{https://doi.org/10.1109/ROMAN.2012.6343857}
\showDOI{\tempurl}


\bibitem[Choi et~al\mbox{.}(2012)]%
        {choi2012affective}
\bibfield{author}{\bibinfo{person}{Ahyoung Choi}, \bibinfo{person}{Celso~De Melo}, \bibinfo{person}{Woontack Woo}, {and} \bibinfo{person}{Jonathan Gratch}.} \bibinfo{year}{2012}\natexlab{}.
\newblock \showarticletitle{Affective engagement to emotional facial expressions of embodied social agents in a decision-making game}.
\newblock \bibinfo{journal}{\emph{Computer Animation and Virtual Worlds}} (\bibinfo{year}{2012}).
\newblock


\bibitem[Choudhary et~al\mbox{.}(2023)]%
        {GerdIeeeVR2023}
\bibfield{author}{\bibinfo{person}{Zubin Choudhary}, \bibinfo{person}{Nahal Norouzi}, \bibinfo{person}{Austin Erickson}, \bibinfo{person}{Ryan Schubert}, \bibinfo{person}{Gerd Bruder}, {and} \bibinfo{person}{Gregory~F. Welch}.} \bibinfo{year}{2023}\natexlab{}.
\newblock \showarticletitle{Exploring the Social Influence of Virtual Humans Unintentionally Conveying Conflicting Emotions}. In \bibinfo{booktitle}{\emph{2023 IEEE Conference Virtual Reality and 3D User Interfaces (VR)}}. \bibinfo{pages}{571--580}.
\newblock
\urldef\tempurl%
\url{https://doi.org/10.1109/VR55154.2023.00072}
\showDOI{\tempurl}


\bibitem[Clarke(2010)]%
        {clarke2010emotional}
\bibfield{author}{\bibinfo{person}{Nicholas Clarke}.} \bibinfo{year}{2010}\natexlab{}.
\newblock \showarticletitle{Emotional intelligence abilities and their relationships with team processes}.
\newblock \bibinfo{journal}{\emph{Team Performance Management: An International Journal}} \bibinfo{volume}{16}, \bibinfo{number}{1/2} (\bibinfo{year}{2010}), \bibinfo{pages}{6--32}.
\newblock


\bibitem[Conover and Iman(1979)]%
        {conover1979multiple}
\bibfield{author}{\bibinfo{person}{William~Jay Conover} {and} \bibinfo{person}{Ronald~L Iman}.} \bibinfo{year}{1979}\natexlab{}.
\newblock \showarticletitle{On multiple-comparisons procedures}.
\newblock \bibinfo{journal}{\emph{Los Alamos Sci. Lab. Tech. Rep. LA-7677-MS}}  \bibinfo{volume}{1} (\bibinfo{year}{1979}), \bibinfo{pages}{14}.
\newblock


\bibitem[Cooper and Boice-Pardee(2011)]%
        {cooper2011managing}
\bibfield{author}{\bibinfo{person}{Mary-Beth Cooper} {and} \bibinfo{person}{Heath Boice-Pardee}.} \bibinfo{year}{2011}\natexlab{}.
\newblock \showarticletitle{Managing conflict from the middle}.
\newblock \bibinfo{journal}{\emph{New directions for student services}} \bibinfo{volume}{2011}, \bibinfo{number}{136} (\bibinfo{year}{2011}), \bibinfo{pages}{35--42}.
\newblock


\bibitem[Cottrell et~al\mbox{.}(1968)]%
        {cottrell1968social}
\bibfield{author}{\bibinfo{person}{Nickolas~B Cottrell}, \bibinfo{person}{Dennis~L Wack}, \bibinfo{person}{Gary~J Sekerak}, {and} \bibinfo{person}{Robert~H Rittle}.} \bibinfo{year}{1968}\natexlab{}.
\newblock \showarticletitle{Social facilitation of dominant responses by the presence of an audience and the mere presence of others.}
\newblock \bibinfo{journal}{\emph{Journal of personality and social psychology}} \bibinfo{volume}{9}, \bibinfo{number}{3} (\bibinfo{year}{1968}), \bibinfo{pages}{245}.
\newblock


\bibitem[{Crazy Minnow Studio}(2014)]%
        {SALSA}
\bibfield{author}{\bibinfo{person}{{Crazy Minnow Studio}}.} \bibinfo{year}{2014}\natexlab{}.
\newblock \bibinfo{title}{SALSA LipSync}.
\newblock
\newblock
\urldef\tempurl%
\url{https://crazyminnowstudio.com/}
\showURL{%
\tempurl}
\newblock
\shownote{Accessed oct. 14, 2022}.


\bibitem[Daher et~al\mbox{.}(2020)]%
        {daher2020embodied}
\bibfield{author}{\bibinfo{person}{Karl Daher}, \bibinfo{person}{Zeno Bardelli}, \bibinfo{person}{Jacky Casas}, \bibinfo{person}{Elena Mugellini}, \bibinfo{person}{Omar~Abou Khaled}, {and} \bibinfo{person}{Denis Lalanne}.} \bibinfo{year}{2020}\natexlab{}.
\newblock \showarticletitle{Embodied conversational agent for emotional recognition training}. In \bibinfo{booktitle}{\emph{Proceedings of ThinkMind, ACHI 2020: The Thirteenth International Conference on Advances in Computer-Human Interactions, 21-25 November 2020, Valencia, Spain}}. 21-25 November 2020.
\newblock


\bibitem[De~Melo et~al\mbox{.}(2011)]%
        {de2011impact}
\bibfield{author}{\bibinfo{person}{Celso~M De~Melo}, \bibinfo{person}{Peter Carnevale}, {and} \bibinfo{person}{Jonathan Gratch}.} \bibinfo{year}{2011}\natexlab{}.
\newblock \showarticletitle{The impact of emotion displays in embodied agents on emergence of cooperation with people}.
\newblock \bibinfo{journal}{\emph{Presence: teleoperators and virtual environments}} \bibinfo{volume}{20}, \bibinfo{number}{5} (\bibinfo{year}{2011}), \bibinfo{pages}{449--465}.
\newblock


\bibitem[de~Melo et~al\mbox{.}(2015)]%
        {celsoDecisionMaking}
\bibfield{author}{\bibinfo{person}{Celso~M. de Melo}, \bibinfo{person}{Jonathan Gratch}, {and} \bibinfo{person}{Peter~J. Carnevale}.} \bibinfo{year}{2015}\natexlab{}.
\newblock \showarticletitle{Humans versus Computers: Impact of Emotion Expressions on People's Decision Making}.
\newblock \bibinfo{journal}{\emph{IEEE Transactions on Affective Computing}} \bibinfo{volume}{6}, \bibinfo{number}{2} (\bibinfo{year}{2015}), \bibinfo{pages}{127--136}.
\newblock
\urldef\tempurl%
\url{https://doi.org/10.1109/TAFFC.2014.2332471}
\showDOI{\tempurl}


\bibitem[DeVault et~al\mbox{.}(2014)]%
        {devault2014simsensei}
\bibfield{author}{\bibinfo{person}{David DeVault}, \bibinfo{person}{Ron Artstein}, \bibinfo{person}{Grace Benn}, \bibinfo{person}{Teresa Dey}, \bibinfo{person}{Ed Fast}, \bibinfo{person}{Alesia Gainer}, \bibinfo{person}{Kallirroi Georgila}, \bibinfo{person}{Jon Gratch}, \bibinfo{person}{Arno Hartholt}, \bibinfo{person}{Margaux Lhommet}, {et~al\mbox{.}}} \bibinfo{year}{2014}\natexlab{}.
\newblock \showarticletitle{SimSensei Kiosk: A virtual human interviewer for healthcare decision support}. In \bibinfo{booktitle}{\emph{Proceedings of the 2014 international conference on Autonomous agents and multi-agent systems}}. \bibinfo{pages}{1061--1068}.
\newblock


\bibitem[Divekar et~al\mbox{.}(2019)]%
        {divekar2019you}
\bibfield{author}{\bibinfo{person}{Rahul~R Divekar}, \bibinfo{person}{Jeffrey~O Kephart}, \bibinfo{person}{Xiangyang Mou}, \bibinfo{person}{Lisha Chen}, {and} \bibinfo{person}{Hui Su}.} \bibinfo{year}{2019}\natexlab{}.
\newblock \showarticletitle{You talkin’to me? A practical attention-aware embodied agent}. In \bibinfo{booktitle}{\emph{Human-Computer Interaction--INTERACT 2019: 17th IFIP TC 13 International Conference, Paphos, Cyprus, September 2--6, 2019, Proceedings, Part III 17}}. Springer, \bibinfo{pages}{760--780}.
\newblock


\bibitem[Elkin et~al\mbox{.}(2021)]%
        {ART-C}
\bibfield{author}{\bibinfo{person}{Lisa~A. Elkin}, \bibinfo{person}{Matthew Kay}, \bibinfo{person}{James~J. Higgins}, {and} \bibinfo{person}{Jacob~O. Wobbrock}.} \bibinfo{year}{2021}\natexlab{}.
\newblock \showarticletitle{An Aligned Rank Transform Procedure for Multifactor Contrast Tests}. In \bibinfo{booktitle}{\emph{The 34th Annual ACM Symposium on User Interface Software and Technology}} (Virtual Event, USA) \emph{(\bibinfo{series}{UIST '21})}. \bibinfo{publisher}{Association for Computing Machinery}, \bibinfo{address}{New York, NY, USA}, \bibinfo{pages}{754–768}.
\newblock
\showISBNx{9781450386357}
\urldef\tempurl%
\url{https://doi.org/10.1145/3472749.3474784}
\showDOI{\tempurl}


\bibitem[Elkins and Derrick(2013)]%
        {elkins2013sound}
\bibfield{author}{\bibinfo{person}{Aaron~C Elkins} {and} \bibinfo{person}{Douglas~C Derrick}.} \bibinfo{year}{2013}\natexlab{}.
\newblock \showarticletitle{The sound of trust: voice as a measurement of trust during interactions with embodied conversational agents}.
\newblock \bibinfo{journal}{\emph{Group decision and negotiation}} \bibinfo{volume}{22}, \bibinfo{number}{5} (\bibinfo{year}{2013}), \bibinfo{pages}{897--913}.
\newblock


\bibitem[Faul et~al\mbox{.}(2009)]%
        {faul2009statistical}
\bibfield{author}{\bibinfo{person}{Franz Faul}, \bibinfo{person}{Edgar Erdfelder}, \bibinfo{person}{Axel Buchner}, {and} \bibinfo{person}{Albert-Georg Lang}.} \bibinfo{year}{2009}\natexlab{}.
\newblock \showarticletitle{Statistical power analyses using G* Power 3.1: Tests for correlation and regression analyses}.
\newblock \bibinfo{journal}{\emph{Behavior research methods}} \bibinfo{volume}{41}, \bibinfo{number}{4} (\bibinfo{year}{2009}), \bibinfo{pages}{1149--1160}.
\newblock


\bibitem[Fischer et~al\mbox{.}(2016)]%
        {fischer2016social}
\bibfield{author}{\bibinfo{person}{Agneta~H Fischer}, \bibinfo{person}{Antony~SR Manstead}, {et~al\mbox{.}}} \bibinfo{year}{2016}\natexlab{}.
\newblock \showarticletitle{Social functions of emotion and emotion regulation}.
\newblock \bibinfo{journal}{\emph{Handbook of emotions}}  \bibinfo{volume}{4} (\bibinfo{year}{2016}), \bibinfo{pages}{424--439}.
\newblock


\bibitem[Fox et~al\mbox{.}(2015)]%
        {fox2015avatars}
\bibfield{author}{\bibinfo{person}{Jesse Fox}, \bibinfo{person}{Sun~Joo Ahn}, \bibinfo{person}{Joris~H Janssen}, \bibinfo{person}{Leo Yeykelis}, \bibinfo{person}{Kathryn~Y Segovia}, {and} \bibinfo{person}{Jeremy~N Bailenson}.} \bibinfo{year}{2015}\natexlab{}.
\newblock \showarticletitle{Avatars versus agents: a meta-analysis quantifying the effect of agency on social influence}.
\newblock \bibinfo{journal}{\emph{Human--Computer Interaction}} \bibinfo{volume}{30}, \bibinfo{number}{5} (\bibinfo{year}{2015}), \bibinfo{pages}{401--432}.
\newblock


\bibitem[Friedman(1940)]%
        {friedman1940comparison}
\bibfield{author}{\bibinfo{person}{Milton Friedman}.} \bibinfo{year}{1940}\natexlab{}.
\newblock \showarticletitle{A comparison of alternative tests of significance for the problem of m rankings}.
\newblock \bibinfo{journal}{\emph{The annals of mathematical statistics}} \bibinfo{volume}{11}, \bibinfo{number}{1} (\bibinfo{year}{1940}), \bibinfo{pages}{86--92}.
\newblock


\bibitem[Gatto et~al\mbox{.}(2022)]%
        {gatto2022met}
\bibfield{author}{\bibinfo{person}{Luigi Gatto}, \bibinfo{person}{Giuseppe~Fulvio Gaglio}, \bibinfo{person}{Agnese Augello}, \bibinfo{person}{Giuseppe Caggianese}, \bibinfo{person}{Luigi Gallo}, {and} \bibinfo{person}{Marco La~Cascia}.} \bibinfo{year}{2022}\natexlab{}.
\newblock \showarticletitle{MET-iquette: enabling virtual agents to have a social compliant behavior in the Metaverse}. In \bibinfo{booktitle}{\emph{2022 16th International Conference on Signal-Image Technology \& Internet-Based Systems (SITIS)}}. IEEE, \bibinfo{pages}{394--401}.
\newblock


\bibitem[Grandey(2000)]%
        {grandey2000emotional}
\bibfield{author}{\bibinfo{person}{Alicia~A Grandey}.} \bibinfo{year}{2000}\natexlab{}.
\newblock \showarticletitle{Emotional regulation in the workplace: A new way to conceptualize emotional labor.}
\newblock \bibinfo{journal}{\emph{Journal of occupational health psychology}} \bibinfo{volume}{5}, \bibinfo{number}{1} (\bibinfo{year}{2000}), \bibinfo{pages}{95}.
\newblock


\bibitem[Gratch et~al\mbox{.}(2007)]%
        {gratch2007can}
\bibfield{author}{\bibinfo{person}{Jonathan Gratch}, \bibinfo{person}{Ning Wang}, \bibinfo{person}{Anna Okhmatovskaia}, \bibinfo{person}{Francois Lamothe}, \bibinfo{person}{Mathieu Morales}, \bibinfo{person}{Rick~J van~der Werf}, {and} \bibinfo{person}{Louis-Philippe Morency}.} \bibinfo{year}{2007}\natexlab{}.
\newblock \showarticletitle{Can virtual humans be more engaging than real ones?}. In \bibinfo{booktitle}{\emph{Human-Computer Interaction. HCI Intelligent Multimodal Interaction Environments: 12th International Conference, HCI International 2007, Beijing, China, July 22-27, 2007, Proceedings, Part III 12}}. Springer, \bibinfo{pages}{286--297}.
\newblock


\bibitem[Gross and Levenson(1993)]%
        {gross1993emotional}
\bibfield{author}{\bibinfo{person}{James~J Gross} {and} \bibinfo{person}{Robert~W Levenson}.} \bibinfo{year}{1993}\natexlab{}.
\newblock \showarticletitle{Emotional suppression: physiology, self-report, and expressive behavior.}
\newblock \bibinfo{journal}{\emph{Journal of personality and social psychology}} \bibinfo{volume}{64}, \bibinfo{number}{6} (\bibinfo{year}{1993}), \bibinfo{pages}{970}.
\newblock


\bibitem[Hall(1966)]%
        {hall1966hidden}
\bibfield{author}{\bibinfo{person}{ET Hall}.} \bibinfo{year}{1966}\natexlab{}.
\newblock \showarticletitle{The Hidden Dimension: Man’s Use of Space in Public and Private (London, The Bodley Head)}.
\newblock  (\bibinfo{year}{1966}).
\newblock


\bibitem[Hall and Watson(1970)]%
        {nasasurvivalTask}
\bibfield{author}{\bibinfo{person}{Jay Hall} {and} \bibinfo{person}{Wilfred~Harvey Watson}.} \bibinfo{year}{1970}\natexlab{}.
\newblock \showarticletitle{The effects of a normative intervention on group decision-making performance}.
\newblock \bibinfo{journal}{\emph{Human relations}} \bibinfo{volume}{23}, \bibinfo{number}{4} (\bibinfo{year}{1970}), \bibinfo{pages}{299--317}.
\newblock


\bibitem[Harkins(1987)]%
        {harkins1987social}
\bibfield{author}{\bibinfo{person}{Stephen~G Harkins}.} \bibinfo{year}{1987}\natexlab{}.
\newblock \showarticletitle{Social loafing and social facilitation}.
\newblock \bibinfo{journal}{\emph{Journal of experimental social psychology}} \bibinfo{volume}{23}, \bibinfo{number}{1} (\bibinfo{year}{1987}), \bibinfo{pages}{1--18}.
\newblock


\bibitem[Harms and Biocca(2006)]%
        {Harms2006InternalCA}
\bibfield{author}{\bibinfo{person}{Chad Harms} {and} \bibinfo{person}{Frank Biocca}.} \bibinfo{year}{2006}\natexlab{}.
\newblock \showarticletitle{Internal Consistency and Reliability of the Networked Minds Social Presence Measure}.
\newblock


\bibitem[Hart(2006)]%
        {hart2006nasa}
\bibfield{author}{\bibinfo{person}{Sandra~G Hart}.} \bibinfo{year}{2006}\natexlab{}.
\newblock \showarticletitle{NASA-task load index (NASA-TLX); 20 years later}. In \bibinfo{booktitle}{\emph{Proceedings of the human factors and ergonomics society annual meeting}}, Vol.~\bibinfo{volume}{50}. Sage publications Sage CA: Los Angeles, CA, \bibinfo{pages}{904--908}.
\newblock


\bibitem[Hiltz et~al\mbox{.}(1991)]%
        {hiltz1991group}
\bibfield{author}{\bibinfo{person}{Starr~Roxanne Hiltz}, \bibinfo{person}{Kenneth Johnson}, {and} \bibinfo{person}{Murray Turoff}.} \bibinfo{year}{1991}\natexlab{}.
\newblock \showarticletitle{Group decision support: The effects of designated human leaders and statistical feedback in computerized conferences}.
\newblock \bibinfo{journal}{\emph{Journal of Management Information Systems}} \bibinfo{volume}{8}, \bibinfo{number}{2} (\bibinfo{year}{1991}), \bibinfo{pages}{81--108}.
\newblock


\bibitem[Hinkin and Schriesheim(1994)]%
        {hinkin1994examination}
\bibfield{author}{\bibinfo{person}{Timothy~R Hinkin} {and} \bibinfo{person}{Chester~A Schriesheim}.} \bibinfo{year}{1994}\natexlab{}.
\newblock \showarticletitle{An examination of subordinate-perceived relationships between leader reward and punishment behavior and leader bases of power}.
\newblock \bibinfo{journal}{\emph{Human Relations}} \bibinfo{volume}{47}, \bibinfo{number}{7} (\bibinfo{year}{1994}), \bibinfo{pages}{779--800}.
\newblock


\bibitem[Hoque et~al\mbox{.}(2013)]%
        {hoque2013mach}
\bibfield{author}{\bibinfo{person}{Mohammed Hoque}, \bibinfo{person}{Matthieu Courgeon}, \bibinfo{person}{Jean-Claude Martin}, \bibinfo{person}{Bilge Mutlu}, {and} \bibinfo{person}{Rosalind~W Picard}.} \bibinfo{year}{2013}\natexlab{}.
\newblock \showarticletitle{Mach: My automated conversation coach}. In \bibinfo{booktitle}{\emph{Proceedings of the 2013 ACM international joint conference on Pervasive and ubiquitous computing}}. \bibinfo{pages}{697--706}.
\newblock


\bibitem[Horstmann et~al\mbox{.}(2021)]%
        {horstmann2021just}
\bibfield{author}{\bibinfo{person}{Aike~C Horstmann}, \bibinfo{person}{Jonathan Gratch}, {and} \bibinfo{person}{Nicole~C Kraemer}.} \bibinfo{year}{2021}\natexlab{}.
\newblock \showarticletitle{I just wanna blame somebody, not something! Reactions to a computer agent giving negative feedback based on the instructions of a person}.
\newblock \bibinfo{journal}{\emph{International Journal of Human-Computer Studies}}  \bibinfo{volume}{154} (\bibinfo{year}{2021}), \bibinfo{pages}{102683}.
\newblock


\bibitem[Huang et~al\mbox{.}(2011)]%
        {huang2011virtual}
\bibfield{author}{\bibinfo{person}{Lixing Huang}, \bibinfo{person}{Louis-Philippe Morency}, {and} \bibinfo{person}{Jonathan Gratch}.} \bibinfo{year}{2011}\natexlab{}.
\newblock \showarticletitle{Virtual Rapport 2.0}. In \bibinfo{booktitle}{\emph{International workshop on intelligent virtual agents}}.
\newblock


\bibitem[Jian et~al\mbox{.}(2000)]%
        {jian2000foundations}
\bibfield{author}{\bibinfo{person}{Jiun-Yin Jian}, \bibinfo{person}{Ann~M Bisantz}, {and} \bibinfo{person}{Colin~G Drury}.} \bibinfo{year}{2000}\natexlab{}.
\newblock \showarticletitle{Foundations for an empirically determined scale of trust in automated systems}.
\newblock \bibinfo{journal}{\emph{International journal of cognitive ergonomics}} \bibinfo{volume}{4}, \bibinfo{number}{1} (\bibinfo{year}{2000}), \bibinfo{pages}{53--71}.
\newblock


\bibitem[Johnson and Johnson(1991)]%
        {johnson1991joining}
\bibfield{author}{\bibinfo{person}{David~W Johnson} {and} \bibinfo{person}{Frank~P Johnson}.} \bibinfo{year}{1991}\natexlab{}.
\newblock \bibinfo{booktitle}{\emph{Joining together: Group theory and group skills}}.
\newblock \bibinfo{publisher}{Prentice-Hall, Inc}.
\newblock


\bibitem[Jones et~al\mbox{.}(2014)]%
        {jones2014expressing}
\bibfield{author}{\bibinfo{person}{Haza{\"e}l Jones}, \bibinfo{person}{Mathieu Chollet}, \bibinfo{person}{Magalie Ochs}, \bibinfo{person}{Nicolas Sabouret}, {and} \bibinfo{person}{Catherine Pelachaud}.} \bibinfo{year}{2014}\natexlab{}.
\newblock \showarticletitle{Expressing social attitudes in virtual agents for social coaching}. In \bibinfo{booktitle}{\emph{Workshop Affects, Compagnons Artificiels et Interaction}}.
\newblock


\bibitem[Kazemi and Sullivan(2014)]%
        {kazemi2014one}
\bibfield{author}{\bibinfo{person}{Vahid Kazemi} {and} \bibinfo{person}{Josephine Sullivan}.} \bibinfo{year}{2014}\natexlab{}.
\newblock \showarticletitle{One millisecond face alignment with an ensemble of regression trees}. In \bibinfo{booktitle}{\emph{Proceedings of the IEEE conference on computer vision and pattern recognition}}. \bibinfo{pages}{1867--1874}.
\newblock


\bibitem[Kenny et~al\mbox{.}(2007)]%
        {kenny2007building}
\bibfield{author}{\bibinfo{person}{Patrick Kenny}, \bibinfo{person}{Arno Hartholt}, \bibinfo{person}{Jonathan Gratch}, \bibinfo{person}{William Swartout}, \bibinfo{person}{David Traum}, \bibinfo{person}{Stacy Marsella}, {and} \bibinfo{person}{Diane Piepol}.} \bibinfo{year}{2007}\natexlab{}.
\newblock \showarticletitle{Building interactive virtual humans for training environments}. In \bibinfo{booktitle}{\emph{Proceedings of i/itsec}}, Vol.~\bibinfo{volume}{174}. \bibinfo{pages}{911--916}.
\newblock


\bibitem[Kim et~al\mbox{.}(2021)]%
        {hanseob2021visualeffect}
\bibfield{author}{\bibinfo{person}{Hanseob Kim}, \bibinfo{person}{Myungho Lee}, \bibinfo{person}{Gerard~J. Kim}, {and} \bibinfo{person}{Jae-In Hwang}.} \bibinfo{year}{2021}\natexlab{}.
\newblock \showarticletitle{The Impacts of Visual Effects on User Perception With a Virtual Human in Augmented Reality Conflict Situations}.
\newblock \bibinfo{journal}{\emph{IEEE Access}}  \bibinfo{volume}{9} (\bibinfo{year}{2021}), \bibinfo{pages}{35300--35312}.
\newblock
\urldef\tempurl%
\url{https://doi.org/10.1109/ACCESS.2021.3062037}
\showDOI{\tempurl}


\bibitem[Kim et~al\mbox{.}({[n.\,d.]})]%
        {8613756}
\bibfield{author}{\bibinfo{person}{Kangsoo Kim}, \bibinfo{person}{Luke Boelling}, \bibinfo{person}{Steffen Haesler}, \bibinfo{person}{Jeremy Bailenson}, \bibinfo{person}{Gerd Bruder}, {and} \bibinfo{person}{Greg~F. Welch}.} \bibinfo{year}{[n.\,d.]}\natexlab{}.
\newblock \showarticletitle{Does a Digital Assistant Need a Body? The Influence of Visual Embodiment and Social Behavior on the Perception of Intelligent Virtual Agents in AR}. In \bibinfo{booktitle}{\emph{2018 IEEE International Symposium on Mixed and Augmented Reality}}.
\newblock
\urldef\tempurl%
\url{https://doi.org/10.1109/ISMAR.2018.00039}
\showDOI{\tempurl}


\bibitem[Kim et~al\mbox{.}(2020)]%
        {kangsooReducing}
\bibfield{author}{\bibinfo{person}{Kangsoo Kim}, \bibinfo{person}{Celso~M. de Melo}, \bibinfo{person}{Nahal Norouzi}, \bibinfo{person}{Gerd Bruder}, {and} \bibinfo{person}{Gregory~F. Welch}.} \bibinfo{year}{2020}\natexlab{}.
\newblock \showarticletitle{Reducing Task Load with an Embodied Intelligent Virtual Assistant for Improved Performance in Collaborative Decision Making}. In \bibinfo{booktitle}{\emph{2020 IEEE Conference on Virtual Reality and 3D User Interfaces (VR)}}. \bibinfo{pages}{529--538}.
\newblock
\urldef\tempurl%
\url{https://doi.org/10.1109/VR46266.2020.00074}
\showDOI{\tempurl}


\bibitem[Kim et~al\mbox{.}(2019)]%
        {kim2019effects}
\bibfield{author}{\bibinfo{person}{Kangsoo Kim}, \bibinfo{person}{Nahal Norouzi}, \bibinfo{person}{Tiffany Losekamp}, \bibinfo{person}{Gerd Bruder}, \bibinfo{person}{Mindi Anderson}, {and} \bibinfo{person}{Gregory Welch}.} \bibinfo{year}{2019}\natexlab{}.
\newblock \showarticletitle{Effects of patient care assistant embodiment and computer mediation on user experience}. In \bibinfo{booktitle}{\emph{2019 IEEE International Conference on Artificial Intelligence and Virtual Reality}}. \bibinfo{pages}{17--177}.
\newblock


\bibitem[Kimmel et~al\mbox{.}(2023)]%
        {kimmel2023let}
\bibfield{author}{\bibinfo{person}{Simon Kimmel}, \bibinfo{person}{Frederike Jung}, \bibinfo{person}{Andrii Matviienko}, \bibinfo{person}{Wilko Heuten}, {and} \bibinfo{person}{Susanne Boll}.} \bibinfo{year}{2023}\natexlab{}.
\newblock \showarticletitle{Let’s Face It: Influence of Facial Expressions on Social Presence in Collaborative Virtual Reality}. In \bibinfo{booktitle}{\emph{Proceedings of the 2023 CHI Conference on Human Factors in Computing Systems}}. \bibinfo{pages}{1--16}.
\newblock


\bibitem[King(2009)]%
        {king2009dlib}
\bibfield{author}{\bibinfo{person}{Davis~E King}.} \bibinfo{year}{2009}\natexlab{}.
\newblock \showarticletitle{Dlib-ml: A machine learning toolkit}.
\newblock \bibinfo{journal}{\emph{The Journal of Machine Learning Research}}  \bibinfo{volume}{10} (\bibinfo{year}{2009}), \bibinfo{pages}{1755--1758}.
\newblock


\bibitem[Knierim et~al\mbox{.}(2017)]%
        {knierim2017emotion}
\bibfield{author}{\bibinfo{person}{Michael~T Knierim}, \bibinfo{person}{Anuja Hariharan}, \bibinfo{person}{Verena Dorner}, {and} \bibinfo{person}{Christof Weinhardt}.} \bibinfo{year}{2017}\natexlab{}.
\newblock \showarticletitle{Emotion feedback in small group collaboration: A research agenda for group emotion management support systems}. In \bibinfo{booktitle}{\emph{Proceedings of the International Conference on Group Decision and Negotiation}}. \bibinfo{pages}{1--12}.
\newblock


\bibitem[Koval and Hansen(2021)]%
        {Togetheralone}
\bibfield{author}{\bibinfo{person}{Olena Koval} {and} \bibinfo{person}{Håvard Hansen}.} \bibinfo{year}{2021}\natexlab{}.
\newblock \showarticletitle{Together alone: complex tourism decisions in couples and random groups}.
\newblock \bibinfo{journal}{\emph{Anatolia}} (\bibinfo{year}{2021}), \bibinfo{pages}{1--11}.
\newblock


\bibitem[Kramer and Hess(2002)]%
        {kramer2002communication}
\bibfield{author}{\bibinfo{person}{Michael~W Kramer} {and} \bibinfo{person}{Jon~A Hess}.} \bibinfo{year}{2002}\natexlab{}.
\newblock \showarticletitle{Communication rules for the display of emotions in organizational settings}.
\newblock \bibinfo{journal}{\emph{Management Communication Quarterly}} \bibinfo{volume}{16}, \bibinfo{number}{1} (\bibinfo{year}{2002}), \bibinfo{pages}{66--80}.
\newblock


\bibitem[Kulms et~al\mbox{.}(2011)]%
        {gendervirtualagent}
\bibfield{author}{\bibinfo{person}{Philipp Kulms}, \bibinfo{person}{Nicole~C Kr{\"a}mer}, \bibinfo{person}{Jonathan Gratch}, {and} \bibinfo{person}{Sin-Hwa Kang}.} \bibinfo{year}{2011}\natexlab{}.
\newblock \showarticletitle{It’s in their eyes: A study on female and male virtual humans’ gaze}. In \bibinfo{booktitle}{\emph{Intelligent Virtual Agents: 10th International Conference, IVA 2011, Reykjavik, Iceland, September 15-17, 2011. Proceedings 11}}. Springer, \bibinfo{pages}{80--92}.
\newblock


\bibitem[Kum and Lee(2022)]%
        {kum2022can}
\bibfield{author}{\bibinfo{person}{Junyeong Kum} {and} \bibinfo{person}{Myungho Lee}.} \bibinfo{year}{2022}\natexlab{}.
\newblock \showarticletitle{Can Gestural Filler Reduce User-Perceived Latency in Conversation with Digital Humans?}
\newblock \bibinfo{journal}{\emph{Applied Sciences}} \bibinfo{volume}{12}, \bibinfo{number}{21} (\bibinfo{year}{2022}), \bibinfo{pages}{10972}.
\newblock


\bibitem[Lafferty et~al\mbox{.}(1974)]%
        {desertSurvivalTask}
\bibfield{author}{\bibinfo{person}{J.~C. Lafferty}, \bibinfo{person}{Eady}, {and} \bibinfo{person}{J. Elmers}.} \bibinfo{year}{1974}\natexlab{}.
\newblock \bibinfo{booktitle}{\emph{{The desert survival problem}}}.
\newblock \bibinfo{address}{Plymouth, Michigan: Experimental Learning Methods}.
\newblock


\bibitem[Lee et~al\mbox{.}(2016)]%
        {myunho2016wobblytable}
\bibfield{author}{\bibinfo{person}{Myungho Lee}, \bibinfo{person}{Kangsoo Kim}, \bibinfo{person}{Salam Daher}, \bibinfo{person}{Andrew Raij}, \bibinfo{person}{Ryan Schubert}, \bibinfo{person}{Jeremy Bailenson}, {and} \bibinfo{person}{Greg Welch}.} \bibinfo{year}{2016}\natexlab{}.
\newblock \showarticletitle{The wobbly table: Increased social presence via subtle incidental movement of a real-virtual table}. In \bibinfo{booktitle}{\emph{2016 IEEE Virtual Reality (VR)}}. \bibinfo{pages}{11--17}.
\newblock
\urldef\tempurl%
\url{https://doi.org/10.1109/VR.2016.7504683}
\showDOI{\tempurl}


\bibitem[Lucas et~al\mbox{.}(2014)]%
        {lucas2014s}
\bibfield{author}{\bibinfo{person}{Gale~M Lucas}, \bibinfo{person}{Jonathan Gratch}, \bibinfo{person}{Aisha King}, {and} \bibinfo{person}{Louis-Philippe Morency}.} \bibinfo{year}{2014}\natexlab{}.
\newblock \showarticletitle{It’s only a computer: Virtual humans increase willingness to disclose}.
\newblock \bibinfo{journal}{\emph{Computers in Human Behavior}}  \bibinfo{volume}{37} (\bibinfo{year}{2014}), \bibinfo{pages}{94--100}.
\newblock


\bibitem[Lucas et~al\mbox{.}(2017)]%
        {lucas2017reporting}
\bibfield{author}{\bibinfo{person}{Gale~M Lucas}, \bibinfo{person}{Albert Rizzo}, \bibinfo{person}{Jonathan Gratch}, \bibinfo{person}{Stefan Scherer}, \bibinfo{person}{Giota Stratou}, \bibinfo{person}{Jill Boberg}, {and} \bibinfo{person}{Louis-Philippe Morency}.} \bibinfo{year}{2017}\natexlab{}.
\newblock \showarticletitle{Reporting mental health symptoms: breaking down barriers to care with virtual human interviewers}.
\newblock \bibinfo{journal}{\emph{Frontiers in Robotics and AI}}  \bibinfo{volume}{4} (\bibinfo{year}{2017}), \bibinfo{pages}{51}.
\newblock


\bibitem[Lucey et~al\mbox{.}({[n.\,d.]})]%
        {lucey2010extended}
\bibfield{author}{\bibinfo{person}{Patrick Lucey}, \bibinfo{person}{Jeffrey~F Cohn}, \bibinfo{person}{Takeo Kanade}, \bibinfo{person}{Jason Saragih}, \bibinfo{person}{Zara Ambadar}, {and} \bibinfo{person}{Iain Matthews}.} \bibinfo{year}{[n.\,d.]}\natexlab{}.
\newblock \showarticletitle{The extended cohn-kanade dataset (ck+): A complete dataset for action unit and emotion-specified expression}. In \bibinfo{booktitle}{\emph{2010 ieee computer society conference on computer vision and pattern recognition-workshops}}.
\newblock


\bibitem[Maran et~al\mbox{.}(2021)]%
        {visualattentionInGroup}
\bibfield{author}{\bibinfo{person}{Thomas Maran}, \bibinfo{person}{Marco Furtner}, \bibinfo{person}{Simon Liegl}, \bibinfo{person}{Theo Ravet-Brown}, \bibinfo{person}{Lucas Haraped}, {and} \bibinfo{person}{Pierre Sachse}.} \bibinfo{year}{2021}\natexlab{}.
\newblock \showarticletitle{Visual Attention in Real-World Conversation: Gaze Patterns Are Modulated by Communication and Group Size}.
\newblock \bibinfo{journal}{\emph{Applied Psychology}} \bibinfo{volume}{70}, \bibinfo{number}{4} (\bibinfo{year}{2021}), \bibinfo{pages}{1602--1627}.
\newblock


\bibitem[Martin et~al\mbox{.}(2016)]%
        {martin2016leader}
\bibfield{author}{\bibinfo{person}{Robin Martin}, \bibinfo{person}{Yves Guillaume}, \bibinfo{person}{Geoff Thomas}, \bibinfo{person}{Allan Lee}, {and} \bibinfo{person}{Olga Epitropaki}.} \bibinfo{year}{2016}\natexlab{}.
\newblock \showarticletitle{Leader--member exchange (LMX) and performance: A meta-analytic review}.
\newblock \bibinfo{journal}{\emph{Personnel psychology}} \bibinfo{volume}{69}, \bibinfo{number}{1} (\bibinfo{year}{2016}), \bibinfo{pages}{67--121}.
\newblock


\bibitem[Matsuyama et~al\mbox{.}(2015)]%
        {matsuyama2015four}
\bibfield{author}{\bibinfo{person}{Yoichi Matsuyama}, \bibinfo{person}{Iwao Akiba}, \bibinfo{person}{Shinya Fujie}, {and} \bibinfo{person}{Tetsunori Kobayashi}.} \bibinfo{year}{2015}\natexlab{}.
\newblock \showarticletitle{Four-participant group conversation: A facilitation robot controlling engagement density as the fourth participant}.
\newblock \bibinfo{journal}{\emph{Computer Speech \& Language}} \bibinfo{volume}{33}, \bibinfo{number}{1} (\bibinfo{year}{2015}), \bibinfo{pages}{1--24}.
\newblock


\bibitem[Mauchly(1940)]%
        {mauchly1940significance}
\bibfield{author}{\bibinfo{person}{John~W Mauchly}.} \bibinfo{year}{1940}\natexlab{}.
\newblock \showarticletitle{Significance test for sphericity of a normal n-variate distribution}.
\newblock \bibinfo{journal}{\emph{The Annals of Mathematical Statistics}} \bibinfo{volume}{11}, \bibinfo{number}{2} (\bibinfo{year}{1940}), \bibinfo{pages}{204--209}.
\newblock


\bibitem[McHugh et~al\mbox{.}(2016)]%
        {mchugh2016collective}
\bibfield{author}{\bibinfo{person}{Kristie~A McHugh}, \bibinfo{person}{Francis~J Yammarino}, \bibinfo{person}{Shelley~D Dionne}, \bibinfo{person}{Andra Serban}, \bibinfo{person}{Hiroki Sayama}, {and} \bibinfo{person}{Subimal Chatterjee}.} \bibinfo{year}{2016}\natexlab{}.
\newblock \showarticletitle{Collective decision making, leadership, and collective intelligence: Tests with agent-based simulations and a Field study}.
\newblock \bibinfo{journal}{\emph{The Leadership Quarterly}} \bibinfo{volume}{27}, \bibinfo{number}{2} (\bibinfo{year}{2016}), \bibinfo{pages}{218--241}.
\newblock


\bibitem[Melo et~al\mbox{.}(2010)]%
        {melo2010influence}
\bibfield{author}{\bibinfo{person}{Celso M~de Melo}, \bibinfo{person}{Peter Carnevale}, {and} \bibinfo{person}{Jonathan Gratch}.} \bibinfo{year}{2010}\natexlab{}.
\newblock \showarticletitle{The influence of emotions in embodied agents on human decision-making}. In \bibinfo{booktitle}{\emph{International conference on intelligent virtual agents}}. Springer, \bibinfo{pages}{357--370}.
\newblock


\bibitem[Meslec and Cur{\c{s}}eu(2013)]%
        {meslec2013too}
\bibfield{author}{\bibinfo{person}{Nicoleta Meslec} {and} \bibinfo{person}{Petru~Lucian Cur{\c{s}}eu}.} \bibinfo{year}{2013}\natexlab{}.
\newblock \showarticletitle{Too close or too far hurts: cognitive distance and group cognitive synergy}.
\newblock \bibinfo{journal}{\emph{Small Group Research}} \bibinfo{volume}{44}, \bibinfo{number}{5} (\bibinfo{year}{2013}), \bibinfo{pages}{471--497}.
\newblock


\bibitem[Mixamo(2008)]%
        {Mixamo}
\bibfield{author}{\bibinfo{person}{Mixamo}.} \bibinfo{year}{2008}\natexlab{}.
\newblock \bibinfo{title}{Mixamo}.
\newblock
\newblock
\urldef\tempurl%
\url{https://www.mixamo.com}
\showURL{%
\tempurl}
\newblock
\shownote{Accessed oct. 14, 2022}.


\bibitem[Moridis and Economides(2012)]%
        {moridis2012affective}
\bibfield{author}{\bibinfo{person}{Christos~N Moridis} {and} \bibinfo{person}{Anastasios~A Economides}.} \bibinfo{year}{2012}\natexlab{}.
\newblock \showarticletitle{Affective learning: Empathetic agents with emotional facial and tone of voice expressions}.
\newblock \bibinfo{journal}{\emph{IEEE Transactions on Affective Computing}} \bibinfo{volume}{3}, \bibinfo{number}{3} (\bibinfo{year}{2012}), \bibinfo{pages}{260--272}.
\newblock


\bibitem[M{\"u}ller et~al\mbox{.}(2018)]%
        {muller2018robust}
\bibfield{author}{\bibinfo{person}{Philipp M{\"u}ller}, \bibinfo{person}{Michael~Xuelin Huang}, \bibinfo{person}{Xucong Zhang}, {and} \bibinfo{person}{Andreas Bulling}.} \bibinfo{year}{2018}\natexlab{}.
\newblock \showarticletitle{Robust eye contact detection in natural multi-person interactions using gaze and speaking behaviour}. In \bibinfo{booktitle}{\emph{Proceedings of the 2018 ACM Symposium on Eye Tracking Research \& Applications}}. \bibinfo{pages}{1--10}.
\newblock


\bibitem[Nam et~al\mbox{.}(2022)]%
        {nam2022emotionally}
\bibfield{author}{\bibinfo{person}{Hyeongil Nam}, \bibinfo{person}{Chanhee Kim}, \bibinfo{person}{Kangsoo Kim}, \bibinfo{person}{Ji-Young Yeo}, {and} \bibinfo{person}{Jong-Il Park}.} \bibinfo{year}{2022}\natexlab{}.
\newblock \showarticletitle{An Emotionally Responsive Virtual Parent for Pediatric Nursing Education: A Framework for Multimodal Momentary and Accumulated Interventions}. In \bibinfo{booktitle}{\emph{IEEE International Symposium on Mixed and Augmented Reality}}.
\newblock


\bibitem[{Naver Cloud}(2019)]%
        {NaverCloudAPI}
\bibfield{author}{\bibinfo{person}{{Naver Cloud}}.} \bibinfo{year}{2019}\natexlab{}.
\newblock \bibinfo{title}{Naver Cloud Platform AI Services}.
\newblock
\newblock
\urldef\tempurl%
\url{https://www.ncloud.com/product/aiService/}
\showURL{%
\tempurl}
\newblock
\shownote{Accessed oct. 14, 2022}.


\bibitem[Nayak and Turk(2005)]%
        {nayak2005emotional}
\bibfield{author}{\bibinfo{person}{Vishal Nayak} {and} \bibinfo{person}{Matthew Turk}.} \bibinfo{year}{2005}\natexlab{}.
\newblock \showarticletitle{Emotional expression in virtual agents through body language}. In \bibinfo{booktitle}{\emph{International Symposium on Visual Computing}}. Springer, \bibinfo{pages}{313--320}.
\newblock


\bibitem[Nemiroff and Pasmore(2001)]%
        {nemiroff2001lost}
\bibfield{author}{\bibinfo{person}{Paul~M Nemiroff} {and} \bibinfo{person}{William~A Pasmore}.} \bibinfo{year}{2001}\natexlab{}.
\newblock \showarticletitle{Lost at sea: A consensus-seeking task}.
\newblock \bibinfo{journal}{\emph{The Pfeiffer book of successful team-building tools: Best of the annuals}} (\bibinfo{year}{2001}), \bibinfo{pages}{165--172}.
\newblock


\bibitem[Paakkanen et~al\mbox{.}(2021)]%
        {paakkanen2021responding}
\bibfield{author}{\bibinfo{person}{Miia~A Paakkanen}, \bibinfo{person}{Frank Martela}, {and} \bibinfo{person}{Anne~B Pessi}.} \bibinfo{year}{2021}\natexlab{}.
\newblock \showarticletitle{Responding to positive emotions at work--the four steps and potential benefits of a validating response to coworkers’ positive experiences}.
\newblock \bibinfo{journal}{\emph{Frontiers in Psychology}} (\bibinfo{year}{2021}), \bibinfo{pages}{4553}.
\newblock


\bibitem[Paiva et~al\mbox{.}(2017)]%
        {PaivaEtAl2017_EmpathyInVirtualAgents}
\bibfield{author}{\bibinfo{person}{Ana Paiva}, \bibinfo{person}{Iolanda Leite}, \bibinfo{person}{Hana Boukricha}, {and} \bibinfo{person}{Ipke Wachsmuth}.} \bibinfo{year}{2017}\natexlab{}.
\newblock \showarticletitle{Empathy in Virtual Agents and Robots: A Survey}.
\newblock \bibinfo{journal}{\emph{ACM Trans. Interact. Intell. Syst.}} \bibinfo{volume}{7}, \bibinfo{number}{3}, Article \bibinfo{articleno}{11} (\bibinfo{date}{sep} \bibinfo{year}{2017}), \bibinfo{numpages}{40}~pages.
\newblock
\showISSN{2160-6455}
\urldef\tempurl%
\url{https://doi.org/10.1145/2912150}
\showDOI{\tempurl}


\bibitem[Pauw et~al\mbox{.}(2022)]%
        {pauw2022avatar}
\bibfield{author}{\bibinfo{person}{Lisanne~S Pauw}, \bibinfo{person}{Disa~A Sauter}, \bibinfo{person}{Gerben~A van Kleef}, \bibinfo{person}{Gale~M Lucas}, \bibinfo{person}{Jonathan Gratch}, {and} \bibinfo{person}{Agneta~H Fischer}.} \bibinfo{year}{2022}\natexlab{}.
\newblock \showarticletitle{The avatar will see you now: Support from a virtual human provides socio-emotional benefits}.
\newblock \bibinfo{journal}{\emph{Computers in Human Behavior}}  \bibinfo{volume}{136} (\bibinfo{year}{2022}), \bibinfo{pages}{107368}.
\newblock


\bibitem[Pham(2006)]%
        {pham2006pyaudio}
\bibfield{author}{\bibinfo{person}{Hubert Pham}.} \bibinfo{year}{2006}\natexlab{}.
\newblock \showarticletitle{Pyaudio: Portaudio v19 python bindings}.
\newblock \bibinfo{journal}{\emph{URL: https://people. csail. mit. edu/hubert/pyaudio}} (\bibinfo{year}{2006}).
\newblock


\bibitem[Phelps et~al\mbox{.}(2006)]%
        {phelps2006helping}
\bibfield{author}{\bibinfo{person}{Fiona~G Phelps}, \bibinfo{person}{Gwyneth Doherty-Sneddon}, {and} \bibinfo{person}{Hannah Warnock}.} \bibinfo{year}{2006}\natexlab{}.
\newblock \showarticletitle{Helping children think: Gaze aversion and teaching}.
\newblock \bibinfo{journal}{\emph{British journal of developmental psychology}} \bibinfo{volume}{24}, \bibinfo{number}{3} (\bibinfo{year}{2006}), \bibinfo{pages}{577--588}.
\newblock


\bibitem[P{\"u}tten et~al\mbox{.}(2010)]%
        {putten2010our}
\bibfield{author}{\bibinfo{person}{Astrid~M P{\"u}tten}, \bibinfo{person}{Nicole~C Kr{\"a}mer}, {and} \bibinfo{person}{Jonathan Gratch}.} \bibinfo{year}{2010}\natexlab{}.
\newblock \showarticletitle{How our personality shapes our interactions with virtual characters-implications for research and development}. In \bibinfo{booktitle}{\emph{International Conference on Intelligent Virtual Agents}}. Springer, \bibinfo{pages}{208--221}.
\newblock


\bibitem[Qu et~al\mbox{.}(2014)]%
        {qu2014conversations}
\bibfield{author}{\bibinfo{person}{Chao Qu}, \bibinfo{person}{Willem-Paul Brinkman}, \bibinfo{person}{Yun Ling}, \bibinfo{person}{Pascal Wiggers}, {and} \bibinfo{person}{Ingrid Heynderickx}.} \bibinfo{year}{2014}\natexlab{}.
\newblock \showarticletitle{Conversations with a virtual human: Synthetic emotions and human responses}.
\newblock \bibinfo{journal}{\emph{Computers in Human Behavior}}  \bibinfo{volume}{34} (\bibinfo{year}{2014}), \bibinfo{pages}{58--68}.
\newblock


\bibitem[Randhavane et~al\mbox{.}(2019)]%
        {randhavane2019eva}
\bibfield{author}{\bibinfo{person}{Tanmay Randhavane}, \bibinfo{person}{Aniket Bera}, \bibinfo{person}{Kyra Kapsaskis}, \bibinfo{person}{Rahul Sheth}, \bibinfo{person}{Kurt Gray}, {and} \bibinfo{person}{Dinesh Manocha}.} \bibinfo{year}{2019}\natexlab{}.
\newblock \showarticletitle{Eva: Generating emotional behavior of virtual agents using expressive features of gait and gaze}. In \bibinfo{booktitle}{\emph{ACM symposium on applied perception 2019}}. \bibinfo{pages}{1--10}.
\newblock


\bibitem[Rehm and Andr{\'e}(2005)]%
        {rehm2005they}
\bibfield{author}{\bibinfo{person}{Matthias Rehm} {and} \bibinfo{person}{Elisabeth Andr{\'e}}.} \bibinfo{year}{2005}\natexlab{}.
\newblock \showarticletitle{Where do they look? Gaze behaviors of multiple users interacting with an embodied conversational agent}. In \bibinfo{booktitle}{\emph{International Workshop on Intelligent Virtual Agents}}. Springer, \bibinfo{pages}{241--252}.
\newblock


\bibitem[Reimers and Gurevych(2019)]%
        {reimers-2019-sentence-bert}
\bibfield{author}{\bibinfo{person}{Nils Reimers} {and} \bibinfo{person}{Iryna Gurevych}.} \bibinfo{year}{2019}\natexlab{}.
\newblock \showarticletitle{Sentence-BERT: Sentence Embeddings using Siamese BERT-Networks}. In \bibinfo{booktitle}{\emph{Proceedings of the 2019 Conference on Empirical Methods in Natural Language Processing}}. \bibinfo{publisher}{Association for Computational Linguistics}.
\newblock
\urldef\tempurl%
\url{https://arxiv.org/abs/1908.10084}
\showURL{%
\tempurl}


\bibitem[Rosenberg-Kima et~al\mbox{.}(2020)]%
        {rosenberg2020robot}
\bibfield{author}{\bibinfo{person}{Rinat~B Rosenberg-Kima}, \bibinfo{person}{Yaacov Koren}, {and} \bibinfo{person}{Goren Gordon}.} \bibinfo{year}{2020}\natexlab{}.
\newblock \showarticletitle{Robot-supported collaborative learning (RSCL): Social robots as teaching assistants for higher education small group facilitation}.
\newblock \bibinfo{journal}{\emph{Frontiers in Robotics and AI}}  \bibinfo{volume}{6} (\bibinfo{year}{2020}), \bibinfo{pages}{148}.
\newblock


\bibitem[Rudman and Goodwin(2004)]%
        {rudman2004gender}
\bibfield{author}{\bibinfo{person}{Laurie~A Rudman} {and} \bibinfo{person}{Stephanie~A Goodwin}.} \bibinfo{year}{2004}\natexlab{}.
\newblock \showarticletitle{Gender differences in automatic in-group bias: Why do women like women more than men like men?}
\newblock \bibinfo{journal}{\emph{Journal of personality and social psychology}} \bibinfo{volume}{87}, \bibinfo{number}{4} (\bibinfo{year}{2004}), \bibinfo{pages}{494}.
\newblock


\bibitem[Sajjadi et~al\mbox{.}(2020)]%
        {sajjadi2020effect}
\bibfield{author}{\bibinfo{person}{Pejman Sajjadi}, \bibinfo{person}{Jiayan Zhao}, \bibinfo{person}{Jan~O Wallgr{\"u}n}, \bibinfo{person}{Tanya Furman}, \bibinfo{person}{Peter~C La~Femina}, \bibinfo{person}{Alex Fatemi}, \bibinfo{person}{Zachary~E Zidik}, {and} \bibinfo{person}{Alexander Klippel}.} \bibinfo{year}{2020}\natexlab{}.
\newblock \showarticletitle{The effect of virtual agent gender and embodiment on the experiences and performance of students in Virtual Field Trips}. In \bibinfo{booktitle}{\emph{2020 IEEE International Conference on Teaching, Assessment, and Learning for Engineering}}.
\newblock


\bibitem[Scharlemann et~al\mbox{.}(2001)]%
        {scharlemann2001value}
\bibfield{author}{\bibinfo{person}{J{\"o}rn~PW Scharlemann}, \bibinfo{person}{Catherine~C Eckel}, \bibinfo{person}{Alex Kacelnik}, {and} \bibinfo{person}{Rick~K Wilson}.} \bibinfo{year}{2001}\natexlab{}.
\newblock \showarticletitle{The value of a smile: Game theory with a human face}.
\newblock \bibinfo{journal}{\emph{Journal of Economic Psychology}} \bibinfo{volume}{22}, \bibinfo{number}{5} (\bibinfo{year}{2001}), \bibinfo{pages}{617--640}.
\newblock


\bibitem[{Science of People}(2022)]%
        {BodyLanguage}
\bibfield{author}{\bibinfo{person}{{Science of People}}.} \bibinfo{year}{2022}\natexlab{}.
\newblock \bibinfo{title}{The Ultimate Guide to Body Language}.
\newblock
\newblock
\urldef\tempurl%
\url{https://www.scienceofpeople.com/category/body-language/}
\showURL{%
\tempurl}


\bibitem[Sebo et~al\mbox{.}(2020)]%
        {sebo2020robots}
\bibfield{author}{\bibinfo{person}{Sarah Sebo}, \bibinfo{person}{Brett Stoll}, \bibinfo{person}{Brian Scassellati}, {and} \bibinfo{person}{Malte~F Jung}.} \bibinfo{year}{2020}\natexlab{}.
\newblock \showarticletitle{Robots in groups and teams: a literature review}.
\newblock \bibinfo{journal}{\emph{Proceedings of the ACM on Human-Computer Interaction}} \bibinfo{volume}{4}, \bibinfo{number}{CSCW2} (\bibinfo{year}{2020}), \bibinfo{pages}{1--36}.
\newblock


\bibitem[Senju and Johnson(2009)]%
        {senju2009eye}
\bibfield{author}{\bibinfo{person}{Atsushi Senju} {and} \bibinfo{person}{Mark~H Johnson}.} \bibinfo{year}{2009}\natexlab{}.
\newblock \showarticletitle{The eye contact effect: mechanisms and development}.
\newblock \bibinfo{journal}{\emph{Trends in cognitive sciences}} \bibinfo{volume}{13}, \bibinfo{number}{3} (\bibinfo{year}{2009}).
\newblock


\bibitem[SHAPIRO and WILK(1965)]%
        {shapiro}
\bibfield{author}{\bibinfo{person}{S.~S. SHAPIRO} {and} \bibinfo{person}{M.~B. WILK}.} \bibinfo{year}{1965}\natexlab{}.
\newblock \showarticletitle{{An analysis of variance test for normality (complete samples)}}.
\newblock \bibinfo{journal}{\emph{Biometrika}} \bibinfo{volume}{52}, \bibinfo{number}{3-4} (\bibinfo{year}{1965}).
\newblock
\showISSN{0006-3444}


\bibitem[Sidner et~al\mbox{.}(2004)]%
        {sidner2004look}
\bibfield{author}{\bibinfo{person}{Candace~L Sidner}, \bibinfo{person}{Cory~D Kidd}, \bibinfo{person}{Christopher Lee}, {and} \bibinfo{person}{Neal Lesh}.} \bibinfo{year}{2004}\natexlab{}.
\newblock \showarticletitle{Where to look: a study of human-robot engagement}. In \bibinfo{booktitle}{\emph{Proceedings of the 9th international conference on Intelligent user interfaces}}. \bibinfo{pages}{78--84}.
\newblock


\bibitem[Singh et~al\mbox{.}(2021)]%
        {singh2021beholden}
\bibfield{author}{\bibinfo{person}{Ranjit~Konrad Singh}, \bibinfo{person}{Birgit~Johanna Voggeser}, {and} \bibinfo{person}{Anja~Simone G{\"o}ritz}.} \bibinfo{year}{2021}\natexlab{}.
\newblock \showarticletitle{Beholden: the emotional effects of having eye contact while breaking social norms}.
\newblock \bibinfo{journal}{\emph{Frontiers in psychology}}  \bibinfo{volume}{12} (\bibinfo{year}{2021}), \bibinfo{pages}{545268}.
\newblock


\bibitem[Takano et~al\mbox{.}(2012)]%
        {takano2012tablets}
\bibfield{author}{\bibinfo{person}{Kentaro Takano}, \bibinfo{person}{Hirohito Shibata}, \bibinfo{person}{Kengo Omura}, \bibinfo{person}{Junko Ichino}, \bibinfo{person}{Tomonori Hashiyama}, {and} \bibinfo{person}{Shun'ichi Tano}.} \bibinfo{year}{2012}\natexlab{}.
\newblock \showarticletitle{Do tablets really support discussion? Comparison between paper, tablet, and laptop PC used as discussion tools}. In \bibinfo{booktitle}{\emph{Proceedings of the 24th Australian Computer-Human Interaction Conference}}. \bibinfo{pages}{562--571}.
\newblock


\bibitem[ter Heijden and Brinkman(2011)]%
        {ter2011design}
\bibfield{author}{\bibinfo{person}{Niels ter Heijden} {and} \bibinfo{person}{Willem-Paul Brinkman}.} \bibinfo{year}{2011}\natexlab{}.
\newblock \showarticletitle{Design and evaluation of a virtual reality exposure therapy system with automatic free speech interaction}.
\newblock \bibinfo{journal}{\emph{Journal of CyberTherapy and Rehabilitation}} \bibinfo{volume}{4}, \bibinfo{number}{1} (\bibinfo{year}{2011}).
\newblock


\bibitem[ter Stal et~al\mbox{.}(2020)]%
        {ageprefer2020}
\bibfield{author}{\bibinfo{person}{Silke ter Stal}, \bibinfo{person}{Monique Tabak}, \bibinfo{person}{Harm op~den Akker}, \bibinfo{person}{Tessa Beinema}, {and} \bibinfo{person}{Hermie Hermens}.} \bibinfo{year}{2020}\natexlab{}.
\newblock \showarticletitle{Who Do You Prefer? The Effect of Age, Gender and Role on Users’ First Impressions of Embodied Conversational Agents in eHealth}.
\newblock \bibinfo{journal}{\emph{International Journal of Human–Computer Interaction}} \bibinfo{volume}{36}, \bibinfo{number}{9} (\bibinfo{year}{2020}).
\newblock


\bibitem[Torre et~al\mbox{.}(2019)]%
        {torre2019effect}
\bibfield{author}{\bibinfo{person}{Ilaria Torre}, \bibinfo{person}{Emma Carrigan}, \bibinfo{person}{Rachel McDonnell}, \bibinfo{person}{Katarina Domijan}, \bibinfo{person}{Killian McCabe}, {and} \bibinfo{person}{Naomi Harte}.} \bibinfo{year}{2019}\natexlab{}.
\newblock \showarticletitle{The effect of multimodal emotional expression and agent appearance on trust in human-agent interaction}. In \bibinfo{booktitle}{\emph{Proceedings of the 12th ACM SIGGRAPH Conference on Motion, Interaction and Games}}. \bibinfo{pages}{1--6}.
\newblock


\bibitem[Torre et~al\mbox{.}(2020)]%
        {torre2020if}
\bibfield{author}{\bibinfo{person}{Ilaria Torre}, \bibinfo{person}{Jeremy Goslin}, {and} \bibinfo{person}{Laurence White}.} \bibinfo{year}{2020}\natexlab{}.
\newblock \showarticletitle{If your device could smile: People trust happy-sounding artificial agents more}.
\newblock \bibinfo{journal}{\emph{Computers in Human Behavior}}  \bibinfo{volume}{105} (\bibinfo{year}{2020}), \bibinfo{pages}{106215}.
\newblock


\bibitem[Trevors et~al\mbox{.}(2016)]%
        {trevors2016identity}
\bibfield{author}{\bibinfo{person}{Gregory~J Trevors}, \bibinfo{person}{Krista~R Muis}, \bibinfo{person}{Reinhard Pekrun}, \bibinfo{person}{Gale~M Sinatra}, {and} \bibinfo{person}{Philip~H Winne}.} \bibinfo{year}{2016}\natexlab{}.
\newblock \showarticletitle{Identity and epistemic emotions during knowledge revision: A potential account for the backfire effect}.
\newblock \bibinfo{journal}{\emph{Discourse Processes}}  \bibinfo{volume}{53} (\bibinfo{year}{2016}).
\newblock


\bibitem[{Unterweger}(2022)]%
        {Unterweger2022_EmotionalInvolvement}
\bibfield{author}{\bibinfo{person}{A.J. {Unterweger}}.} \bibinfo{year}{2022}\natexlab{}.
\newblock \bibinfo{title}{Emotional involvement in group discussions : an explorative study}.
\newblock
\newblock


\bibitem[van~de Kerkhof(2006)]%
        {Kerkhof2006_MakingADifference}
\bibfield{author}{\bibinfo{person}{Marleen van~de Kerkhof}.} \bibinfo{year}{2006}\natexlab{}.
\newblock \showarticletitle{Making a Difference: On the Constraints of Consensus Building and the Relevance of Deliberation in Stakeholder Dialogues}.
\newblock \bibinfo{journal}{\emph{Policy Sciences}} \bibinfo{volume}{39}, \bibinfo{number}{3} (\bibinfo{year}{2006}), \bibinfo{pages}{279--299}.
\newblock


\bibitem[Van~Kleef et~al\mbox{.}(2004)]%
        {van2004interpersonal}
\bibfield{author}{\bibinfo{person}{Gerben~A Van~Kleef}, \bibinfo{person}{Carsten~KW De~Dreu}, {and} \bibinfo{person}{Antony~SR Manstead}.} \bibinfo{year}{2004}\natexlab{}.
\newblock \showarticletitle{The interpersonal effects of anger and happiness in negotiations.}
\newblock \bibinfo{journal}{\emph{Journal of personality and social psychology}} \bibinfo{volume}{86}, \bibinfo{number}{1} (\bibinfo{year}{2004}), \bibinfo{pages}{57}.
\newblock


\bibitem[Veletsianos(2012)]%
        {veletsianos2012learners}
\bibfield{author}{\bibinfo{person}{George Veletsianos}.} \bibinfo{year}{2012}\natexlab{}.
\newblock \showarticletitle{How do learners respond to pedagogical agents that deliver social-oriented non-task messages? Impact on student learning, perceptions, and experiences}.
\newblock \bibinfo{journal}{\emph{Computers in Human Behavior}} \bibinfo{volume}{28}, \bibinfo{number}{1} (\bibinfo{year}{2012}), \bibinfo{pages}{275--283}.
\newblock


\bibitem[Volonte et~al\mbox{.}(2022)]%
        {blendshapereference}
\bibfield{author}{\bibinfo{person}{Matias Volonte}, \bibinfo{person}{Eyal Ofek}, \bibinfo{person}{Ken Jakubzak}, \bibinfo{person}{Shawn Bruner}, {and} \bibinfo{person}{Mar Gonzalez-Franco}.} \bibinfo{year}{2022}\natexlab{}.
\newblock \showarticletitle{HeadBox: A Facial Blendshape Animation Toolkit for the Microsoft Rocketbox Library}. In \bibinfo{booktitle}{\emph{2022 IEEE Conference on Virtual Reality and 3D User Interfaces Abstracts and Workshops (VRW)}}. \bibinfo{pages}{39--42}.
\newblock
\urldef\tempurl%
\url{https://doi.org/10.1109/VRW55335.2022.00015}
\showDOI{\tempurl}


\bibitem[Von~der P{\"u}tten et~al\mbox{.}(2010)]%
        {von2010doesn}
\bibfield{author}{\bibinfo{person}{Astrid~M Von~der P{\"u}tten}, \bibinfo{person}{Nicole~C Kr{\"a}mer}, \bibinfo{person}{Jonathan Gratch}, {and} \bibinfo{person}{Sin-Hwa Kang}.} \bibinfo{year}{2010}\natexlab{}.
\newblock \showarticletitle{“It doesn’t matter what you are!” Explaining social effects of agents and avatars}.
\newblock \bibinfo{journal}{\emph{Computers in Human Behavior}} \bibinfo{volume}{26}, \bibinfo{number}{6} (\bibinfo{year}{2010}), \bibinfo{pages}{1641--1650}.
\newblock


\bibitem[Walker et~al\mbox{.}(2019)]%
        {walker2019influence}
\bibfield{author}{\bibinfo{person}{Michael~E Walker}, \bibinfo{person}{Daniel Szafir}, {and} \bibinfo{person}{Irene Rae}.} \bibinfo{year}{2019}\natexlab{}.
\newblock \showarticletitle{The influence of size in augmented reality telepresence avatars}. In \bibinfo{booktitle}{\emph{2019 IEEE Conference on Virtual Reality and 3D User Interfaces (VR)}}. IEEE, \bibinfo{pages}{538--546}.
\newblock


\bibitem[Wang et~al\mbox{.}(2019)]%
        {wang2019exploring}
\bibfield{author}{\bibinfo{person}{Isaac Wang}, \bibinfo{person}{Jesse Smith}, {and} \bibinfo{person}{Jaime Ruiz}.} \bibinfo{year}{2019}\natexlab{}.
\newblock \showarticletitle{Exploring virtual agents for augmented reality}. In \bibinfo{booktitle}{\emph{Proceedings of the 2019 CHI Conference on Human Factors in Computing Systems}}. \bibinfo{pages}{1--12}.
\newblock


\bibitem[Wobbrock et~al\mbox{.}(2011)]%
        {ARTmethodCitation}
\bibfield{author}{\bibinfo{person}{Jacob~O. Wobbrock}, \bibinfo{person}{Leah Findlater}, \bibinfo{person}{Darren Gergle}, {and} \bibinfo{person}{James~J. Higgins}.} \bibinfo{year}{2011}\natexlab{}.
\newblock \showarticletitle{The Aligned Rank Transform for Nonparametric Factorial Analyses Using Only Anova Procedures}. In \bibinfo{booktitle}{\emph{Proceedings of the SIGCHI Conference on Human Factors in Computing Systems}} (Vancouver, BC, Canada). \bibinfo{pages}{143–146}.
\newblock
\showISBNx{9781450302289}


\bibitem[Wong et~al\mbox{.}(2023)]%
        {wong2023comparing}
\bibfield{author}{\bibinfo{person}{Sai-Keung Wong}, \bibinfo{person}{Matias Volonte}, \bibinfo{person}{Kuan-Yu Liu}, \bibinfo{person}{Elham Ebrahimi}, {and} \bibinfo{person}{Sabarish~V Babu}.} \bibinfo{year}{2023}\natexlab{}.
\newblock \showarticletitle{Comparing Visual Attention with Leading and Following Virtual Agents in a Collaborative Perception-Action Task in VR}. In \bibinfo{booktitle}{\emph{2023 IEEE Conference Virtual Reality and 3D User Interfaces (VR)}}. IEEE, \bibinfo{pages}{152--162}.
\newblock


\bibitem[Xavier(2005)]%
        {xavier2005you}
\bibfield{author}{\bibinfo{person}{Stephen Xavier}.} \bibinfo{year}{2005}\natexlab{}.
\newblock \showarticletitle{Are you at the top of your game? Checklist for effective leaders}.
\newblock \bibinfo{journal}{\emph{Journal of Business Strategy}} \bibinfo{volume}{26}, \bibinfo{number}{3} (\bibinfo{year}{2005}), \bibinfo{pages}{35--42}.
\newblock


\bibitem[Yukl(1989)]%
        {yukl1989managerial}
\bibfield{author}{\bibinfo{person}{Gary Yukl}.} \bibinfo{year}{1989}\natexlab{}.
\newblock \showarticletitle{Managerial leadership: A review of theory and research}.
\newblock \bibinfo{journal}{\emph{Journal of management}} \bibinfo{volume}{15}, \bibinfo{number}{2} (\bibinfo{year}{1989}), \bibinfo{pages}{251--289}.
\newblock


\bibitem[Yumak et~al\mbox{.}(2014a)]%
        {yumak2014modelling}
\bibfield{author}{\bibinfo{person}{Zerrin Yumak}, \bibinfo{person}{Jianfeng Ren}, \bibinfo{person}{Nadia~Magnenat Thalmann}, {and} \bibinfo{person}{Junsong Yuan}.} \bibinfo{year}{2014}\natexlab{a}.
\newblock \showarticletitle{Modelling multi-party interactions among virtual characters, robots, and humans}.
\newblock \bibinfo{journal}{\emph{Presence: Teleoperators and Virtual Environments}} \bibinfo{volume}{23}, \bibinfo{number}{2} (\bibinfo{year}{2014}).
\newblock


\bibitem[Yumak et~al\mbox{.}(2014b)]%
        {yumak2014tracking}
\bibfield{author}{\bibinfo{person}{Zerrin Yumak}, \bibinfo{person}{Jianfeng Ren}, \bibinfo{person}{Nadia~Magnenat Thalmann}, {and} \bibinfo{person}{Junsong Yuan}.} \bibinfo{year}{2014}\natexlab{b}.
\newblock \showarticletitle{Tracking and fusion for multiparty interaction with a virtual character and a social robot}.
\newblock In \bibinfo{booktitle}{\emph{SIGGRAPH Asia 2014 Autonomous Virtual Humans and Social Robot for Telepresence}}. \bibinfo{pages}{1--7}.
\newblock


\bibitem[Zeman and Garber(1996)]%
        {zeman1996display}
\bibfield{author}{\bibinfo{person}{Janice Zeman} {and} \bibinfo{person}{Judy Garber}.} \bibinfo{year}{1996}\natexlab{}.
\newblock \showarticletitle{Display rules for anger, sadness, and pain: It depends on who is watching}.
\newblock \bibinfo{journal}{\emph{Child development}} \bibinfo{volume}{67}, \bibinfo{number}{3} (\bibinfo{year}{1996}), \bibinfo{pages}{957--973}.
\newblock


\bibitem[Zhou et~al\mbox{.}(2020)]%
        {zhou2020emotional}
\bibfield{author}{\bibinfo{person}{Wencang Zhou}, \bibinfo{person}{Zhu Zhu}, {and} \bibinfo{person}{Donald Vredenburgh}.} \bibinfo{year}{2020}\natexlab{}.
\newblock \showarticletitle{Emotional intelligence, psychological safety, and team decision making}.
\newblock \bibinfo{journal}{\emph{Team Performance Management: An International Journal}} (\bibinfo{year}{2020}).
\newblock


\end{thebibliography}


\newpage
\appendix
\onecolumn
\section{Appendix}

\begin{table}[h]
\caption{The sentiment analysis results of the statements spoken by the virtual agents (VAs) for each task.~\label{tableAppendixSentiment}}
\Description{The sentiment analysis results of the statements spoken by the virtual agents (VAs) for each task.
}
\centering
\resizebox{1\linewidth}{!}{
\renewcommand{\arraystretch}{1.05}
\begin{tabular}{llcccccccc}
\toprule[1pt]\midrule[0.3pt]
\multicolumn{1}{l}{\multirow{3}{*}{\textbf{Task}}}   & \multirow{3}{*}{\begin{tabular}[c]{@{}c@{}}\textbf{Aspect}\\ \textbf{of Statements}\end{tabular}} & \multirow{3}{*}{\begin{tabular}[c]{@{}c@{}}\textbf{Number}\\ \textbf{of Sentences}\end{tabular}} & \multicolumn{1}{c}{\multirow{3}{*}{\begin{tabular}[c]{@{}c@{}}\textbf{Length}\\ (Mean) \end{tabular}}} & \multicolumn{6}{c}{\textbf{Sentiment} (\%)}                                                                                                                                                      \\ \cmidrule(lr){5-10}
\multicolumn{1}{l}{}                        &                            &                      & \multicolumn{1}{c}{}                                                                         & \multicolumn{2}{c}{\textbf{Neutral}}                                         & \multicolumn{2}{c}{\textbf{Positive}}                                     & \multicolumn{2}{c}{\textbf{Negative}}                 \\ 
\cmidrule(lr){5-6}
\cmidrule(lr){7-8}
\cmidrule(lr){9-10}
\multicolumn{1}{l}{}                        &                            &                      & \multicolumn{1}{c}{}                                                                         & \multicolumn{1}{c}{Mean (SD)}    & \multicolumn{1}{c}{Min-Max}     & \multicolumn{1}{c}{Mean (SD)}   & \multicolumn{1}{c}{Min-Max}   & \multicolumn{1}{c}{Mean (SD)}    & Min-Max   \\ 
\midrule[1pt]
\textbf{NASA}   & Agreement                  & 105                  & \multicolumn{1}{c}{148.30}                                                                   & \multicolumn{1}{c}{99.92 (0.00)} & \multicolumn{1}{c}{99.77-99.99} & \multicolumn{1}{c}{0.04 (0.00)} & \multicolumn{1}{c}{0.00-0.16} & \multicolumn{1}{c}{0.03 (0.00)} & 0.00-0.09 \\
\multicolumn{1}{l}{}                        & Disagreement               & 105                  & \multicolumn{1}{c}{146.66}                                                                   & \multicolumn{1}{c}{99.77 (0.03)} & \multicolumn{1}{c}{99.20-99.99} & \multicolumn{1}{c}{0.14 (0.02)} & \multicolumn{1}{c}{0.00-0.75} & \multicolumn{1}{c}{0.08 (0.00)} & 0.00-0.30 \\ \midrule
\textbf{WINTER} & Agreement                  & 105                  & \multicolumn{1}{c}{148.09}                                                                   & \multicolumn{1}{c}{99.75 (0.06)} & \multicolumn{1}{c}{99.05-99.98} & \multicolumn{1}{c}{0.08 (0.01)} & \multicolumn{1}{c}{0.00-0.90} & \multicolumn{1}{c}{0.16 (0.05)} & 0.00-0.75 \\
\multicolumn{1}{l}{}                        & Disagreement               & 105                  & \multicolumn{1}{c}{143.70}                                                                   & \multicolumn{1}{c}{99.80 (0.04)} & \multicolumn{1}{c}{99.14-99.87} & \multicolumn{1}{c}{0.11 (0.03)} & \multicolumn{1}{c}{0.00-0.77} & \multicolumn{1}{c}{0.07 (0.00)} & 0.00-0.32 \\ \midrule
\textbf{DESERT} & Agreement                  & 105                  & \multicolumn{1}{c}{144.98}                                                                   & \multicolumn{1}{c}{99.91 (0.00)} & \multicolumn{1}{c}{99.34-99.98} & \multicolumn{1}{c}{0.04 (0.00)} & \multicolumn{1}{c}{0.00-0.27} & \multicolumn{1}{c}{0.03 (0.00)} & 0.00-0.62 \\
\multicolumn{1}{l}{}                        & Disagreement               & 105                  & \multicolumn{1}{c}{147.12}                                                                   & \multicolumn{1}{c}{99.90 (0.02)} & \multicolumn{1}{c}{99.13-99.99} & \multicolumn{1}{c}{0.04 (0.00)} & \multicolumn{1}{c}{0.00-0.51} & \multicolumn{1}{c}{0.05 (0.01)} & 0.00-0.82 \\ \midrule
\textbf{SEA}    & Agreement                  & 105                  & \multicolumn{1}{c}{146.54}                                                                   & \multicolumn{1}{c}{99.84 (0.01)} & \multicolumn{1}{c}{99.54-99.98} & \multicolumn{1}{c}{0.04 (0.00)} & \multicolumn{1}{c}{0.00-0.32} & \multicolumn{1}{c}{0.10 (0.01)} & 0.01-0.37 \\
\multicolumn{1}{l}{}                        & Disagreement               & 105                  & \multicolumn{1}{c}{143.44}                                                                   & \multicolumn{1}{c}{99.91 (0.01)} & \multicolumn{1}{c}{99.05-99.99} & \multicolumn{1}{c}{0.05 (0.01)} & \multicolumn{1}{c}{0.00-0.75} & \multicolumn{1}{c}{0.03 (0.00)} & 0.00-0.18 \\ 
\midrule[0.3pt]\bottomrule[1pt]
\end{tabular}
}
\end{table}

\begin{table}[!h]
\centering
\caption{
Two types of questionnaires were utilized: the post-treatment questionnaire, administered to participants following each treatment session, which included a range of questions on a 7-point Likert scale, and the post-experiment questionnaire, consisting of five ranking questions, provided upon the completion of all treatments.
}
\Description{Two types of questionnaires were utilized: the post-treatment questionnaire, administered to participants following each treatment session, which included a range of questions on a 7-point Likert scale, and the post-experiment questionnaire, consisting of five ranking questions, provided upon the completion of all treatments.}
~\label{TableAppendixQuestionnaire}
\renewcommand{\arraystretch}{1}
\resizebox{1\linewidth}{!}{
\begin{tabular}{lll}
\toprule[1pt]\midrule[0.3pt]
\textbf{Factor}                    & \textit{\textbf{abberv.}} & \textbf{Question}          \\ \midrule[1pt]
Engagement                & EG-1       & I felt that the VA was focusing on me and one other participant during the discussion.      \\
                                   & EG-2       & I felt that the VA was actively involved in the discussion.                                  \\ \midrule
Affectiveness             & AF-1       & I felt that the VA interacted emotionally with me and one other participant during the discussion.  \\
                                   & AF-2       & I felt that the VA was emotionally involved in the discussion.                               \\ \midrule
\multirow{2}{*}{\begin{tabular}[c]{@{}l@{}}Social Presence\\ (Continuity)\end{tabular}}  & SP-C1                  & I desired to prolong the conversation with the VA.                       \\
                                                                                       & SP-C2                     &  I desired to establish a relationship with the VA.                  \\ \midrule
\multirow{2}{*}{\begin{tabular}[c]{@{}l@{}}Social Presence\\ (Responsiveness)\end{tabular}}  & SP-R1              &   I felt that the VA's words and actions were natural in context.                         \\
                                                                                       & SP-R2                     &  I felt that the VA actually reacted to me.  
                          \\ \midrule
\multirow{2}{*}{\begin{tabular}[c]{@{}l@{}}Social Presence\\ (Membership)\end{tabular}}  & SP-M1                   & During the discussion, I participated by looking at and speaking to the VA.                           \\
                                                                                       & SP-M2                     & During the discussion, I felt the VA was one member of a three-member group.                                \\ \midrule                                   
\multirow{2}{*}{\begin{tabular}[c]{@{}l@{}}Social Influence\\ (Positive)\end{tabular}} & SI-P1                     & My decision was influenced by the opinion of the VA.                             \\
                                                                                       & SI-P2                     & I was persuaded by the VA and thus, I accepted the VA's opinion.                           \\ \midrule
\multirow{2}{*}{\begin{tabular}[c]{@{}l@{}}Social Influence\\ (Negative)\end{tabular}} & SI-N1                     & I felt like I had to agree with the VA's opinion during the discussion.                              \\
                                                                                       & SI-N2                     & I was not persuaded by VA's opinion, but I accepted the VA's opinion.                                       \\ \midrule
\multirow{2}{*}{\begin{tabular}[c]{@{}l@{}}Trustworthiness\\ (Positive)\end{tabular}} & TN-P1                     & I trusted the intentions, thoughts, and conversations of VA.                            \\
                                                                                       & TN-P2     &  Regardless of the VA's opinion, I felt reliable of her based on her tone and behavior.                          \\ \midrule
\multirow{2}{*}{\begin{tabular}[c]{@{}l@{}}Trustworthiness\\ (Negative)\end{tabular}} & TN-N1      &  I doubt the intentions, thoughts, and conversations of VA.                               \\
            & TN-N2     &    Regardless of the VA's opinion, I felt suspicious of her based on her tone and behavior.                    \\ \midrule
Task Load (Mental demand)                    & TL-1         & I felt mental and perceptual efforts due to the VA.                           \\
Task Load (Performance)                      & TL-2         & I felt that the problem had been resolved successfully or accurately due to the VA.                           \\
Task Load (Frustration)                      & TL-3         & I felt perplexed due to the VA.                           \\ \midrule[1pt]
Ranking (Task Difficulty)                                                                        &   -                        & Please rank the assigned tasks in order of their level of difficulty.                   \\
Ranking (VA's Expertise)                                                                         &   -                        & Please rank the VAs in order of perceived professionalism in their speech.                    \\
Ranking (VA's Engagement)                                                                        &   -                        & Please rank the VAs in order of their level of engagement in the discussion.                   \\
Ranking (VA's Affectiveness)                                                                     &   -                        & Please rank the VAs in order of their level of emotional expression in the discussion.       \\

Ranking (Preference)                                                                     &   -                        & Please rank the VAs as team members in the order of preference during the group discussion.
\\\midrule[0.3pt]\bottomrule[1pt]
\end{tabular}
}
\end{table}

\end{document}